\newcommand{\bpusc}[2]{\bar{u}^{#1}_{#2}}
\newcommand{\bpu}{\bar{u}}

\newcommand{\da}{{\dot\alpha}}
\newcommand{\db}{{\dot\beta}}

\newcommand{\be}{\begin{equation}}
\newcommand{\ee}{\end{equation}}
\newcommand{\bea}{\begin{eqnarray}}
\newcommand{\eaa}{\end{eqnarray}}

\newcommand{\genb}{g'}

\newcommand{\nn}{\nonumber}

\renewcommand{\a}{\alpha}

\newcommand{\pa}{\partial}

\renewcommand{\b}{\beta}

\newcommand{\la}{\lambda}

\newcommand{\q}{\theta}
\newcommand{\bq}{\bar\theta}
\newcommand{\bu}{\bar u}

\newcommand{\ep}{\epsilon}
\newcommand{\m}{\mu}
\newcommand{\n}{\nu}
\newcommand{\tz}{{\Upsilon}}

\newcommand{\cN}{{\cal N}}

\newcommand{\p}[1]{(\ref{#1})}
\newcommand{\bt}[1]{{\bar t}}
\newcommand{\ts}{\textstyle}

\newcommand{\half}{{\ts \frac{1}{2}}}

\newcommand{\N}{\mathcal{N}}

\newcommand{\FIz}[1]{\mathcal{F}_{#1}}
\newcommand{\FIOz}[1]{\mathcal{F}_{#1}^{\text{open}}}

\documentclass[12pt,a4paper]{article}
\input epsf
\usepackage{amsmath,amsfonts,epsfig,color,latexsym}
\usepackage{amssymb}
\usepackage[english, USenglish]{babel}
\usepackage{epsfig}
\usepackage{a4wide}
\usepackage{rotating}
\usepackage{slashed}
\csname @addtoreset\endcsname{equation}{section}
\author{\\[-0.3cm]\Large I.~Antoniadis\footnote{\tt ignatios.antoniadis@cern.ch}
\footnote{On leave from CPHT (UMR CNRS 7644) Ecole Polytechnique, F-91128 Palaiseau}~, S.~Hohenegger\footnote{{\tt stefanh@itp.phys.ethz.ch}}~, K.S.~Narain\footnote{{\tt narain@ictp.trieste.it}}~, E.~Sokatchev\footnote{{\tt emeri.sokatchev@cern.ch}}
}
\title{\begin{flushright}{\vspace{-0.8cm}\small CERN-PH-TH/2009-057\\[-0.2cm] LAPTH-1328/09}\end{flushright}
\vspace{0.8cm}
\bf{A New Class of $\N=2$ Topological Amplitudes}}
\dateUSenglish
\date{}
\graphicspath{{figures/}}
\begin{document}
\begin{titlepage}
%\title{\bf{A New Class of $\N=2$ Topological Amplitudes }}
%\dateUSenglish
%\date{\today}
%\graphicspath{{figures/}}

\maketitle
\begin{center}
\renewcommand{\thefootnote}{\fnsymbol{footnote}}\vspace{-0.5cm}
%\footnotemark[1]
\footnotemark[1]Department of Physics, CERN - Theory Division, CH-1211 Geneva 23, Switzerland\\[0.5cm]
\vspace{-0.3cm}
\footnotemark[3]Institut f\"ur Theoretische Physik, ETH
  Z\"urich, 
CH-8093 Z\"urich, Switzerland\\[0.5cm]
\vspace{-0.3cm}
\footnotemark[4]High Energy Section, The Abdus Salam International Center for Theoretical Physics,
\\Strada Costiera, 11-34014 Trieste, Italy\\[0.5cm]
\vspace{-0.3cm}\footnotemark[5] LAPTH\footnote[6]{Laboratoire d'Annecy-le-Vieux de Physique Th\'{e}orique, UMR 5108},   Universit\'{e} de Savoie, CNRS, 
B.P. 110,  F-74941 Annecy-le-Vieux, France\\[0.5cm]
\end{center}
\begin{abstract}
We describe a new class of $\N=2$ topological amplitudes that compute a particular class
of  BPS terms in the low energy effective supergravity action. Specifically they compute the coupling $F^2 (\la\la)^{g-2} (\partial \phi)^2$ where $F$, $\la$ and $\phi$ are gauge field strengths, gaugino
and holomorphic vector multiplet scalars. The novel feature of these terms is that they depend both on the
vector and hypermultiplet moduli. The BPS nature of these terms implies that they satisfy a holomorphicity condition with respect to vector moduli
and a harmonicity condition as well as a second order differential equation with respect to
hypermultiplet moduli. We study these conditions explicitly in heterotic string theory and show that
they are indeed satisfied up to anomalous boundary terms
in the world-sheet moduli space. We also analyze the boundary terms in the holomorphicity and harmonicity equations at a generic point in the vector and hyper moduli space. In
particular we show that the obstruction
to the
holomorphicity arises from the one loop threshold correction to the gauge
couplings and we argue that this
is due to the contribution of non-holomorphic couplings to the connected
graphs via elimination of the auxiliary fields.
\end{abstract}
\thispagestyle{empty}
\end{titlepage}

\tableofcontents
%\pagebreak

%%%%%%%%%%%%%%%%%%%%%%%%%%%%%%%%%%%%%%%%%%%%%%
\section{Intoduction}\label{Sect:Introduction}
A special role in extended supersymmetric theories is played by 1/2-BPS couplings that depend only on half of the superspace, generalizing chiral $\N=1$ supersymmetric F-terms. Usually, such interactions are easier to study because they are subject to non-renormalization theorems, while they have a variety of interesting physical applications varying from the vacuum structure all the way up to properties of supersymmetric black-holes. Moreover, in string effective field theory, these couplings are expected to be computed by topological amplitudes, depending only on the zero-mode structure of the compactification space~\cite{Witten:1992fb, Antoniadis:1993ze, Bershadsky:1993cx}. An interesting property is that the half-BPS structure of these terms is broken at the quantum level. On the topological side, this breaking is due to a violation in the conservation of the BRST current described by an anomaly equation~\cite{Bershadsky:1993cx, Cecotti:1992qh, Bershadsky:1993ta}, while on the string side it is understood from the difference between the Wilsonian and `physical' effective action that includes also the contribution of massless degrees of freedom~\cite{Antoniadis:1993ze}.

The first instance of well studied 1/2-BPS couplings in $\N=2$ supersymmetry is the series $F_g W^{2g}$, where $W$ is the chiral (self-dual) gravitational Weyl superfield and the coefficients $F_g$ depend on the vector multiplet moduli in the Coulomb phase of the theory~\cite{Antoniadis:1993ze, Bershadsky:1993cx}. $F_g$'s are computed by the genus $g$ topological partition function of an $\N=2$ twisted $\sigma$-model on the six-dimensional Calabi-Yau compactification manifold of type II string theory in four dimensions, subject to a holomorphic anomaly equation that takes the form of a recursion relation. Moreover, the independence of $F_g$'s from hypermultiplets, which include the string dilaton, implies a non-renormalization theorem for their form. These results have been generalized to $\N=4$ supersymmetric compactifications of type II string on $K3\times T^2$, where two series of higher order terms were identified, computed by topological amplitudes: $F^{(1)}_g{\bar K}^2K^{2g}$ and $F^{(3)}_{g-1}{ K}^{2g}$, where $K$ is a superdescendent of the $\N=4$ Weyl superfield~\cite{Antoniadis:2006mr, Antoniadis:2007ta}. The half-BPS property leads to a harmonicity equation for the moduli dependence of the couplings~\cite{Berkovits:1994vy, Ooguri:1995cp, Antoniadis:2007cw}, generalizing $\N=2$ holomorphicity, up to anomalous contributions from boundary terms~\cite{Antoniadis:2007cw}. Despite the bigger supersymmetry, the analysis is more involved than in the case of $\N=2$ vector multiplets, since the lack of an ordinary superspace description implies the use of on-shell harmonic superspace~\cite{Galperin:1984av, Galperin:1984bu, Hartwell:1994rp}.

A different question is to study the corresponding couplings when one reduces the supersymmetry by half. On the string side, this can be done in two ways that are dual to each other. Either by considering the `semi-topological' theory obtained by twisting the supersymmetric left-movers of the heterotic string~\cite{Antoniadis:1996qg, Beasley:2005iu}, or by applying a world-sheet involution on the type II amplitudes that introduces open string boundaries~\cite{Antoniadis:2005sd}. In the case of $F_g$'s, this generates an $\N=1$ series of higher order F-terms of the form $F_g^{N=1}{\cal W}^{2g}$, where ${\cal W}$ is now the gauge superfield with the gauge indices contracted in an appropriate way~\cite{Antoniadis:1996qg}. The holomorphic anomaly equation however does not close on $F_g^{N=1}$'s; it brings new objects that give rise to a double series $F_{g,n}^{N=1}{\cal W}^{2g}\Pi^n$, where $\Pi$ denotes generically a chiral projection of a real function of chiral superfields. On the topological side, the same results are obtained upon introducing world-sheet boundaries.\footnote{It would be interesting to understand the relation of the string effective action with the open topological amplitudes of ref.~\cite{Walcher:2007tp} which seem to avoid the appearance of new objects in the holomorphic anomaly equation.}

In this work, we apply the above reduction mechanism to the $\N=4$ topological amplitudes and obtain a new series of higher order 1/2-BPS terms with $\N=2$ supersymmetry. The novel feature of these terms is that they {\em mix} $\N=2$ vector multiplets with neutral hypermultiplets, despite the common wisdom. Indeed, starting with $F^{(3)}_g$, one generates the series ${\hat F}^{(3)}_{g-1}{\hat K}^{2g}$, where ${\hat K}$ is now a superdescendent of an $\N=2$ vector superfield (with the gauge indices contracted appropriately, as before). The coupling coefficients ${\hat F}^{(3)}_g$ depend in this case on both analytic vector multiplet as well as on (Grassmann analytic) hypermultiplet moduli, as dictated by the half-BPS structure. Moreover, these coupling share similar properties at the same time with the $\N=4$ topological couplings and with the $\N=1$ series. More precisely, the appropriate formalism for their study is again (on-shell) harmonic superspace, which complicates the analysis compared to the $\N=2$ $F_g$'s. On the other hand, quantum corrections violate both the holomorphicity condition with respect to the vector moduli, and the harmonicity with respect to the hypermultiplets. Furthermore, the anomaly equation does not closes on ${\hat F}^{(3)}_g$'s; it brings new objects generating the double series $P\left({\hat F}_{g,n}{\hat K}^{2(g-1)}{\hat{\bar{K}}}^{2(n-1)}\right)$, where $P$ is an appropriate $\N=2$ half-BPS projection.

The organization of the paper and the outline of the results obtained are described below. The next two sections contain the string computation of the new topological amplitudes. In Section~\ref{Sect:TypeIamp}, we compute the special type of $\N=2$ topological amplitudes ${\hat F}^{(3)}_g$ in type I open string theory, from the $\N=4$ topological amplitudes $F^{(3)}_g$, by applying a $\mathbb{Z}_2$ world-sheet involution.\footnote{For notational simplicity, we will drop in the text the hats introduced above, as well as the superscripts of the $\N=4$ topological amplitudes.} In fact, we evaluate a physical amplitude involving two gauge field strengths, two vector multiplet scalars (with one derivative each) and $2(g-1)$ gauginos with the same four-dimensional chirality, $F^2 (\la\la)^{g-1} (\partial \phi)^2$, on a world-sheet with $2(g+1)$ boundaries, and we show that it is reduced to a topological expression within the twisted $\sigma$-model on $K3\times T^2$. Then, in Section~\ref{Sect:hetF3}, we compute the same amplitudes on the heterotic side (compactified on $K3\times T^2$), which turns out to be easier for our subsequent analysis because of the absence of the problematic Ramond-Ramond sector, exploiting heterotic -- type I duality. Again, the physical amplitude is expressed as a semi-topological expression, i.e. only for the (supersymmetric) left-movers, while the bosonic part provides the gauge indices appropriately contracted (we are essentially taking products of differences of gauge groups with no charged massless matter).

These two sections are complemented by three appendices. In Appendix~\ref{append:SCA}, we review the main properties of the $\N=2$ and $\N=4$ world-sheet superconformal algebras, Appendix~\ref{App:VertexOperators} contains the expressions of the three main vertex operators we use, while Appendix~\ref{App:mathstruct} contains the definitions of the theta-functions and prime forms. 

The following section contain the effective field theory description of the topological amplitudes and the study of the generalized analyticity relations and anomaly equations. In Section~\ref{Sect:HarmonicDescription}, we study the interpretation of the string results, obtained in Sections~\ref{Sect:TypeIamp} and \ref{Sect:hetF3}, in the context of the effective supergravity. As mentioned above, the appropriate formulation is in terms of the $\N=2$ harmonic superspace (for a review see~\cite{Galperin:2001uw}). We first make an analysis in global supersymmetry (subsection~\ref{global}), introduce the $SU(2)$ harmonic variables, define the series of the effective interaction terms and derive the conditions on the moduli dependence of the couplings ${\hat F}_g$ from their half-BPS structure. These are the usual holomorphicity with respect to the vector multiplet moduli, while the hypermultiplet moduli dependence is subject to two differential constraints, in close analogy with the equations found for the $\N=4$ terms: the so-called harmonicity condition, expressing the property that only one combination of the four components of the hypermultiplets enter in the coupling, as well as a second-order constraint. We then study the effects of the curvature of the hypermultiplet scalar manifold (subsection~\ref{coset}), considering as an example the coset $SO(4,n)/SO(4)\times SO(n)$ for $n$ hypermultiplets (using the harmonic description of~\cite{Galperin:1992pj}). We show in particular that the second-order differential equation is modified by an additional term linear in ${\hat F}_g$ and proportional to a $U(1)$ R-charge $(g-1)$. The generalization to $\N=2$ (conformal) supergravity is done in subsection~\ref{css}, where the full covariantized expressions of the effective operators are obtained, as well as of the differential equations they obey.

In Section~\ref{App:RelTopPhys}, we present a different derivation of the equivalence between string and topological amplitudes which is free of an ambiguity that appears in the computation we perform in Sections~\ref{Sect:TypeIamp} and \ref{Sect:hetF3}. This is achieved by evaluation of a different amplitude related by supersymmetry to the previous one, containing only fermions: two chiral and two antichiral hyperfermions, besides the gauginos. We also generalize the computation from orbifolds considered in the text, to the most general $\N=4$ superconformal theory.

In Section~\ref{Sect:StringHarmonicity}, we verify explicitly the analyticity equations in string theory, on the heterotic side. Moreover, we evaluate the world-sheet boundary contributions for the holomorphicity and harmonicity equations that give rise to anomalous terms. In contrast to the familiar $\N=2$ $F_g$'s and their $\N=4$ generalizations computed by closed topological amplitudes, the anomalous terms do {\it not} generate recursion relations for the non holomorphic/harmonic dependence of the same couplings, because they involve new objects. This is similar to the case encountered in $\N=1$ topological amplitudes, irrespectively on which string framework they are defined (heterotic or type I). The new objects involve chiral/half-BPS projections of general non-holomorphic/harmonic functions and generate a double series of higher-dimensional operators with moduli-dependent coefficients ${\hat F}_{g,n}$, described above. In both equations, the new quantities are proportional to the one-loop threshold corrections to the gauge couplings, on the heterotic side. We argue that the
non-holomorphicity appears due to the contribution
to the string amplitude (which computes the sum of all connected graphs)
from $P\left({\hat F}_{g-1,1}{\hat K}^{2(g-2)}\right)$  via the
elimination of
the auxiliary fields. This section is supplemented by Appendix \ref{App:RelTopPhysD4}, where we explicitly compute the string amplitudes generating the double series described above in a generic Calabi-Yau compactification.
%%%%%%%%%%%%%%%%%%%%%%%%%%%%%%%%%%%%%%%%%%%%%%
\section{Type I open topological amplitudes}\label{Sect:TypeIamp}
In this section we will calculate a special type of topological amplitudes in type I open string theory. They are related to similar objects in the type II theory (see \cite{Antoniadis:2006mr,Antoniadis:2007cw,Antoniadis:2007ta}) via a $\mathbb{Z}_2$ world-sheet involution \cite{Bianchi:1988fr,Blau:1987pn} which we will describe in detail first.
%%%%%%%%%%%%%%%%%%%%%%%%%%%%%%%%%%%%%%%%%%%
\subsection{$\mathbb{Z}_2$ world-sheet involutions}\label{Sect:Involution}
\begin{figure}[htb]
\begin{center}
\epsfig{file=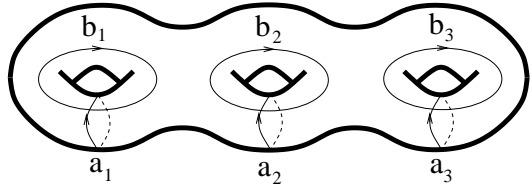, width=7cm}
\caption{Compact Riemann surface of genus $g=3$ with canonical basis of homology cycles $(\mathbf{a_i},\mathbf{b_i})$.}
\label{fig:compRiem}
\end{center}
\end{figure}

In the type II theory, the world-sheet corresponding to a $g$ loop scattering amplitude is a compact Riemann surface $\Sigma_g$ of genus $g$. This surface can be endowed with a canonical homology basis of 1-cycles $(\mathbf{a}_i,\mathbf{b}_i)$, with $i=1,\ldots,g$ (an example for $g=3$ is depicted in figure \ref{fig:compRiem}).
The surface can furthermore be equipped with a set of $g$ holomorphic 1-differentials $\omega_i$, whose integrals over the homology cycles is given by
\begin{align}
&\int_{\textbf{a}_i}\omega_j=\delta_{ij},&&\text{and} &&\int_{\textbf{b}_i}\omega_j=\tau_{ij}\,.\label{periodmatrix}
\end{align}
Here the symmetric $g\times g$ matrix $\tau_{ij}$ is called the period matrix and it encodes all the information about the shape and size of the surface $\Sigma_g$.\\

\begin{figure}[htbp]
\begin{center}
\epsfig{file=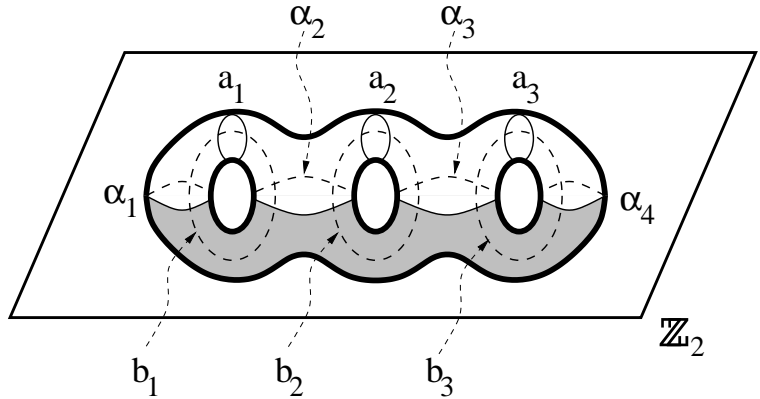, width=10cm}
\caption{$\mathbb{Z}_2$ involution of a genus 3 Riemann surface leading to an open surface with four boundaries $\alpha_I$.}
\label{fig:twoinvolutions1}
\end{center}
\end{figure}

\noindent
By viewing $\Sigma_g$ as a double cover we can construct an open surface by taking the quotient with respect to some involution which we will denote $I$ in the following. $I$ acts linearly on the homology cycles and we will focus on the special case
\begin{align}
&I_*\mathbf{a}_i=\Gamma_{ij}\mathbf{a}_j\,, &&\text{and} &&I_*\mathbf{b}_i=-\Gamma_{ij}\mathbf{b}_j\ \,.\label{identificationcycle}
\end{align}
Here $\Gamma$ is a matrix that enjoys the following properties
\begin{align}
&\Gamma^2=1\!\!1\,, &&\text{and} &&\text{det}\Gamma=\pm 1\,.
\end{align}
The action of $I$ on the $\omega$-differentials reads
\begin{align}
&I^*\omega_i=\Gamma_{ij}\bar{\omega}_j\,,
\end{align}
and the period matrix has to satisfy
\begin{align}
\tau=-\Gamma^T\bar{\tau}\Gamma\,.
\end{align}
The quotient $\Sigma_g/I$ constructed from the prescription (\ref{identificationcycle}) is an open Riemann surface and the boundaries are given by the fixed points of $I$. The case which will be most important for us in the following is to choose $I$ in such a way to create as large a number of boundaries as possible (see also \cite{Antoniadis:2005sd}), which is obviously
\begin{align}
\Gamma=1\!\!1\,.
\end{align}
In this way the boundaries are given by (combinations of) the $\mathbf{a}$-cycles of the original Riemann surface $\Sigma_g$ (Returning to the genus $g=3$ example the involution then acts as displayed in figure \ref{fig:twoinvolutions1}, creating a surface with 4 boundaries).
${}$\\

\noindent
In order to calculate (open)string correlation functions on the quotient $\Sigma_g/I$ we still need to specify boundary conditions which are necessarily the same for all boundary components. In this work we will focus on the simplest case and choose either Dirichlet or Neumann conditions, in which case the open correlators on $\Sigma_g/I$ are the square root of the closed correlators on $\Sigma_g$ up to the following multiplicative correction factor
\begin{align}
R=\left\{\begin{array}{ccl}\text{det}(\text{Im}\tau) &\ldots& \text{Dirichlet conditions} \\ \frac{1}{\text{det}(\text{Im}\tau)} &\ldots& \text{Neumann conditions}\end{array}\right.\ .\label{boundarycorrectionffactor}
\end{align}
%%%%%%%%%%%%%%%%%%%%%%%%%%%%%%%%%%%%%%%%%%%%%%%%%%%%%%%%%%%%%%%%%
\subsection{Involution of $\N=4$ topological amplitudes}\label{Sect:TypeIAmplitudes}
The techniques described in the previous subsection have been used in \cite{Antoniadis:2005sd} to compute open topological amplitudes in $\N=1$ type I string theory as involutions of the familiar $\N=2$ topological amplitudes $F_g$ (see \cite{Antoniadis:1993ze,Bershadsky:1993cx}). In this work, we wish to generalize the computations of \cite{Antoniadis:2005sd} and apply the method of $\mathbb{Z}_2$ world-sheet involutions to the $\N=4$ topological amplitudes of type II string theory on $K3\times T^2$ studied in \cite{Antoniadis:2006mr,Antoniadis:2007cw} (see also \cite{Antoniadis:2007ta}). We hope to find open topological amplitudes in type I string theory preserving $\N=2$ space-time supersymmetry.\\

\noindent
To be more precise, there are two different types of topological amplitudes in the $\N=4$ theory. In this work we will focus exclusively on the involution of one of them, which was called $\mathcal{F}_g^{(3)}$ in \cite{Antoniadis:2006mr}.\footnote{In order to save writing we will call them simply $\FIz{g}$ in this work.} We recall that $\FIz{g}$ was shown in \cite{Antoniadis:2006mr} to be computed by a $g$-loop type~II string amplitude with two graviton, two graviscalar and $2g-4$ graviphoton insertions. Starting again from the corresponding genus $g$ (closed) world-sheet $\Sigma_g$, the quotient $\Sigma_{g}/I$ has $(g+1)$ boundaries. In the process of calculating this involution we will consider the special case that none of the vertex operators is inserted in the bulk but all of them will be pairwise distributed over the $g+1$ boundary components.\footnote{In fact one boundary will stay ``empty`` in the sense that no vertex operator will be inserted.} Moreover, we will choose Neumann boundary conditions for all the components. In this way we will compute a type~I correlator of two gauge fields, two boundary scalars and $2g-4$ gauginos. We will consider a torus-orbifold realization of $K3$ in which case we can use a free-field representation for all vertex operators. The precise helicity combinations can then be displayed in the following table (the $\phi$ in the last five columns denote the $U(1)$ charges in the various $\mathbb{C}_2$ planes)
\begin{center}
\begin{tabular}{|c|c|c||c|c||c|c|c|}\hline
\textbf{field} & \textbf{pos.} & \textbf{number} & \parbox{0.5cm}{\vspace{0.2cm}$\phi_1$\vspace{0.2cm}}& \parbox{0.5cm}{\vspace{0.2cm}$\phi_2$\vspace{0.2cm}} & \parbox{0.5cm}{\vspace{0.2cm}$\phi_3$\vspace{0.2cm}} & \parbox{0.5cm}{\vspace{0.2cm}$\phi_4$\vspace{0.2cm}} & \parbox{0.5cm}{\vspace{0.2cm}$\phi_5$\vspace{0.2cm}} \\\hline
gauge field & \parbox{0.35cm}{\vspace{0.2cm}$z_1$\vspace{0.2cm}} & 1 & $+1$ & $+1$ & $0$ & $0$ & $0$ \\\hline
& \parbox{0.35cm}{\vspace{0.2cm}$z_2$\vspace{0.2cm}} & 1 & $-1$ & $-1$ & $0$ & $0$ & $0$ \\\hline
scalar & \parbox{0.35cm}{\vspace{0.2cm}$z_3$\vspace{0.2cm}} & 1 & $0$ & $+1$ & $+1$ & $0$ & $0$ \\\hline
& \parbox{0.35cm}{\vspace{0.2cm}$z_4$\vspace{0.2cm}} & $1$ & $0$ & $-1$ & $+1$ & $0$ & $0$ \\\hline\hline
gaugino & \parbox{0.35cm}{\vspace{0.2cm}$x_i$\vspace{0.2cm}} & $g-2$ & \parbox{0.7cm}
{\vspace{0.2cm}$+\frac{1}{2}$\vspace{0.2cm}} & \parbox{0.7cm}
{\vspace{0.2cm}$+\frac{1}{2}$\vspace{0.2cm}} & \parbox{0.7cm}
{\vspace{0.2cm}$+\frac{1}{2}$\vspace{0.2cm}} & \parbox{0.7cm}
{\vspace{0.2cm}$+\frac{1}{2}$\vspace{0.2cm}} & \parbox{0.7cm}
{\vspace{0.2cm}$+\frac{1}{2}$\vspace{0.2cm}} \\\hline
 & \parbox{0.35cm}{\vspace{0.2cm}$y_i$\vspace{0.2cm}} & $g-2$ & \parbox{0.7cm}
{\vspace{0.2cm}$-\frac{1}{2}$\vspace{0.2cm}} & \parbox{0.7cm}
{\vspace{0.2cm}$-\frac{1}{2}$\vspace{0.2cm}} & \parbox{0.7cm}
{\vspace{0.2cm}$+\frac{1}{2}$\vspace{0.2cm}} & \parbox{0.7cm}
{\vspace{0.2cm}$+\frac{1}{2}$\vspace{0.2cm}} & \parbox{0.7cm}
{\vspace{0.2cm}$+\frac{1}{2}$\vspace{0.2cm}} \\\hline\hline
PCO & \parbox{0.77cm}{\vspace{0.2cm}$\{s_3\}$\vspace{0.2cm}} & $g$ & $0$ & $0$ & $-1$ & $0$ & $0$ \\\hline
& \parbox{0.77cm}{\vspace{0.2cm}$\{s_4\}$\vspace{0.2cm}} & $g-2$ & $0$ & $0$ & $0$ & $-1$ & $0$ \\\hline
& \parbox{0.77cm}{\vspace{0.2cm}$\{s_5\}$\vspace{0.2cm}} & $g-2$ & $0$ & $0$ & $0$ & $0$ & $-1$ \\\hline
\end{tabular}
\end{center}
${}$\\[10pt]
Here it is understood that the PCO are also inserted on the boundary. This amplitude is precisely equal to the left-moving contribution of the corresponding type~II correlator, which has already been computed in \cite{Antoniadis:2006mr}. Therefore, instead of repeating the calculation again we will only state the final result
\begin{align}
\FIOz{g}=&\frac{\langle\prod_{a}^{\{s_3\}}\bar{\psi}_3\partial X_3(r_a)\psi_3(\alpha)\rangle\cdot\langle\prod_{a}^{\{s_4\}}\bar{\psi}_4\partial X_4(r_a)\bar{\psi}_4(z_4)\rangle\cdot\langle\prod_{a}^{\{s_5\}}\bar{\psi}_5\partial X_5(r_a)\bar{\psi}_5(z_4)\rangle}{\langle\prod_{a=1}^{3\genb-4}b(r_a)b(z_4)\rangle}\cdot\nonumber\\
&\cdot \text{det}\omega_i(x_i,z_1,z_3)\text{det}\omega_i(y_i,z_2,z_4)\ .\label{PhysAmplitudeOpen}
\end{align}
Here $\alpha$ is an arbitrary position on the boundaries of the world-sheet. The correlator in the denominator stems from the $bc$-ghost system while the correlators of the numerator involve the free fermions $\psi_3$ and bosons $X_3$ living on $T^2$ and their respective counterparts $\psi_{4,5}$ and $X_{4,5}$ coming from $K3$. Notice that no factors of $\text{det}(\text{Im}\tau)$ appear in this expression since they have all been cancelled by the correction factor (\ref{boundarycorrectionffactor}) for Neumann boundary conditions. At this stage we can use the free-field representation of the $\N=2$ and $\N=4$ super-conformal algebra (see Appendix \ref{append:SCA}) to write
\begin{align}
\FIOz{g}=\text{det}\omega_i(x_i,z_1,z_3)\text{det}\omega_i(y_i,z_2,z_4)\, A(r_a,z_4)\,,\label{intermediateopen}
\end{align}
where we have introduced the following shorthand notation
\begin{align}
A(r_a,z_4)=\frac{\langle\prod_{a}^{\{s_3\}}G^-_{T^2}(r_a)\psi_3(\alpha)\rangle\cdot\langle\prod_{a}^{\{s_4,s_5\}}G^-_{K3}(r_a)J^{--}_{K3}(z_4)\rangle}{\langle\prod_{a=1}^{3g-4}b(r_a)b(z_4)\rangle}\,.\label{ExpressAOpen}
\end{align}
Notice that  $G^-_{T^2}$ and $G^-_{K3}$ are the supercurrents and $J^{--}_{K3}$ the $SU(2)$ Kac-Moody current in the {\em twisted} internal theory and therefore are dimension 2 operators. Thus $A(r_a, z_4)$ is a meromorphic scalar function of all its arguments. As two $r_a$ approach each other both the numerator and the denominator have a first order zero and hence $A$ has no singularity in this limit. However, when $z_4$ approaches any $r_a$ of the set $\{s_3\}$ the numerator is finite and non-zero (as the first correlator is independent of $z_4$), while the denominator vanishes. This means that $A$ has a pole as $z_4$ approaches any of the $r_a$ of the set $\{s_3\}$. On the other hand this singularity should not be present as picture changing operators sitting at $r_a$ must have no singularity with the physical vertex operator at $z_4$. The reason for this apparent singularity is that we have not included the full physical vertex operator at $z_4$ (as well as at $z_1$, $z_2$ and $z_3$). We have only considered the fermion bilinear part of the physical vertex operators in the zero ghost picture that comes with one power of momenta. However, the full vertex operators  also include  $\partial X^{\mu}$ at $z_1$ and $z_2$ and $\partial X_3$ at $z_3$ and $z_4$ (see Appendix \ref{App:VertexOperators}). If we include all these extra terms the apparent singularity in $A$ as $z_4 $ approaches any of the $r_a$ must disappear. This subtle point was overlooked in \cite{Antoniadis:2006mr}. It is however very difficult to include all the possible terms, indeed it
is not clear if we can choose a gauge condition for the positions of the
PCO's such that the superghost $\vartheta$ function cancels with one of the
space-time $\vartheta$ functions simultaneously for all these different terms,
which could enable us to do the spin structure sum explicitly. This is a
problem in the RNS formulation that we are using here. 

In Section~\ref{App:RelTopPhys}, we show that one can compute another amplitude in
the RNS formulation, involving $2g$ gauginos and two chiral and two anti-chiral hyperfermions. From the discussion of the effective field theory in Section~\ref{Sect:HarmonicDescription} it will become clear that this new amplitude is supersymmetrically related to the one considered
here, more precisely the new amplitude is given by four derivatives of the
latter with respect to hypermultiplet moduli. Moreover, it
turns out that in this new amplitude,  all the PCO's contribute only the
supercurrents of the internal theory, which will allow us to compute it explicitly in Section \ref{App:RelTopPhys} and proof that it is indeed given by four derivatives of the
topological expression given below in eq.~(\ref{top}). It will be
interesting to see if the amplitude considered in the current section can be
directly calculated in some other formalism, such as pure spinor
formalism. In the following we assume that including all the other
terms in the vertex operators amounts to antisymmetrizing $z_4$ with
all the $r_a$ in the numerator of  eq.(\ref{ExpressAOpen}).

Next we remember that this expression is still to be multiplied by $3g-3$ Beltrami differentials folded with the $b$-ghosts to provide the correct measure for the integration over the moduli space $\mathcal{M}_{(0,g+1)}$ of a Riemann surface with $g+1$ boundaries and no handles. However, with our above assumption it is possible to transmute all $r_a$ and $z_4$ inside the expression $A(r_a,z_4)$ of (\ref{ExpressAOpen}) to the positions of these ghost fields folded with the Beltrami differentials. Put differently, we can write
\begin{align}
A(r_a,z_4)(\mu\cdot b)^{3g-3}=(\mu\cdot G^-_{T^2})^{g}(\mu\cdot G^-_{K3})^{2g-4}(\mu\cdot J^{--}_{K3})\psi_3(\alpha)\,,
\label{top}
\end{align}
where $\mu = \sum_{a=1}^{3g-3} dm^a \mu_a$ with $m^a$ being some local coordinates in the moduli space $\mathcal{M}_{(0,g+1)}$ and $\mu_a$ the corresponding Beltrami differentials. Besides that we have introduced the following notation
\begin{align}
(\mu\cdot G^-)=\sum_{a=1}^{3g-3} dm^a \int_{\Sigma_{(0,g+1)}} \mu_a G^-\,,\label{TopNotat}
\end{align}
which is therefore a one form in $\mathcal{M}_{(0,g+1)}$. The final step is the integration of the insertion points $x_i$, $y_i$, $z_{1,2,3,4}$ over the boundaries. This can be performed explicitly and yields, given the fact that the boundary components are just $\mathbf{a}$-cycles on which the $\omega$ are normalized, just a numerical factor, which we drop since it will be of no interest to us. Therefore we can write for the final amplitude
\begin{align}
\FIOz{g}=\int_{\mathcal{M}_{(0,g+1)}}\langle (\mu\cdot G_{T^2}^-)^g(\mu\cdot G_{K3}^-)^{2g-4}(\mu\cdot J^{--}_{K3})\psi_3(\alpha)\rangle\ .
\end{align}
Notice that this expression is purely topological holding only information about the number of boundaries of the world-sheet. We have therefore succeeded in linking the physical amplitude (\ref{PhysAmplitudeOpen}) to a topological theory.
%%%%%%%%%%%%%%%%%%%%%%%%%%%%%%%%%%%%%%%%%%%%%%%%%%%%%%%%%%%%%%%%
%%%%%%%%%%%%%%%%%%%%%%%%%%%%%%%%%%%%%%%%%%%%%%%%%%%%%%%%%%%%%%%%

%%%%%%%%%%%%%%%%%%%%%%%%%%%%%%%%%%%%%%%%%%%%%%%%%%%%%%%%%%%%%%%%%%%%%%%%%%%%%%%%
%%%%%%%%%%%%%%%%%%%%%%%%%%%%%%%%%%%%%%%%%%%%%%%%%%%%%%%%%%%%%%%%%%%%%%%%%%%%%%%%
\section{Topological amplitudes in heterotic orbifold compactifications}\label{Sect:hetF3}
In Section \ref{Sect:TypeIamp} we have been considering topological amplitudes in the type~I theory. However, in order not having to deal with the problem of open string moduli (see e.g. \cite{Walcher:2007tp}), we rather prefer to transfer the problem to a dual setup, in which the topological amplitudes again compute closed string correlators. One possibility is to exploit the duality between type~I and heterotic string theory. Since this duality is perturbative in nature we expect to recover  $\mathcal{F}_{g+1}^{\text{open}}$ of the type~I theory at the (closed) $g$ loop level in the heterotic theory. Since in the heterotic theory the bosonic right moving sector needs a slightly different treatment than the supersymmetric left moving sector, we will present the computation of this amplitude in somewhat more detail.

The field insertions we consider for the heterotic $g$-loop correlator are two gauge fields, two scalar fields and $(2g-4)$ gauginos. The helicity setup we use is identical to the type I setup, and therefore, upon using the vertex operators of Appendix \ref{App:VertexOperators}, the amplitude, which we have to compute is given by
\begin{align}
&\mathcal{F}_{g}=\nonumber\\
&=\langle\prod_{i=1}^{g-2}e^{-\frac{\varphi}{2}}e^{\frac{i}{2}(\phi_1+\phi_2+\phi_3+\phi_4+\phi_5)}(x_i)\prod_{i=1}^{g-2}e^{-\frac{\varphi}{2}}e^{\frac{i}{2}(-\phi_1-\phi_2+\phi_3+\phi_4+\phi_5)}(y_i)e^{i(\phi_1+\phi_2)}(z_1)e^{-i(\phi_1+\phi_2)}(z_2)\cdot\nonumber\\
&\cdot e^{i(\phi_2+\phi_3)}(z_3)e^{-i(\phi_2-\phi_3)}(z_4)\prod_a^{\{s_3\}}e^{\varphi}e^{-i\phi_3}\partial X_3(r_a)\prod_a^{\{s_4\}}e^{\varphi}e^{-i\phi_4}\partial X_4(r_a)\prod_a^{\{s_5\}}e^{\varphi}e^{-i\phi_5}\partial X_5(r_a)\rangle\cdot\nonumber\\
&\cdot\langle\prod_i\bar{J}_{I_i}(\bar{x}_i)\bar{J}_{K_i}(\bar{y}_i)\bar{J}_{I_{g-1}}(\bar{z}_1)\bar{J}_{K_{g-1}}(\bar{z}_2)\bar{J}_{I_g}(\bar{z}_3)\bar{J}_{K_g}(\bar{z}_4)\rangle\,.
\end{align}
The right moving contribution consists just of a correlator of currents, where the subscripts $I_a$ and $K_a$ with $a=1,...,g$ label the vector multiplets. These subscripts are also implicitly present on $\FIz{g}$, but we shall not write them explicitly. We will leave this right moving correlator for the moment as it is and stick to the left moving part. Here we can perform the contractions of the various fields to obtain
{\allowdisplaybreaks
\begin{align}
&\mathcal{F}_{g}=\nonumber\\
&=\frac{\vartheta_s\!\!\left(\frac{1}{2}\sum_{i}(x_i-y_i)+z_1-z_2\right)\vartheta_s\!\!\left(\frac{1}{2}\sum_{i}(x_i-y_i)+z_1-z_2+z_3-z_4\right)}{\vartheta_s\!\!\left(\frac{1}{2}\sum_{i}(x_i+y_i)-\sum_a^{3g-4}r_a+2\Delta\right)\prod_{a<b}^{3g-4}E(r_a,r_b)\prod_a^{3g-4}\!\sigma^2(r_a)\prod_i E(x_i,z_2)E(y_i,z_1)}\cdot\nonumber\\
&\cdot \frac{\vartheta_s\!\!\left(\frac{1}{2}\sum_{i}(x_i+y_i)+z_3+z_4-\sum_a^{\{s_3\}}r_a\right)\prod_{I=4}^5\vartheta_{s,h_I}\!\!\left(\frac{1}{2}\sum_{i}(x_i+y_i)-\sum_a^{\{s_I\}}r_a\right)}{E^2(z_1,z_2)E(z_1,z_4)E(z_2,z_3)}\cdot\nonumber\\
&\cdot \prod_{i<j}E(x_i,x_j)E(y_i,y_j)\prod_iE(x_i,z_1)E(x_i,z_3)E(y_i,z_2)E(y_i,z_4)E(z_1,z_3)E(z_2,z_4)\cdot\nonumber\\
&\cdot \prod_i\sigma(x_i)\sigma(y_i)\prod_{a<b}^{\{s_3\}}E(r_a,r_b)\prod_{a<b}^{\{s_4\}}E(r_a,r_b)\prod_{a<b}^{\{s_5\}}E(r_a,r_b)\prod_a^{\{s_3\}}\partial X_3(r_a)\prod_a^{\{s_4\}}\partial X_4(r_a)\prod_a^{\{s_5\}}\partial X_5(r_a)\cdot\nonumber\\
&\cdot\langle\prod_i\bar{J}_{I_i}(\bar{x}_i)\bar{J}_{K_i}(\bar{y}_i)\bar{J}_{I_{g-1}}(\bar{z}_1)\bar{J}_{K_{g-1}}(\bar{z}_2)\bar{J}_{I_g}(\bar{z}_3)\bar{J}_{K_g}(\bar{z}_4)\rangle\,.\nonumber
\end{align}}
At this stage we can use the gauge-fixing condition
\begin{align}
&\frac{1}{2}\sum_i(x_i-y_i)+z_1-z_2=\frac{1}{2}\sum_i(x_i+y_i)-\sum_a^{3g-4}r_a+2\Delta\,,\nonumber\\
&\Rightarrow \sum_a^{3g-4}r_a=\sum_iy_i-z_1+z_2+2\Delta\,,
\end{align}
which reduces the relevant part for the spin-structure sum to
\begin{align}
&\vartheta_s\!\!\left(\frac{1}{2}\sum_{i}(x_i-y_i)+z_1-z_2+z_3-z_4\right)\vartheta_s\!\!\left(\frac{1}{2}\sum_{i}(x_i+y_i)+z_3+z_4-\sum_a^{\{s_3\}}r_a\right)\cdot\nonumber\\
&\cdot \prod_{I=4}^5\vartheta_{s,h_I}\!\!\left(\frac{1}{2}\sum_{i}(x_i+y_i)-\sum_a^{\{s_I\}}r_a\right)\,.\nonumber
\end{align}
By Riemann addition theorem, the arguments of the summed functions read
{\allowdisplaybreaks
\begin{align}
&\bullet\ ++++:\ \sum_{i}x_i+\frac{1}{2}\sum_iy_i+z_3+\frac{1}{2}(z_1-z_2)-\frac{1}{2}\sum_a^{3g-4}r_a=\nonumber\\*
&\hspace{3cm}=\sum_{i}x_i+z_1+z_3-z_2-\Delta\,,\nonumber\\
&\nonumber\\
&\bullet\ --++:\ \frac{1}{2}\sum_iy_i-z_3-\frac{1}{2}(z_1-z_2)+\frac{1}{2}\sum_a^{\{s_3\}}r_a-\frac{1}{2}\sum_a^{\{s_4,s_5\}}r_a=\sum_a^{\{s_3\}}r_a-z_3-\Delta\,,\nonumber\\
&\nonumber\\
&\bullet\ -+-+:\ \frac{1}{2}\sum_iy_i+z_4-\frac{1}{2}(z_1-z_2)+\frac{1}{2}\sum_a^{\{s_4\}}r_a-\frac{1}{2}\sum_a^{\{s_3,s_5\}}r_a=\sum_a^{\{s_4\}}r_a+z_4-\Delta\,,\nonumber\\
&\nonumber\\
&\bullet\ -++-:\ \frac{1}{2}\sum_iy_i+z_4-\frac{1}{2}(z_1-z_2)+\frac{1}{2}\sum_a^{\{s_5\}}r_a-\frac{1}{2}\sum_a^{\{s_3,s_4\}}r_a=\sum_a^{\{s_5\}}r_a+z_4-\Delta\,.\nonumber
\end{align}}
Multiplying the amplitude furthermore by
\begin{align}
1=\frac{\vartheta\!\!\left(\sum_iy_i+z_2+z_4-z_1-\Delta\right)}{\vartheta\!\!\left(\sum_a^{3g-4}r_a+z_4-3\Delta\right)}\,,
\end{align}
it takes the form
\begin{align}
&\mathcal{F}_{g}=\frac{\vartheta\!\!\left(\sum_{i}x_i+z_1+z_3-z_2-\Delta\right)\vartheta\!\!\left(\sum_iy_i+z_2+z_4-z_1-\Delta\right)\vartheta\!\!\left(\sum_a^{\{s_3\}}r_a-z_3-\Delta\right)}{\vartheta\!\!\left(\sum_a^{3g-4}r_a+z_4-3\Delta\right)\prod_{a<b}^{3g-4}E(r_a,r_b)\prod_a^{3g-4}\!\sigma^2(r_a)\prod_iE(x_i,z_2)E(y_i,z_1)}\cdot\nonumber\\
&\cdot\frac{\vartheta_{-h_4}\!\!\left(\sum_a^{\{s_4\}}r_a+z_4-\Delta\right)\vartheta_{-h_5}\!\!\left(\sum_a^{\{s_5\}}r_a+z_4-\Delta\right)\prod_{i<j}E(x_i,x_j)E(y_i,y_j)\prod_i\sigma(x_i)\sigma(y_i)}{E^2(z_1,z_2)E(z_1,z_4)E(z_2,z_3)}\cdot\nonumber\\
&\cdot \prod_iE(x_i,z_1)E(x_i,z_3)E(y_i,z_2)E(y_i,z_4)E(z_1,z_3)E(z_2,z_4)\prod_{a<b}^{\{s_3\}}E(r_a,r_b)\prod_{a<b}^{\{s_4\}}E(r_a,r_b)\cdot\nonumber\\
&\cdot\prod_{a<b}^{\{s_5\}}E(r_a,r_b)\prod_a^{\{s_3\}}\partial X_3(r_a)\prod_a^{\{s_4\}}\partial X_4(r_a)\prod_a^{\{s_5\}}\partial X_5(r_a)\cdot\nonumber\\
&\cdot\langle\prod_i\bar{J}_{I_i}(\bar{x}_i)\bar{J}_{K_i}(\bar{y}_i)\bar{J}_{I_{g-1}}(\bar{z}_1)\bar{J}_{K_{g-1}}(\bar{z}_2)\bar{J}_{I_g}(\bar{z}_3)\bar{J}_{K_g}(\bar{z}_4)\rangle\,.\nonumber
\end{align}
Here we use bosonization identities \cite{Verlinde:1986kw} in the following way
\begin{align}
&\frac{\vartheta\!\!\left(\sum_{i}x_i+z_1+z_3-z_2-\Delta\right)\prod_{i<j}E(x_i,x_j)\prod_iE(x_i,z_1)E(x_i,z_3)E(z_1,z_3)}{\prod_iE(x_i,z_2)E(z_1,z_2)E(z_2,z_3)\sigma(z_2)}\cdot\nonumber\\
&\hspace{1cm}\cdot\prod_i\sigma(x_i)\sigma(z_1)\sigma(z_3)=\text{det}\omega_i(x_j,z_1,z_3)\,,\\
&\frac{\vartheta\!\!\left(\sum_{i}y_i+z_2+z_4-z_1-\Delta\right)\prod_{i<j}E(y_i,y_j)\prod_iE(y_i,z_2)E(y_i,z_4)E(z_2,z_4)}{\prod_iE(y_i,z_1)E(z_1,z_2)E(z_4,z_1)\sigma(z_1)}\cdot\nonumber\\
&\hspace{1cm}\cdot\prod_i\sigma(y_i)\sigma(z_2)\sigma(z_4)=\text{det}\omega_i(y_j,z_2,z_4)
\end{align}
and write the remaining expression in terms of correlators of the internal theory to get
\begin{align}
\mathcal{F}_{g}=&\frac{\text{det}\omega_i(x_i,z_1,z_3)\text{det}\omega_i(y_i,z_2,z_4)\langle\prod_a^{\{s_3\}}\bar{\psi}_3\partial X_3(r_a)\psi_3(x)\rangle\langle\prod_a^{\{s_4\}}\bar{\psi}_4\partial X_4(r_a)\bar{\psi}_4(z_4)\rangle}{(\text{det}(\text{Im}\tau))^2\langle\prod_a^{3g-4}b(r_a)b(z_4)\rangle}\cdot\nonumber\\
&\cdot \langle\prod_a^{\{s_5\}}\bar{\psi}_5\partial X_5(r_a)\bar{\psi}_5(z_4)\rangle\cdot\langle\prod_i\bar{J}_{I_i}(\bar{x}_i)\bar{J}_{K_i}(\bar{y}_i)\bar{J}_{I_{g-1}}(\bar{z}_1)\bar{J}_{K_{g-1}}(\bar{z}_2)\bar{J}_{I_g}(\bar{z}_3)\bar{J}_{K_g}(\bar{z}_4)\rangle\,.\nonumber
\end{align}
Here we have split off $\psi_3$ which has dimension zero and was originally at $z_3$ and moved it to some arbitrary point $x$ since $\psi_3$ only provides a constant zero mode. Including all the possible distributions of the positions $r_a$ in the sets $\{s_3\}$, $\{s_4\}$ and $\{s_5\}$ we
find:
\begin{align}
\mathcal{F}_{g}=&\frac{\text{det}\omega_i(x_i,z_1,z_3)\text{det}\omega_i(y_i,z_2,z_4)}{(\text{det}(\text{Im}\tau))^2}\cdot A(r_a, z_4)\cdot\nonumber\\
&\cdot\langle\prod_i\bar{J}_{I_i}(\bar{x}_i)\bar{J}_{K_i}(\bar{y}_i)\bar{J}_{I_{g-1}}(\bar{z}_1)\bar{J}_{K_{g-1}}(\bar{z}_2)\bar{J}_{I_g}(\bar{z}_3)\bar{J}_{K_g}(\bar{z}_4)\rangle\,,
\end{align}
where
\begin{equation}
A(r_a, z_4) = \frac{\langle \prod_{a=1}^{3g-4}G^-(r_a) J^{--}_{K3} (z_4) \psi_3(x)\rangle}{\langle \prod_{a=1}^{3g-4} b(r_a) b(z_4)\rangle}\,.
\end{equation}
Here  $G^-$ are the supercurrents and $J^{--}_{K3}$ the $SU(2)$ current in the twisted internal theory and therefore
are dimension 2 operators. Similar to the expression of (\ref{ExpressAOpen}) also here in the heterotic case $A(r_a, z_4)$ is a meromorphic scalar function of all its arguments which develops a pole when $z_a$ approaches any of the $r_a$. In this case the numerator is finite and non-zero when the corresponding $G^-$ contributes the torus part $G^-_{T^2}$ while, however, the denominator vanishes. This problem is again due to the fact that we have not included all the other possible terms in vertex operators. However, this can be resolved in precisely the same manner as in the type~I case (see also Section~\ref{App:RelTopPhys}). As in the open string case we assume that including all the other
terms in the vertex operators amounts to an antisymmetrization of
$z_4$ with all the $r_a$ in the numerator of the above equation
which then cancels the zero coming from the denominator as $z_4$
approaches any of the $r_a$.

The remainder of the argument also follows similarly to the computation in Section \ref{Sect:TypeIAmplitudes}. We note that as a function of any $r_a$ or $z_4$ both the numerator as well as the denominator in $A$ are sections of the line bundle of quadratic differentials and have no poles or zeroes at the remaining $3g-4$ points. Both the numerator and denominator must have $g$ additional zeroes as the degree of the divisor class of quadratic differentials is $4g-4$ but by the Abel map generically these g
additional zeroes are uniquely fixed. This implies that $A$ has no zero or pole as a function of $r_a$ and $z_4$.
Therefore $A$ must be a constant since $A$ is a meromorphic scalar function of its arguments and as a result
\begin{equation}
A(r_a, z_4) (\mu\cdot b)^{3g-3} = (\mu\cdot G^-)^{3g-4} (\mu\cdot J^{--}_{K3})\psi_3(x)\,,
\end{equation}
where we have used the same notation as in (\ref{TopNotat}). We therefore find
\begin{align}
\mathcal{F}_{g}=&\langle(\mu\cdot G^-)^{3g-4}(\mu\cdot J^{--}_{K3})\psi_3(x)
\rangle\cdot\frac{\text{det}\omega_i(x_i,z_1,z_3)\text{det}
\omega_i(y_i,z_2,z_4)}{(\text{det}(\text{Im}\tau))^2}\cdot\nonumber\\
&\cdot\langle\prod_i\bar{J}_{I_i}(\bar{x}_i)\bar{J}_{K_i}(\bar{y}_i)\bar{J}_{I_{g-1}}(\bar{z}_1)\bar{J}_{K_{g-1}}(\bar{z}_2)\bar{J}_{I_g}(\bar{z}_3)\bar{J}_{K_g}(\bar{z}_4)\rangle\,.
\end{align}
In principle, the right moving correlator would contribute arbitrary contractions of the non-abelian currents. In order to simplify our computation we consider differences of various gauge groups, which precisely cancels all contractions among the currents. The contribution we then get from each current is
\begin{equation}
\bar{J}_I\sim \bar{\omega}_i Q^i_I\,.
\end{equation}
The simplification consists in the fact that the holomorphic
$\omega_i$ are
supplemented by precisely the correct anti-holomorphic differentials
$\bar{\omega}_i$, to provide the correct
$(\text{det}(\text{Im}\tau))^2$ after
the integral. Finally, we therefore find
\begin{equation}
\mathcal{F}_{g}=\int_{\mathcal{M}_{g}}\langle(\mu\cdot G^-)^{3g-4}(\mu\cdot J^{--}_{K3})\psi_3(x)(\text{det}Q_i)(\text{det}Q_j)\rangle\,,
\end{equation}
which is indeed a topological correlator. The splitting into torus and $K3$ contribution takes the following form
\begin{align}
\mathcal{F}_{g}=\int_{\mathcal{M}_{g}}&\langle(\mu\cdot G_{T^2}^-)^g(\mu\cdot G_{K3}^-)^{2g-4}(\mu\cdot J^{--}_{K3})\psi_3(x)\rangle\cdot(\text{det}Q_i)(\text{det}Q_j)\,.\label{topamp2}
\end{align}
%%%%%%%%%%%%%%%%%%%%%%%%%%%%%%%%%%%%%%%%%%%%%%%%%%%%%%%%%%%%%%%%
%%%%%%%%%%%%%%%%%%%%%%%%%%%%%%%%%%%%%%%%%%%%%%%%%%%%%%%%%%%%%%%%
\section{Harmonic description and harmonicity relations}\label{Sect:HarmonicDescription}
In the previous Section we have considered particular amplitudes in heterotic string theory which are captured by correlation functions in a twisted two-dimensional theory. In this Section we would like to understand which terms in the heterotic effective action these amplitudes correspond to and whether they have any interesting properties with respect to their moduli dependence. It turns out that the effective action is best formulated in $\cN=2$ harmonic superspace and we will begin by constructing this space explicitly.
%%%%%%%%%%%%%%%%%%%%%%%%%%%%%%%%%%%%%%%%%%%%%%%%%%%%%%%%%%%%%%%%
\subsection{Global $\cN=2$ supersymmetry}\label{global}

\subsubsection{$SU(2)$ harmonic variables}

We consider $\cN=2$ supersymmetry in four dimensions whose automorphism group is
$SU(2)$. We introduce harmonic variables \cite{Galperin:1984av} on the coset $SU(2)/U(1)$ in the
form of matrices $(u^+_i, \, u^-_i) \in SU(2)$. They have an index
$i=1,2$ transforming under the fundamental representation of $SU(2)$ and $U(1)$ charges $\pm 1$. Together with their
complex conjugates $\bu^i_+ = \overline{(u^+_i)}, \, \bu^i_- =
\overline{(u^-_i)}$  they satisfy the unitarity conditions
\begin{align}
&u^+_i\, \bu^i_+ = u^-_i\, \bu^i_- = 1 \,, && u^+_i\, \bu^i_- = u^-_i\, \bu^i_+ = 0\,, &&u^+_i\, \bu^j_+ + u^-_i\, \bu^j_- = \delta^j_i\label{12'}
\end{align}
and the unit determinant condition
\begin{align}
&\ep^{ij} u^+_i u^-_j = 1\ , &&u^+_i u^-_j - u^-_i u^+_j = \ep_{ij}\,,\label{12}
\end{align}
(with $\ep^{12} = -\ep_{12} = 1$).

The harmonic functions have harmonic expansions homogeneous under the action of the
subgroup $U(1)$. The harmonic expansions are organized in irreps of
$SU(2)$, keeping the balance of projected indices so that the overall $U(1)$ charge is always the same.
An example of a harmonic function which we shall frequently encounter is $f^+(u) = f^i u^+_i + f^{ijk} u^+_i u^+_j u^-_k + \cdots$. The first component in this expansion is a doublet of $SU(2)$. The higher components give rise to higher-dimensional irreps, but we shall not need them
here.

The harmonic derivatives can be viewed as the covariant derivatives on the harmonic coset  $SU(2)/U(1)$, or equivalently, as the generators of the algebra of $SU(2)$ written down in an $U(1)$ basis  (see Section \ref{coset}). This means that they are invariant under the left action of the group $SU(2)$, but covariant under the right action of the subgroup $U(1)$. They can be split into generators of the subalgebra $U(1)$:
\begin{equation}\label{subhd}
   D_0 = u^+_i \frac{\pa}{\pa u^+_i } - \bu_+^i \frac{\pa}{\pa \bu_+^i}
\end{equation}
and of the coset:
\begin{align}
&D_+{}^- = u^-_i \frac{\pa}{\pa u^+_i }  = \bu_+^i \frac{\pa}{\pa \bu_-^i}\,, &&\text{and} &&D_-{}^+ = u^+_i \frac{\pa}{\pa u^-_i } = \bu_-^i \frac{\pa}{\pa \bu_+^i} \ . \label{cosethd}
\end{align}
The harmonic derivatives are differential operators preserving the defining algebraic constraints (\ref{12'}) and (\ref{12}).

The derivative (\ref{subhd}) acts homogeneously on the harmonic functions. For instance, the function $f^{+}(u)$ above has $U(1)$ charge $+1$, hence
\begin{equation}\label{exhf}
    D_0  f^{+}(u) = f^{+}(u)\ .
\end{equation}
The harmonic expansion of this function defines an infinitely reducible representation of $SU(2)$. It can be made irreducible by requiring that the raising operator $D_-{}^+$ annihilates the function:
\begin{equation}\label{ropcon}
    D_-{}^+ f^{+}(u) = 0\ \quad \Rightarrow \quad  f^{+}(u) = f^i u^+_i \ .
\end{equation}
In other words, such a function is a highest-weight state of a doublet of $SU(2)$. The irreducibility condition (\ref{ropcon}) is also called a condition for harmonic (H-) analyticity.

\subsubsection{Grassmann analytic on-shell superfields}

The introduction of harmonic variables allows us to define `1/2 BPS short' or Grassmann
(G-) analytic superfields.\footnote{A more systematic derivation of the G-analytic superfields as functions on a coset of the $\cN=2$ superconformal algebra $SU(2,2/2)$ will be given in Section \ref{css}.} \footnote{The notion of Grassmann analyticity (with breaking of the R symmetry) was first proposed in \cite{Galperin:1980fg} in the context of the $\cN=2$ hypermultiplet. Later on it was made R-symmetry covariant in the framework of $\cN=2$ harmonic superspace in \cite{Galperin:1984av}. }  They depend only on half of the Grassmann variables which can
be chosen to be $\q^{+}_\a = \q^i_\a\, u_i^{+}$ and $\bq^\da_{-} =
\bu^i_{-}\,\bq^\da_i $. One such superfield is the {\it linearized  on-shell} hypermultiplet ($\cN=2$ matter multiplet)
\begin{align}
q^{+}(x^\mu,\q^+,\bq_-,u) = f^{i} u^{+}_{i} + \q^{+}_\a\, \chi^{\a} +\bar\psi_{\da}\, \bq^\da_{-} +\text{derivative terms}.\label{01}
\end{align}
Here $f^{i}$ are the two complex scalars,
$\chi^{\a}$ and $\bar\psi_{\da}$ are the two fermions of the on-shell multiplet. To exhibit manifest G-analyticity, one has to choose the appropriate analytic basis in superspace,
\begin{equation}\label{chbashss}
    x^\mu\ \to \ x^\mu + i \q^{+}\sigma^\mu\bar\q_{+} - i \q^{-}\sigma^\mu\bar\q_{-}\ ,
\end{equation}
analogous to the familiar chiral basis. Note that the harmonic dependence here is cut down
to linear. This is typical for on-shell
multiplets which, in addition to the G-analyticity condition, also satisfy the H-analyticity condition
\begin{equation}\label{hansf}
    D_{-}{}^{+} q^{+}(\q^+,\bq_-,u) = 0\ .
\end{equation}
Here the harmonic derivative is supersymmetrized by going to the manifestly G-analytic superspace coordinates (\ref{chbashss}). One can show that the `ultrashort' on-shell superfield (\ref{01}) is the solution to the simultaneous conditions for G- and H-analyticity \cite{Galperin:1984av,Hartwell:1994rp,Andrianopoli:1999vr}.

Note that in the $\cN=2$ G-analytic superspace there exists a special conjugation $\
\widetilde{}\ $ combining complex conjugation with a reflection on the harmonic coset,
such that G-analyticity is preserved. In this sense we can define the conjugate hypermultiplet
\begin{align}
&\tilde q_-(x^\mu,\q^+,\bq_-,u) = \bar f_{i} \bu_-^{i} + \bq_-^{\da}\, \bar\chi_{\da} +\psi^{\a}\, \q^+_\a +\text{derivative terms}.\label{01'}
\end{align}
In what follows it will be convenient to combine the two versions of the hypermultiplet into a doublet of an external $SU(2)$ (not the R symmetry one), $(q^+, \tilde q_-) \ \leftrightarrow \ q^+_a$, $a=1,2$.

Another example of a G-analytic superfield is the {\it linearized}  on-shell vector
multiplet. It is obtained from the off-shell chiral field strength
\begin{equation}\label{14}
    W(\q_\a^i) = \varphi + \q_\a^i  \la^\a_i + \q_\a^i \q_\b^j\, ( \ep_{ij}F^{(\a\b)}_{(+)} + \ep^{\a\b} S_{(ij)}) \ .
\end{equation}
Here $\varphi$ is the complex physical scalar and $F_{(+)}$ is the self-dual part of the gluon field strengths, while $S$ is a triplet of auxiliary fields. On shell the latter must vanish,\footnote{See Section \ref{3.1.3} for a discussion of the proper elimination of this auxiliary field.} hence
the additional constraint
\begin{equation}\label{15}
   \ep_{\a\b} D_i^\a D_j^\b\ W = 0\,.
\end{equation}
Now, define the superfield (a superdescendant of $W$)
\begin{equation}\label{16}
    K_{-}^\a = D_{-}^\a\ W\,,
\end{equation}
where we have projected the $SU(2)$ index of $D^\a_i$ with the harmonic
$\bu_{-}^i$ . This superfield is annihilated by half of the spinor
derivatives and hence is 1/2 BPS short. Indeed, this is true for the projections $\bar
D^{+}_{\dot\b}$ since $\{\bar D^{+}, D_{-}\}=0$ and $\bar D^i_{\dot\b}W=0$ (chirality).
Further, hitting (\ref{16}) with $D_{-}^\b$ we obtain zero as a consequence of
the projection of the on-shell constraint (\ref{15}) with $\bu_{-}^i \bu_{-}^j$.
We conclude that $K_{-}^\a$ satisfies the G-analyticity constraints
\begin{equation}\label{16'}
    D_{-}^\b K_{-}^\a = \bar D^{+}_{\dot\b} K_{-}^\a = 0\,,
\end{equation}
which imply that it depends on half of the $\q$'s:
\begin{equation}\label{17}
    K_{-}^\a(\q^+,\bq_-,u) = \la^\a_i  \bu_{-}^i  + (\sigma^{\m})^{\a\da}\bq_{\da\,-}\ i\pa_\m\varphi + \q^+_\b\, F^{\a\b}_{(+)} +  \mbox{derivative terms}.
\end{equation}
In addition, the harmonic dependence of $K_{-}^\a$ is restricted to be linear. As in (\ref{hansf}), this follows from the condition for H-analyticity
\begin{equation}\label{hwconK}
    D_{-}{}^{+} K_{-}^\a = 0 \ ,
\end{equation}
in turn derived from the harmonic independence of $W$ ($D_{-}{}^{+}  W = 0$) and the commutator $[D_{-}{}^{+} , D_{-}^\a]=0$. This is another example of an ultrashort superfield. Note, however, that it is not a primary object but rather a superdescendant of the chiral  on-shell vector
multiplet.

\subsubsection{Higher-derivative couplings}\label{3.1.3}

After having defined the G-analytic superfields (\ref{01}) and (\ref{17}), we now want to construct the corresponding effective action couplings. In this paper we consider the following term:
\begin{equation}
    S = \int \ d^4x \ du \ d^2\q^+ d^2\bq_{-} \ (K_{-} \cdot K_{-})^{g}\, F^{-2(g-2)}(q^{+}_{{\hat A}\, a},u)\ , \label{19'}
\end{equation}
where $K_{-} \cdot K_{-} \equiv K_{-}^\a\ep_{\a\b} K_{-}^\b$  and  ${\hat A}=1,\ldots ,n$ is an $SO(n)$ vector index
labelling the coordinates of the coset of physical scalars (see Section \ref{coset}).\footnote{The superfield $K_{-}^\a$ being a fermion, one needs a gauge group of sufficiently high rank, so that $(K_{-}^\a\ep_{ab} K_{-}^\b)^{g} \neq 0$.} Moreover, the integer $g$ was chosen in such a way that amplitudes computed from the effective action term (\ref{19'}) correspond precisely to the genus $g$ amplitudes which we have calculated in section \ref{Sect:hetF3}.

We can generalize the above coupling in various ways. The simplest is to add, e.g., a holomorphic dependence on the vector multiplets, $F^{-2(g-2)}(q^+,W;u)$. Being chiral, $W$ are annihilated by $\bar D^+_{\da}$, therefore to make $F$ G-analytic we only need to act with
$(D_- \cdot D_-)$, giving rise to the gaugino term
\begin{equation}\label{gen3}
    (K_{-P} \cdot K_{-Q})\ F_{P,Q}^{-2g}(q^+,W,u)\ .
\end{equation}
Here $P, Q$ are the gauge group indices of the vector multiplets and $F_{P,Q}$ is the second-order derivative of the function $F$ with respect to $W$. Note that the $U(1)$ charge of the function $F$ has changed, to compensate for the charge of the extra factor $K_-\cdot K_-$.

Another way is to let the function $F$ depend on both chiral and antichiral vector multiplets,  $F^{-2(g-2)}(q^+,W, \bar W;u)$. We can make it manifestly G-analytic by acting on it with four spinor derivatives:
\begin{equation}\label{gen1}
    (D_- \cdot D_-) (\bar D^+\cdot \bar D^+)\ F^{-2(g+1)}(q^+,W, \bar W;u)\ .
\end{equation}
Using the (anti)chirality of $W,\bar W$ and the on-shell constraint \p{15}, it is easy to see that the only way to distribute these four derivatives is that one derivative hits one superfield, producing $K^\a_- $ or its conjugate $\tilde K^{+}_{\da}$. This gives rise to the following four-fermion term:
\begin{equation}\label{gen2}
    (\tilde K^{+\bar M} \cdot \tilde K^{+\bar N})\, (K_{-P} \cdot K_{-Q})\ (F^{-2(g+1)})_{\bar M, \bar N, P, Q}(W,\bar W)\ .
\end{equation}
Yet another possibility would be to add  hypermultiplets of the `wrong' analyticity, i.e. $q^-$ and $\tilde q_+$:
\begin{equation}\label{gen4}
(D_- \cdot D_-) (\bar D^+\cdot \bar D^+)\ F^{-2(g+1)}(q^+,W, \bar W,q^-,\tilde q_+; u)\ .
\end{equation}
This time there are many ways we can distribute the four spinor derivatives, among which we find a term of the type
\begin{equation}\label{gen5}
    (\bar D^+\bar W \cdot \bar D^+ q^-)\ (D_- W \cdot D_- W)\,, \ \ \mbox{etc.}
\end{equation}

\subsubsection{The harmonicity condition}

It is important to stress upon two points concerning the effective action term (\ref{19'}):
\begin{itemize}
\item The Grassmann measure is G-analytic, i.e. it involves only half of the projected
$\q$'s, and so must be the integrand, otherwise supersymmetry will be broken. This is
why we have to use the {\it linearized on-shell} superfields $q^+, \tilde q_-$ and $K_-$ which
are G-analytic like the measure.

\item The harmonic integral should produce an $SU(2)$ invariant, i.e. it  picks out the
$SU(2)$ singlet part of the integrand. This is only possible if the latter is a {\it
chargeless} harmonic function. For example, $f(u) = f_0 + f^{ij}u^{+}_{i}u^+_j
 + \cdots $ integrates to $\int du \, f(u) = f_0$, but a charged function
like $f^{+} = f^{i}u^{+}_{i}  + \cdots$ will have a vanishing integral.  Notice that
for this reason the harmonic integral should always be done {last}, after the Grassmann
integrals, since the latter are charged.
\end{itemize}
In our case (\ref{19'}) the function $F$ carries $U(1)$ charge $-2(g-2)$
needed to compensate that of the factor $K$ $(+2g)$ and of the Grassmann measure
$(-4)$. Given the fact that the arguments $q^+_a$ of $F$ have a positive charge, we have
to introduce a set of {\it constant}  $SU(2)$ multispinors
\begin{equation}\label{20'}
     \xi^{-p}(u) = \xi_{(i_1 \cdots i_p)}\  \bu_{+}^{i_1} \cdots \bu_{+}^{i_p} + \ldots
\end{equation}
thus explicitly breaking $SU(2)$.\footnote{The other possibility, which we do not
consider here, would be to use singular functions involving inverse powers of fields.} The dots denote higher-order terms in the harmonic expansion of the coefficients
$\xi(u)$ which will not be of interest for us, see below. Note that the product of harmonics forms an  irreducible representation of
$SU(2)$ of isospin $p$. In what follows this fact will be of crucial importance. So, we
consider the potential ($m=2(g-2)$; the $Sp(n)$ index ${\hat A}$, with ${\hat A}=1,...,2n$ of $F$ are suppressed)
\begin{equation}\label{20}
    F^{-m}(q^+,u) = \sum_{n=0}^\infty  \ \xi^{{-(m+n)}}(u) \ (q^+)^n\ .
\end{equation}
The factors $K$ in (\ref{19'}) contribute, among others, the term
\begin{equation}\label{21}
 %   (\q^+)^2({\bq}_-)^2\ F^2_{(+)} \ F^2_{(-)}\  (\la_-\la_-)^{\frac{m}{2}}\,,   \qquad
(\q^+)^2({\bq}_-)^2\ F^2_{(+)}(\pa\varphi)^2 \    (\la_-\cdot\la_-)^{\frac{m}{2}}\,,
\end{equation}
which is exactly the one which we have considered in section \ref{Sect:hetF3}. The $\q$'s saturate the superspace measure and are integrated out. The
remainder has a harmonic charge,
\begin{equation}\label{22}
        (\la_-\cdot\la_-)^{\frac{m}{2}} = (\la_{i_1}\cdot\la_{i_2})\ \cdots\ (\la_{i_{m-1}}\cdot\la_{i_m})\ \bu^{i_1}_-\cdots \bu^{i_m}_-
\end{equation}
which is compensated by the factor $F$ in order to have a non-vanishing harmonic
integral (i.e., an $SU(2)$ singlet). Clearly, (\ref{22}) is a representation of $SU(2)$ of isospin $m$. This can be reformulated as the highest-weight condition (cf. (\ref{ropcon}))
\begin{equation}\label{hwcon}
    D_{-}{}^{+} (\la_-\cdot\la_-)^{\frac{m}{2}} = 0 \ .
\end{equation}
A similar condition holds for the entire superfield term $(K_{-} \cdot K_{-})^{g}$.

The singlet needed for the harmonic integral is obtained by combining (\ref{22}) with
the matching representation in $F$. Consider the harmonic structure of $F$ (all
$\q=0$):
\begin{align}
F^{-m}(f^+,u) = \sum_{n=0}^\infty  \ \xi_{(i_1 \cdots i_{m+n})}\    \bu_{+}^{i_1} \cdots \bu_{+}^{i_{m+n}}f^{(k_1} \cdots f^{k_{n})}  u^{+}_{k_1} \cdots u^{+}_{k_n} \ .\label{23}
\end{align}
Here we have restricted the harmonic expansion (\ref{20'}) of the coefficient function $\xi^{{-(m+n)}}(u)$ to the lowest-rank $SU(2)$ representation. The higher-rank terms are irrelevant due to the gauge invariance of
the coupling (\ref{19'}). Indeed, consider adding a total supersymmetrized harmonic
derivative $D_{-}{}^{+} \Lambda^{(-2g+2) }(\q^+,\bq_-,u)$ to the potential
$F^{-2(g-2)}$. After
integration by parts (the G-analytic measure allows this), $D_{-}{}^{+}$ annihilates the on-shell superfield $K_-$ (recall (\ref{hwconK})), hence the gauge invariance of (\ref{19'}) with the G-analytic parameter $\Lambda$. By examining the harmonic
expansion of $\Lambda(0,0,u)$ one can show that all the omitted terms in (\ref{23}) can be gauged away.

The gauge-fixed function (\ref{23}) satisfies two differential conditions. The first one expresses the fact that it is a function only of the projection $f^+$ of the $SU(2)$ doublet of physical scalars:
\begin{equation}\label{firstcon}
    \frac{\pa}{\pa f^-} F^{-m} = 0\ .
\end{equation}
This is yet another kind of analyticity condition (S-analyticity), this time with respect to the scalars (which in fact are the coordinates on the curved manifold  $SO(n,4)/SO(n) \times SO(4)$, see Section \ref{coset}).
The second one restricts the harmonic dependence
\begin{equation}\label{secocon}
    D_{+}{}^{-} F^{-m} = f^-\frac{\pa}{\pa f^+} F^{-m}\ .
\end{equation}
Note that if the right-hand side in (\ref{secocon}) vanished, this would be a condition defining a lowest-weight state of $SU(2)$. However, the dependence on the scalars makes the harmonic structure in (\ref{23}) reducible.

{}From (\ref{23}) we have to extract the  irreducible harmonic structure
$\bu_{+}^{i_1} \cdots \bu_{+}^{i_{m}}$ needed to match the conjugate
structure in (\ref{22}). It is obtained by contracting all the $u^{+}$ in (\ref{23}) with
a subset of the $\bu_{+}$, using $\bu_{+}^{i}\; u^{+}_{k} = 1/2\,
\delta^{i}_k + \mbox{traceless}$ (see (\ref{12'})). This confirms that the omitted
terms in the harmonic expansion of $\xi$ in (\ref{23}) cannot contribute - they contain higher-isospin $SU(2)$ irreps. The result is
the {\it relevant part} of the function $F$, or the {\it reduced} function
\begin{equation}\label{24}
    \mathcal{F}_g = \sum_n   \xi_{(i_1  \cdots i_{m+n})}\  \bu_{+}^{i_1}
    \cdots \bu_{+}^{i_{m}}  \ f^{i_{m+1}} \cdots f^{i_{m+n} } \ .
\end{equation}
In fact, this object is the amplitude computed in the heterotic string
theory analysis in section
\ref{Sect:hetF3}. The $U(1)$ charge of $\mathcal{F}_g$ is $-m$ which is identical to the one of $F^{-m}$.
Notice the full symmetrization of the indices of  $\xi$ inherited
from (\ref{23}). As required, the reduced function is manifestly
H-analytic (i.e., $SU(2)$ irreducible),
\begin{equation}\label{mansu4}
    D_{+}{}^{-} \mathcal{F}_g =  0\ .
\end{equation}
However, now the manifest S-analyticity (i.e., the dependence only on $f^+$) of (\ref{23}) is lost.

It should be made clear that (\ref{24}) is just a rearrangement of the harmonic expansion of the gauge-fixed function $F^{-m}$. The information contained in this function is encoded in the fact that the coefficients $\xi_{(i_1 \cdots i_{m+n})}$, which are the same in (\ref{23}) and (\ref{24}), form the $SU(2)$ representation
of isospin $m+n$. This information can be translated into two types of differential constraints on the function $\mathcal{F}_g$. In
general, the harmonic and scalar factors in (\ref{24}) form the reducible representation
$(m)\, \otimes\, \prod_{p=1}^n\otimes\, (1)_p\ \rightarrow \ (m+n) + \ldots$. The
relevant projection $(m+n)$ is obtained by symmetrizing all the indices $i$.  Any
other representation in this tensor product will have a subset of the $i$'s
antisymmetrized. The product of two $u$'s is irreducible, $(1)\otimes(1) \
\rightarrow \ (2)$ as follows form the commuting nature of the $SU(2)$ harmonics
$\bu^i_{+}$.  The antisymmetrization of indices carried by the $\bu$'s and the $f$'s
is ruled out by the so-called {\it harmonicity} condition:
\begin{equation}\label{25}
    \ep^{ij}\frac{\pa}{\pa \bu^i_{+}}\ \frac{\pa}{\pa f^{jA}} \mathcal{F}_g =  0\,,
\end{equation}
where we have combined the $SO(n)\times Sp(1)$ indices ${\hat A}\, a$ into the $Sp(n)$ index ${A}$.  This constraint involves partial derivatives
with respect to $\bu_{+}$. Strictly speaking, such an operation is illegal in the
harmonic formalism, since the variables $u$ are not independent, as can be seen from
(\ref{12'}), (\ref{12}). However the above equation can be rewritten using covariant harmonic derivatives
introduced in (\ref{subhd}) and (\ref{cosethd}) as
\begin{equation}\label{25prime}
    \ep^{ij}\left(u^{-}_i D_{-}{}^{+} - u^{+}_i D_0\right)
\ \frac{\pa}{\pa f^{j A}} \mathcal{F}_g =  0\ .
\end{equation}
Indeed, it is easy to see that this equation reduces to (\ref{25}) since our function $\mathcal{F}_g$ explicitly involves only $\bu_{+}$ harmonics.   In the following however we will continue to write
the formula using
partial derivatives with respect to $\bu_{+}$.

Further, the  antisymmetrization of indices carried by the $f$'s is ruled out by the
constraint
\begin{equation}\label{25'}
   \ep^{ij} \frac{\pa}{\pa f^{iA}} \ \frac{\pa}{\pa f^{jB}}\ {\cal F}_g =  0\ .
\end{equation}
Here we do not take into account the fact that the physical scalars $f$
parametrize a
curved manifold and hence the derivatives in (\ref{25'}) should be
considered covariant
with respect to the metric of the manifold. In Section \ref{coset} we
will consider the curved manifold in a special case namely the coset
space $SO(n,4)/SO(n) \times SO(4)$, where the $Sp(n)$ will be
represented by the an $SO(n)$ vector index ${\hat A}$ and and external
$SU(2)$ index $a$ with $a=1,2$. There  we show
that (\ref{25'}) is modified by a term proportional to
$\epsilon_{ab} \delta_{{\hat A}{\hat B}}$.

\subsubsection{The role of the auxiliary field}\label{3.1.5}

In the discussion above we have always treated the auxiliary field $S_{ij}$ in \p{14} as vanishing on shell. In other words, we have considered a free (flat) kinetic term for the vector multiplets,
\begin{equation}\label{flatkin}
    S_0 = \int d^4x\ d^4 \q \ W^2 + \mbox{c.c.}\ .
\end{equation}
In reality, the spaces we deal with are not flat, they have a non-trivial metric originating from the kinetic term
\begin{equation}\label{curvekin}
    S_0 = \int d^4x\ d^4 \q \ h(W^I) + \mbox{c.c.}\ ,
\end{equation}
where $h(W^I)$ is the  holomorphic ``prepotential". Notice that the auxiliary field is real, as follows from the defining constraint (Bianchi identity) on the vector multiplet $(D_i\cdot D_j) W = (\bar D_i \cdot \bar D_j) \bar W $. Then we can easily work out the part of the action \p{curvekin} involving this auxiliary field:
\begin{equation}\label{aufac}
\half S^{ij\, I} S^{J}_{ij} (h_{IJ} + \bar h_{IJ}) + S^{ij\, I}\left[h_{IKL} (\la_i^K\cdot\la_j^L) + \bar h_{IKL} (\bar\la_i^K\cdot\bar\la_j^L) \right]\,,
\end{equation}
and solve for it:
\begin{equation}\label{solv}
    S^I_{ij} = G^{IJ}\left[h_{JKL} (\la_i^K\cdot\la_j^L) + \bar h_{JKL} (\bar\la_i^K\cdot\bar\la_j^L) \right]\,,
\end{equation}
where the metric $G^{IJ}$ is defined by $G^{IJ}(h_{JK} + \bar h_{JK}) = \delta^I_K$.

It should be pointed out that the same auxiliary field also appears in all of the higher-derivative couplings described in Section \ref{3.1.3}. Consequently, we should modify the expression \p{solv} by terms involving those new couplings. Then plugging this expression for the auxiliary field back into the action will result in terms quadratic (and higher) in the couplings. Since the general case is rather complicated, here we would like to present a simple example where the interaction resulting from the elimination of the auxiliary field is multilinear in the couplings.   We want to show that purely chiral couplings of the type \p{21} can originate from different terms in the effective action, and not only from the obvious term \p{19'}. Consider, for instance, the non-holomorphic term \p{gen1} or, together with the chiral prefactor,
\begin{equation}\label{togchipre}
    (D_- \cdot D_-) (\bar D^+\cdot \bar D^+)\ \left[ (K_{-} \cdot K_{-})^{g-3}\, F^{-2(g-3)}(W, \bar W) \right]\,.
\end{equation}
If we distribute the spinor derivatives as described in Section \ref{3.1.3} (i.e., assuming that the auxiliary field vanishes), we obtain the non-chiral coupling \p{gen2}. However, in the presence of a non-vanishing auxiliary field we have other possibilities. First of all, the two $\bar D^+$ may hit the same $\bar W$, giving an auxiliary field. From its expression \p{solv} we retain only the chiral half, since this is what we wish to reproduce. Further, if a $D_-$ hits a $K_-$, another auxiliary field will appear. In this way we will get a term quadratic in the gauge couplings $h_{IJK}$ but we have decided to keep terms multilinear in the different couplings only. The same will happen if the two $D_-$ hit the same $W$ from the function $F$, so we drop such terms as well. Then the only way the two $D_-$ act is by distributing onto two different $W$, which is another factor of $K_{-} \cdot K_{-}$. The net effect of all this is the purely chiral term
\begin{equation}\label{puchite}
    F^2_{(+)}(\pa\varphi)^2 \ (\la_-\cdot\la_-)^{g-3}\ (\la^P_-\cdot\la^Q_-)\ F^{-2(g-3)}_{PQ;\bar I}(W, \bar W)\ G^{IJ}h_{JKL} (\la^K_-\cdot\la^L_-) \,,
\end{equation}
where the indices $PQ$ of $F_{PQ;\bar I}$ denote derivatives with respect to $W$, and $\bar I$ with respect to $\bar W$.

This term should be compared to a similar one obtained from the coupling \p{gen3}, holomorphic in $W$. In the presence of auxiliary fields it should be rewritten as
\begin{equation}\label{togchipre'}
    (D_- \cdot D_-)\ \left[ (K_{-} \cdot K_{-})^{g-2}\, F^{-2(g-2)}(W) \right]\,.
\end{equation}
If the two derivatives $D_-$ are distributed over the $W$, they produce another factor $\la_-\cdot\la_-$. Otherwise, they produce auxiliary fields, either by hitting a $K_-$ or by acting together on a $W$. The net effect is a term of the same structure as \p{puchite}, however, without any anti-holomorphic dependence on $\bar W$. Putting the two couplings \p{togchipre} and \p{togchipre'} together, we may say that we have generated a ``holomorphic anomaly", as described around eq.~\p{exactlyas}.

\subsection{The coset of physical scalars}\label{coset}

\subsubsection{Quaternionic geometry}

Let us consider\footnote{In this subsection we follow Ref.~\cite{Galperin:1992pj}.} a $4n$ dimensional Riemann manifold $M$ with local coordinates
$\{ x^{M k}\}, \;\;
M=1,\ldots, 2n; \;\;k=1,2$. One of the definitions of
quaternionic geometry \cite{{a9},{a12},{a13}} restricts the holonomy
group to a subgroup of $Sp(n)\times Sp(1)$. Hence we can choose from the
very beginning the tangent group to be $Sp(n)\times Sp(1)$. So,
the tensor fields defined on the manifold $\{ x^{M k}\}$
undergo gauge transformations both in their $Sp(n)$ and $Sp(1)$
indices
\begin{eqnarray}
\delta\phi_{AB\ldots ij\ldots}(x) &\equiv&
 \phi'_{AB\ldots ij\ldots}(x+\delta x)-
\phi_{AB\ldots ij\ldots}(x) \nn \\
 & =&
\tau_A{}^{A'}(x)\phi_{A'B\ldots ij\ldots}(x)+
\tau_B{}^{B'}(x)\phi_{AB'\ldots ij\ldots}(x)+\ldots \nn
\\  & +&
\tau_i{}^{i'}(x)\phi_{AB\ldots i'j\ldots}(x)+
\tau_j{}^{j'}(x)\phi_{AB\ldots ij'\ldots}(x)+\ldots
\nn \\
\delta x^{M i} &=& \tau^{M i}(x)\;. \label{5.2.1}
\end{eqnarray}
Correspondingly, the covariant derivative is given by
\begin{eqnarray}
{\cal D}_{A i} = e^{M k}_{A i}\partial_{M k}-
\omega_{A i(CD)}\Gamma^{(CD)}-
\omega_{A i(lk)}\Gamma^{(lk)} \equiv  \nabla_{A i}-
 \omega_{A i(CD)}\Gamma^{(CD)}-
\omega_{A i(lk)}\Gamma^{(lk)}\;. \label{5.3.1}
\end{eqnarray}
Here $\omega_{A i(CD)}$ and
$\omega_{A i(lk)}$ are the $Sp(n)$ and $Sp(1)$ connections. The  $Sp(n)$ generators
$\Gamma^{(CD)}$ obey the algebra
\begin{eqnarray}
[\Gamma^{(CD)},\Gamma^{(EF)}]  =
{1\over 2}( \Omega^{CE}\Gamma^{(DF)}+
\Omega^{CF}\Gamma^{(DE)}  + \Omega^{DE}
\Gamma^{(CF)}+
\Omega^{DF}\Gamma^{(CE)}) \ ,
\end{eqnarray}
and similarly for the  $Sp(1)$ generators $\Gamma^{(lk)}$, with the $Sp(n)$
invariant tensor $\Omega^{AB}$ replaced by $\epsilon^{lk}$.
As a rule, we  deal with the fundamental spinor representations
of $Sp(n)$ and $Sp(1)$,
\begin{eqnarray}
({\cal D}_{A i})_{B n}{}^{B' n'} =
\delta_B^{B'}\delta_n^{n'}\nabla_{A i} +
\delta_n^{n'}\omega_{A i\;B}{}^{B'} +
\delta_B^{B'}\omega_{A i\;n}{}^{n'}\;. \label{5.3.2}
\end{eqnarray}
The commutator of two covariant derivatives produces
the $Sp(n)$ and $Sp(1)$ components of the curvature tensor,
\begin{eqnarray}
[{\cal D}_{A i},{\cal D}_{B j}]_{C n}{}^{C' n'} =
\delta_n^{n'}
R_{A i\;B j\; C}{}^{C'}+
\delta_C^{C'}R_{A i\;B j\;n}{}^{n'}
\equiv {R_{A i,\;B j\; C n}}^{C' n'}\;. \label{5.4.1}
\end{eqnarray}

Now, the requirement that the holonomy group of this $4n$ dimensional
Riemannian manifold (i.e. the group generated by the Riemann tensor)
belongs to  $Sp(n) \times Sp(1)$  is equivalent to the following
covariant constraints
\begin{eqnarray}
R_{A i\;B j\; C}{}^{C'} &=& \epsilon_{ij}
R_{AB ; C}{}^{C'} \label{5.4.2a}\\
R_{A i\;B j\;n}{}^{n'} &=&
\Omega_{AB}R_{i j\;n}{}^{n'}\;. \label{5.4.2b}
\end{eqnarray}
For the analogous
constraint defining the hyper-K\"ahler manifolds \cite{a2}, the right-hand side of eq.
\p{5.4.2b} vanishes, while
in the quaternionic case it corresponds to the non-vanishing $Sp(1)$
part of the holonomy group.

The Bianchi identities combined with \p{5.4.2a} and \p{5.4.2b} imply
\begin{eqnarray}
R_{(AB)\;(A' B')} &=&R_{(ABA' B')}+
(\Omega_{BA'}\Omega_{AB'}+
\Omega_{AA'}\Omega_{BB'})R \label{5.4.3a} \\
R_{(ij)(kl)}&=&(\epsilon_{ik}\epsilon_{jl}+
\epsilon_{il}\epsilon_{jk})R\;, \;\;\;\ R={\rm constant}\;,\label{5.4.3b}
\end{eqnarray}
where we have introduced
\begin{align}
&R\equiv {1\over 6}R_{(ij)}{}^{(ij)}\,, &&R_{A i\;B j}{}^{A i\;B j}=8n(n+2)R\,. \label{5.4.4}
\end{align}
The constant $R$ can be positive or negative since the constraints (\ref{5.4.2a}) and (\ref{5.4.2b})
do not fix its sign. It is easy to see from eqs. (\ref{5.4.1}) -- (\ref{5.4.3b})
that the quaternionic manifolds are Einstein manifolds (with a cosmological
constant proportional to $R$). Hence the
{homogeneous} quaternionic manifolds are compact in the case $R > 0$
and non-compact if $R < 0$ \cite{besse}. The scalar $Sp(1)$
curvature is by definition positive and is given by $|R_{ij}{}^{ij}| = 6 |R|$.

Thus, irrespective of the value of $n$, the basic constraint
defining the quaternionic geometry can be written as follows
\begin{align}
[{\cal D}_{A i},{\cal D}_{B j}]_{C n}{}^{C' n'} =
-2\;\delta_C^{C'}\Omega_{AB}\;R\;\tz_{(ij)\;n}{}^{n'}
+ \delta_n^{n'}\epsilon_{ij}[ R_{(ABC}{}^{C')}-
R\;(\Omega_{BC}\delta_A^{C'}+
\Omega_{AC}\delta_B^{C'})]\,, \label{5.5.1}
\end{align}
where
\be
\tz_{(ij)\;n}{}^{n'}={1\over 2}(\epsilon_{in}\delta_j^{n'}+
\epsilon_{jn}\delta_i^{n'}) \nn
\ee
are the generators of $Sp(1)$ acting on a field with indices $\phi_{n'\ldots}\ $.

In order to prepare for the harmonic description of the manifold, let us decompose the $Sp(1)$ indices $i,j$ into $U(1)$ charges $\pm$:
\begin{align}
  &[{\cal D}_{A \pm},{\cal D}_{B \pm}]_{C n}{}^{C' n'} = \pm2\delta_C^{C'}\Omega_{AB}\;R\; (\tz_{\pm\pm})_n{}^{n'}\, \label{firstt}\\
  &{[}{\cal D}_{A +},{\cal D}_{B -}{]}_{C n}{}^{C' n'} = -2\delta_C^{C'}\Omega_{AB}\;R\; (\tz_{0})_n{}^{n'} + \delta_n^{n'}[ R_{(ABC}{}^{C')}-
R\;(\Omega_{BC}\delta_A^{C'}+
\Omega_{AC}\delta_B^{C'})]\,, \label{secondt}
\end{align}
where we have introduced the projected quantities
\begin{align}
&\tz_0 \equiv \tz_{+-} = \tz_{-+}=u^i_+ u^j_-\tz_{ij}, &&\text{and} &&  \tz_{\pm\pm}=\mp\frac{1}{2} u^i_\pm u^j_\pm \tz_{ij}.
\end{align}
Notice that the second term on the right-hand side of \p{secondt} acts only on fields with $Sp(n)$ indices $\phi_{C'\ldots}\ $. Applying this algebra to an $Sp(n)$ scalar (no indices $C'\ldots$, but some indices $n'\ldots$), we obtain
\begin{eqnarray}
[{\cal D}_{A \pm},{\cal D}_{B \pm}]_{C n}{}^{C' n'} &=& \pm2\delta_C^{C'}\Omega_{AB}\;R\; (\tz_{\pm\pm})_n{}^{n'}\, \\
  {[}{\cal D}_{A +},{\cal D}_{B -}{]}_{C n}{}^{C' n'} &=& -2\delta_C^{C'}\Omega_{AB}\;R\; (\tz_{0})_n{}^{n'}\,.
\end{eqnarray}

\subsubsection{The coset $SO(n,4)/SO(n)\times SO(4)$}\label{subcoset}

In the special case of the  coset $SO(n,4)/SO(n)\times SO(4)$ the holonomy group $Sp(n)\times Sp(1)$ is reduced to $SO(n)\times SU(2)\times Sp(1) \sim SO(n)\times SO(4)$. So, the $Sp(n)$ indices split into $SO(n)\times SU(2)$ indices, $ A\ \to \ {\hat A}a$, where ${\hat A}=1\ldots n$ is an $SO(n)$ vector index and $a=1,2$ is an $SU(2)$ doublet index. Thus, we have $\Omega_{AB} = \delta_{{\hat A}{\hat B}}\, \ep_{ab}$. With this decomposition it is easy to identify the last term in the right-hand side of \p{secondt} (with $R=-1$) as the generators $Z_{ab}= Z_{ba}$ of $SU(2)$. Further, the curvature term $R_{(ABC}{}^{C')}$ now becomes the generators $M_{{\hat A}{\hat B}}=-M_{{\hat B}{\hat A}}$ of $SO(n)$. Finally, let us denote the covariant derivatives $\cal D$ (i.e., the generators of the coset $SO(n,4)/SO(n)\times SO(4)$) by $L_{{\hat A}\, a\pm}$. Then the algebra of $SO(n,4)$ takes the
form
\begin{align}
&[L_{{\hat A}\, a\pm},L_{{\hat B}\, b\pm}]=\pm\delta_{{\hat A}{\hat B}}\epsilon_{ab}\tz_{\pm\pm}\,,\nn\\
&[L_{{\hat A}\, a+},L_{{\hat B}\, b-}]=M_{{\hat A}{\hat B}}\epsilon_{ab}-\frac{1}{2}(\epsilon_{ab}\tz_{0}+Z_{ab})\delta_{{\hat A}{\hat B}}\,,\nn\\
&[M_{{\hat A}{\hat B}},L_{{\hat C}\, a\pm}]=\delta_{{\hat A}{\hat C}} L_{{\hat B}\, a\pm}-\delta_{{\hat B}{\hat C}} L_{{\hat A}\, a\pm}\,,\nn\\
&[\tz_{\pm\pm},L_{{\hat A}\, a\mp}]=L_{{\hat A}\, a\pm}\,,\nn\\
&[\tz_{0},L_{{\hat A}\, a\pm}]=\pm L_{{\hat A}\, a\pm}\,,\label{algebra}\\
&[Z_{ab},L_{{\hat A}\, c\pm}]=-(\epsilon_{ac} L_{{\hat A}\, b\pm}+\epsilon_{bc} L_{{\hat A}\, a\pm})\,,\nn\\
&[M_{{\hat A}{\hat B}}, M_{{\hat C}\hat D} ]= \delta_{{\hat A}{\hat C}} M_{{\hat B}\hat D} + \text{permutations}\,,\nn\\
&[Z_{ab}, Z_{cd} ] =-\epsilon_{ac}Z_{bd} + \text{permutations}\,,\nn\\
&[\tz_{++},\tz_{--}]=\tz_0\,,\nn\\
&[\tz_{0},\tz_{\pm\pm}]=\pm2\tz_{\pm\pm}\,.\nn
\label{algebra}
\end{align}
Here the first two relations are in fact the commutators of covariant derivatives \p{firstt}, \p{secondt}.

\subsubsection{Harmonic description}

The higher-derivative term (\ref{19'}) involves the function (potential) $F$
defined on the coset of physical scalars. The peculiarity of this function is that it
depends only on a single projection $f^{+}_{{\hat A}\, a}(x,u)$ of the four-vectors of coset coordinates,
obtained with the help of the $SU(2)$ harmonic variables. This is a typical example of
an {\it analytic harmonic realization} of a coset space. Another, very similar example
is that of the $\cN=2$ superconformal group $SU(2,2/2)$ realized on the Grassmann
analytic superfields (\ref{01}) (see Section \ref{css}). Here we explain this coset construction, following closely the case of $\cN=2$ superconformal symmetry and Poincar\'e supergravity \cite{Galperin:1985zv,Galperin:1987ek,Galperin:2001uw} and of $\cN=2$ quaternionic sigma models \cite{Bagger:1987rc,Galperin:1992pj,Ivanov:1999vg}.

Now, we want to realize the algebra \p{algebra} on a coset of the group $SO(n,4)$. The standard
coset $SO(n,4)/SO(n) \times SO(4)$ is obtained by putting the generators $M,Z,\tz$ of $SO(n) \times SO(4)$
in the coset denominator and leaving all the $L$'s in the coset with associated $4n$
coordinates $f$:
\begin{equation}\label{30'}
     \frac{SO(n,4)}{(M,Z,\Upsilon)}\ \sim\ \{ f^{\pm}_{{\hat A}\, a}\}\ .
\end{equation}

We wish to have an alternative {\it S-analytic} coset involving only the coordinates $f^{+}_{{\hat A}\, a}$
associated with the generators $L_{{\hat A}\, a+}$. To this end we have to move the generators
$L_{{\hat A}\, a-}$ to the coset denominator. In doing this we encounter a
problem: The $Sp(1)$ generator  $\tz_{++}$ converts $L_{{\hat A}\, a-}$ into the coset
generator $L_{{\hat A}\, a+}$. In order to avoid this, we proceed to the `harmonization' of the
coset. This means to introduce an additional group $\widehat{SU(2)}$ which we treat as
independent of the $Sp(1)$ from the coset denominator. Let us denote its generators by
$T_{++}$, $T_{--}$, $T_0$. We assume that this extra $\widehat{SU(2)}$ acts as an
external automorphism of  (\ref{algebra}), i.e. $[T,\tz] = \tz$, $[T,L] = L$. Then it is
clear that the combination $\tz_{++} - T_{++}$ commutes with the
generators of (\ref{algebra}), in particular, with  $L_{{\hat A}\, a -}$. So, to avoid the
above problem, we replace $\tz_{++}$ in the coset denominator by this
combination. The group $\widehat{SU(2)}$ is itself realized on the harmonic coset
$\widehat{SU(2)}/\widehat{U(1)}$, which means that we have to add the generator of
the automorphism subgroup $\widehat{U(1)}$ to  the coset denominator. The
result is a particular {\it S-analytic} realization of the coset
\begin{equation}\label{30}
    \frac{SO(n,4)\subset\hskip-11pt\times\ \widehat{SU(2)}}{(M,Z_{ab},L_{{\hat A}\, a-},  \tz_0, \tz_{--}, \tz_{++} - T_{++}, T_0)}\ \sim (f^{+}_{{\hat A}\, a}, w_i^{\pm})
\end{equation}
parametrized by the coordinates $f^{+}_{{\hat A}\, a}$ associated with the $SO(n,4)$  generators
$L_{{\hat A}\, a+}$ and by harmonics $w_i^{\pm}$ (the latter differ from the usual
${SU(2)}$ harmonics $u$ (\ref{12'}), as explained below).

This coset is analytic in the sense that we consider functions $F(f^{+}_{{\hat A}\, a}, w_i^{\pm})$  on it
which are annihilated by the generators $L_{{\hat A}\, a-}$. Then the
algebra  (\ref{algebra}) implies
\begin{equation}\label{31}
    L_{{\hat A}\, a-} F = 0 \ \ \Rightarrow \ \ \tz_{--} F =0\,.
\end{equation}
In addition, we only consider scalar functions under $SO(n)\times SU(2)$, i.e.,  functions which do not carry $SO(n)\times SU(2)$ indices, but can have $U(1)$
charges under both $\tz_0$ and $T_0$. This amounts to the extra constraints
\begin{equation}\label{31'}
    M_{{\hat A}{\hat B}}F = Z_{ab}F=0 \ .
\end{equation}
Finally, we impose the coset defining constraint
\begin{equation}\label{32}
    (\tz_{++} - T_{++}) F = 0 \ .
\end{equation}
It leads to a particular mixing of the coordinates associated with
the $Sp(1)$ generators $\tz$ and with the $\widehat{SU(2)}$ generators $T$. For this
reason (\ref{30}) is a semi-direct product (denoted by $\subset\hskip-11pt\times$ in (\ref{30})) of the two cosets  $SO(n,4)/SO(n) \times SO(4)$ and   $\widehat{SU(2)}/\widehat{U(1)}$.

The actual construction of the coset goes through the following steps. We first
introduce a {\it double harmonic space} involving,  in addition to the $\widehat{SU(2)}$
harmonic variables $u$, harmonics $\kappa_I{}^i$ (with $I=\pm$) on $Sp(1) \sim {SU(2)}$ satisfying the
defining conditions (cf. (\ref{12'}))
\begin{align}
&\kappa_I{}^i \bar\kappa_i{}^J = \delta^J_I\,,&& \bar\kappa_i{}^I \kappa_I{}^j = \delta^j_i\,, && \ep^{IJ}\kappa_I{}^i\kappa_J{}^j = \ep^{ij}\ .\label{33}
\end{align}
They undergo $SU(2)$ transformations of two types: local (in the sense of $SU(2)$ from the coset denominator)  with parameter $\la$ and rigid with parameter
$\sigma$:
\begin{equation}\label{34}
    \delta \kappa_I{}^i = \la_I{}^J \kappa_J{}^i + \kappa_I{}^j \sigma_j{}^i \ .
\end{equation}
The introduction of the new harmonics allows us to realize the $Sp(1)$ generators $\Gamma^{(lk)}$ in (\ref{5.3.1}) as differential operators:
\begin{equation}\label{gdiff}
    \Gamma^{(lk)} = \frac{1}{2}\left(   \kappa_I{}^l\epsilon^{kt}{\partial\over\partial \kappa_I{}^t} +
\kappa_I{}^k\epsilon^{lt}{\partial\over\partial \kappa_I{}^t}  \right)\ .
\end{equation}
Further, the projected covariant derivatives (\ref{5.3.1}), ${\cal D}_{\a I} = \kappa_I{}^i {\cal D}_{\a i}$ now act on fields with projected $Sp(1)$ indices, $\phi_{IJ\ldots K}(x,\kappa) = \kappa_I{}^i \kappa_J{}^j \ldots \kappa_K{}^k \phi_{ij\ldots k}(x)$. They satisfy the algebra (on fields without $Sp(n)$ indices)
\begin{equation}\label{prcod}
    [{\cal D}_{A I},{\cal D}_{B J}]_C{}^{C'}=
-2\;\delta_C{}^{C'}
\Omega_{AB}R\tz_{(IJ)}\;,
\end{equation}
where $\tz_{(IJ)}$ are the
covariant derivatives with respect to the coordinates $v_{a}{}^{i}$, $\tz^{(IJ)} = -\bar\kappa^I_k \bar\kappa^J_l \Gamma^{(kl)}$. It is important to realize that these derivatives do not act on  the $\widehat{SU(2)}$
harmonic variables $u$, but only on the newly introduced $Sp(1)$ harmonics $\kappa$.

Our task now will be to make a change of variables from $\kappa, u$ to $z, w$ which are
inert under the rigid $SU(2)$ and have simple transformation properties under the local
$SU(2)$. This will allow us to impose the coset constraint (\ref{32}) in a covariant
way. We start by projecting the harmonics $\kappa$ with $u, \bu$:
\begin{equation}
    \kappa_{\pm}{}^{\pm} = \bu_{\pm}{}^{I} \kappa_I{}^i u_i{}^{\pm}
\end{equation}
and similarly for the conjugate matrix $\bar\kappa$. Next we make the following non-linear
change of variables:
\begin{align}
& z_-{}^+ = \kappa_-{}^+ (\kappa_+{}^+)^{-1} = - (\bar\kappa_-{}^-)^{-1}\bar\kappa_-{}^+\,, && z_-{}^- = \bar\kappa_-{}^-\,,  \nn \\
& z_+{}^- = \kappa_+{}^- (\kappa_-{}^-)^{-1} = - (\kappa_+{}^+)^{-1}\bar\kappa_+{}^-\,, && z_+{}^+ = \bar\kappa_+{}^+\ .
\end{align}
These new variables satisfy an algebraic constraint following from the fact that $\kappa \in
SU(2)$, i.e. $\det \kappa =1$. It can be used to eliminate, e.g.  $z_-{}^-$ while the
remaining  $z_+{}^+$ can be treated as the coordinate of $U(1) \subset SU(2)$.

It is then not hard to check that the new variables $z$ transform in the following way
under the local $SU(2)$:
\begin{align}
& \delta z_-{}^+ = \hat\la_-{}^+ \,, \quad \delta z_-{}^- = z_-{}^-\hat\la_-{}^-   \,, &&\quad \delta z_+{}^+ = \hat\la_+{}^+z_+{}^+\,, \nn \\
& \delta z_+{}^- = \hat\la_+{}^+z_+{}^- + z_+{}^-\hat\la_-{}^- - \hat\la_+{}^-\,, \label{deltaz}
\end{align}
where $\hat\la_\pm{}^\pm = \bar w_\pm{}^I \la_I{}^J w_J{}^\pm$ and we have
introduced the {\it new harmonics}
\begin{align}
& w_i{}^{+} = u_i{}^{+} + u_i{}^{-}z_{-}{}^{+}\,, &&  w_i{}^{-} = u_i{}^{-}\,, \nn\\
& \bar w_{+}{}^i = \bar u_{+}{}^i\,, &&\bar w_{-}{}^i = \bar u_{-}{}^i - z_{-}{}^{+} \bu_{+}{}^i\,, \label{wharm}
\end{align}
with transformation laws
\begin{align}
& \delta w_i{}^{+} = w_i{}^{-}\hat\la_{-}{}^{+}\,, &&  \delta w_i{}^{-} = 0\,,\nn\\
& \delta \bar w_{+}{}^i = 0\,, &&\delta \bar w_{-}{}^i = -\hat\la_{-}{}^{+} \bar w_{+}{}^i \ . \label{trw}
\end{align}
We point out that these new harmonics are not unitary anymore (i.e., $\bar w$ is not the
conjugate of $w$), but they still satisfy the same algebraic relations as the unitary
harmonics $u$ (\ref{12'}).

What we have achieved is that the new variables do not mix under the local $SU(2)$
transformations with parameters $\hat\la$. This allows us to eliminate  all of the
$z$ variables (with the exception of $z_+{}^+$) in a covariant way, which corresponds to
imposing the $\tz$ coset conditions from (\ref{31}) and the $\tz-T$ condition (\ref{32}).

\subsubsection{Covariant constraints on the function ${F}$}

Now we are able to see how the naive constraint (\ref{25'}) is modified due to
the curvature of the coset space (\ref{30}) on which the reduced function ${\cal F}$ (\ref{24}) lives. The origin of these constraints can be traced back to the S-analyticity conditions satisfied by the gauge-fixed function $F$ (\ref{23}). On the curved manifold they become {covariant} constraints (cf. (\ref{31})):
\begin{equation}\label{covsancon}
    {\cal D}_{{\hat A}\, a-} F^{-2(g-2)} = 0\ .
\end{equation}
Here ${\cal D}_{{\hat A}\, a\pm}$ are covariant derivatives generalizing the flat derivatives
$\pa/\pa f$. They satisfy the same $SO(n,4)$ algebra as the generators $L_{{\hat A}\, a\pm}$.

The second-order derivative in  the constraint  (\ref{25'}) can be rewritten as follows:
\begin{equation}
  ({\cal D}_{{\hat A}\, a+} {\cal D}_{{\hat B}\, b-} - {\cal D}_{{\hat A}\, a-} {\cal D}_{{\hat B}\, b+}) F^{-2(g-2)} = -[{\cal D}_{{\hat A}\, a-} , {\cal D}_{{\hat B}\, b+}]  F^{-2(g-2)} = -\frac{1}{2}\delta_{{\hat A}{\hat B}}\ep_{ab} \tz_0 F^{-2(g-2)} \,, \label{correctcons}
\end{equation}
where we have used the S-analyticity constraints (\ref{covsancon}), the scalar conditions (\ref{31'}) and the algebra (\ref{algebra}). The function $F^{-2(g-2)}$ has two independent $U(1)$ charges, one with respect to the generator $T_0$, $T_0 F^{-2(g-2)} = -2(g-2) F^{-2(g-2)}$ and the
other for $\Upsilon_0$. For a reason which will become clear in the next subsection, the $\tz_0$ charge takes a different value, $\tz_0 F = -2(g-1) F$. Thus, we have
\begin{equation}\label{modconstr}
    \ep^{ij} {\cal D}_{{\hat A}\, ai}{\cal D}_{{\hat B}\, bj} F^{-2(g-2)} =  (g-1)\ \delta_{{\hat A}{\hat B}}\ep_{ab} F^{-2(g-2)}\ .
\end{equation}
We would like to point out that in the string theory analysis given in the
following subsections, the
differential equations are obtained on functions ${\mathcal F}$  which is the
relevant part of $F$ that survives the harmonic space integrals.  Indeed
string theory amplitudes directly see ${\mathcal F}$.
The crucial step used in equation (\ref{correctcons}) was that $F$ does
not depend on two combinations of moduli (projection $a+$) as is expressed
in the S-analyticity constraint (\ref{covsancon}). It is easy to see
that ${\mathcal F}$ does not satisfy this S-analyticity constraint since it is
obtained by making a
certain $SU(2)$ projection on $F$.  Therefore the individual steps in this
derivation cannot be applied to ${\mathcal F}$. However, the second order
differential operators considered here are
not sensitive to any particular $SU(2)$ projection of $F$ and therefore
the final equations are
still true on ${\mathcal F}$.

\subsection{$\cN=2$ conformal supersymmetry and supergravity}\label{css}

\subsubsection{G-analytic coset realization of $SU(2,2/2)$}

Here we show that the realization of G-analytic superfields of the type (\ref{01}) as
functions on a particular coset of the $\cN=2$ conformal superalgebra $SU(2,2/2)$ is
very similar to the bosonic coset construction of the preceding subsection. This algebra
involves the generators of Lorentz transformations (${\cal M}_{\m\n}$), translations ($P_\m$),
conformal boosts ($K^\m$), dilatation ($D$), R symmetry $SU(2)\times U(1)$ ($I_i^j,\ R$), Poincar\'{e}
supersymmetry ($Q^\a_i$ and $\bar Q_\da^i$) and special conformal supersymmetry
($S_\a^i$ and $\bar S^\da_i$). The (anti)commutation relations relevant for our discussion are \footnote{We use the conventions of \cite{Galperin:2001uw}, except from the rescaling of the $SU(2)$ generators $I \to -i/2 I$.}
\begin{eqnarray}
\big\{Q_i^\alpha,S_{\beta }^j\big\} &=&
\delta_i^j(\sigma^{\mu\nu})_{\beta }^{\alpha}{\cal M}_{\mu\nu} - 2
\delta_{\beta }^\alpha I_i^j + 2i\delta_{\beta }^\alpha \delta_i^j
D - 2\delta_{\beta }^\alpha \delta_i^j R\,,
\label{scnfal}\end{eqnarray}
\begin{align}
&[D,Q] = {i\over 2}Q\,,&& [D,\bar Q]={i\over 2}\bar Q\,,&&[D,S]=-{i\over 2}S\,, &&[D,\bar S]=-{i\over 2}\bar S\,,\label{2.21'}
\end{align}
together with the ${SU}(2)$ relations
\begin{align}
&[I^i_j,I^k_l] = 2 (\delta^k_j I^i_l - \delta^i_l I^k_j)\,, && \big[I^i_j, Q_k\big] = -\delta^i_k  Q_j + {1\over 2}\delta^i_j Q_k \,,\label{SUNcomm}
\end{align}
and similarly for $S$. The standard superspace corresponds to the coset
\begin{equation}\label{44}
    \frac{SU(2,2/2)}{({\cal M},K,D,R,S, \bar S, I)}\ \sim (x^\m, \q_\a^i, \bq^\da_i)\,,
\end{equation}
involving all the 8 Grassmann variables associated with the supersymmetry generators.
In order to obtain G-analytic superfields depending on half of these Grassmann variables, we add the  $SU(2)$ harmonic projections of
the $Q$ generators $Q^\a_{-} = \bu^i_{-}\, Q^\a_i $ and $\bar Q_\da^{+} = \bar
Q_\da^i\, u^{+}_i $ to the coset denominator, thus leaving only the odd coordinates
$\q_\a^{+}$ and $\bq^\da_{-}$ in the coset. However, exactly as in the bosonic case
of Section \ref{coset}, the $SU(2)$ generator $I_{++}$ converts $Q_-$ and $\bar Q^+$
from the coset denominator into the coset generators $Q_+$ and $\bar Q^-$. In order to
avoid this, we introduce the external automorphism group $\widehat{SU(2)}$ with
generators $T$. Then the combination $I_{++}-T_{++}$ commutes with all the $Q$'s and
thus can be safely put in the coset denominator:\footnote{Here we follow the formulation of $\cN=2$ conformal supersymmetry of \cite{Galperin:1985zv,Galperin:2001uw}. A somewhat different approach is proposed in \cite{Hartwell:1994rp}.}
\begin{equation}\label{45}
    \frac{SU(2,2/2)\subset\hskip-11pt\times\  \widehat{SU(2)}}{({\cal M},K,D,R,S, \bar S, Q_-, \bar Q^+,  I_0, I_{--}, I_{++}-T_{++}, T_0)}\ \sim (x, \q^+, \bq_-, w)\ .
\end{equation}
Here the harmonics $w$ are defined in exactly the same way as in Section \ref{coset},
eq.~(\ref{wharm}), replacing the $SU(2)$ harmonics $\kappa$ by R-symmetry $SU(2)$ harmonics. They transform as in (\ref{trw}) with the parameter $\hat\lambda$
replaced by the G-analytic superparameter
\begin{equation}\label{46}
\Lambda_{-}{}^{+}(x, \q^+, \bq_-, w) = \bar w_{-}{}^i \lambda_i{}^j w_j{}^{+} + i\q^{+}\sigma^\m \bq_{-} k_\m + i   \bar w_{-}{}^i \eta^\a_i \q^{+}_\a + i \bq_{-}^\da \bar\eta_\da^i w_i{}^{+}\,,
\end{equation}
containing the parameters $\lambda$ of the R-symmetry $SU(2)$, $k$ of conformal boosts and $\eta$ of special conformal supersymmetry.

\subsubsection{G-analytic representations}\label{Ganrep}

The basic G-analytic conformal superfield $q^+(x, \q^+, \bq_-, w)$ (\ref{01}) (with
superconformal harmonics $w$ instead of $u$) describes the hypermultiplet. It
transforms with a G-analytic superconformal weight factor:
\begin{eqnarray}
  &&  \delta q^+ = {q^+}'(x',\q',\bq',w') - q^+(x,\q,\bq,w) = \Lambda q^+ \,,  \label{47}\\
  &&  \Lambda(x, \q^+, \bq_-, w) = (\rho+i\tau) +k_\m x^\m + \bar w_{+}{}^i \lambda_i{}^j w_j{}^{+} + i   \bar w_{+}{}^i \eta^\a_i \q^{+}_\a  + i \bq_{-}^\da \bar\eta_\da^i w_i{}^{+}\,, \nn
\end{eqnarray}
where $\rho$ is the parameter of dilatations and $\tau$ of the $U(1)$ R symmetry.\footnote{It can be shown that $\Lambda_{-}{}^{+} = D_{-}{}^{+}\Lambda$.}

An important property of the `short' (BPS) representations of the $\cN=2$ superconformal group is that their $U(1)$ charge $I_0$ must be equal to their conformal weight \cite{Galperin:2001uw,Ferrara:2000eb}. This follows form the conditions of (super)conformal primarity, e.g. for $q^+$
\begin{equation}\label{scpr}
    Sq^+ = \bar S q^+ = K q^+ = 0\ ,
\end{equation}
together with the half-BPS conditions
\begin{equation}\label{halfbps}
    Q_- q^+ = \bar Q^+ q^+ = 0\ .
\end{equation}
The algebra \p{scnfal} then implies constraints on the quantum numbers of the BPS representation. Consider, for instance, the anticommutator
 \begin{equation}\label{conantic'}
    \{ Q_-^{\a}, S_{\b}^-  \} = {\cal M}^{\b}_{\a} + 2\delta^{\b}_{\a}(I_0 + iD - R)\,,
\end{equation}
obtained by projecting \p{scnfal} with $\bar u_-^i u^-_j$.
Then the consistency conditions for the subset of constraints $S q^+ = Q_- q^+ = 0$ are
\begin{equation}\label{eqcha'}
     {\cal M}^{\b}_{\a} q^+ =(I_0 + iD - R) q^+ =0\ ,
\end{equation}
i.e., such representations must have Lorentz spin $(0,j_2)$ and conformal dimension $d=i_0-r$, where $d$ is the eigenvalue of the dilatation operator $-iD$ \footnote{With this definition of the conformal dimension we achieve conventional values for the weights of, e.g., the (conformal) supercharges $Q$ and $S$ from  \p{2.21'}.} and $r$, $i_0$ are the eigenvalues of the $R$ charge and $U(1)$ charge ($I_0$) generators, correspondingly. Repeating the same analysis, but this time with  the anticommutator
\begin{equation}\label{conantic}
    \{ \bar Q^+_{\da}, \bar S^{\db}_+  \} = {\cal M}^{\da}_{\db} - 2\delta^{\da}_{\db}(I_0 + iD + R)\ ,
\end{equation}
we find the additional conditions
\begin{equation}\label{eqcha''}
     {\cal M}^{\da}_{\db} q^+ =(I_0 + iD + R) q^+ =0\ .
\end{equation}
The combination of \p{eqcha'} and \p{eqcha''} implies that the 1/2-BPS  representations must be Lorentz scalars with vanishing R charge and $d=i_0$. In the case of the hypermultiplet $q^+$ we have $d=i_0=1$.

The analogous statement for a chiral superfield, e.g. for the vector multiplet $W$, annihilated by $\bar Q^+_{\db}W=\bar Q^-_{\db}W=0$, is $I_{++}W = I_{--} W=I_0W=(D-iR)W=0$. This means that it must be a singlet under the R symmetry $SU(2)$. In addition, it must have $d=r$, but the value is not fixed by the superconformal algebra. The conformal dimension of $W$ is  determined from the vector multiplet Lagrangian $\int d^4x d^4\q\ W^2$. The chiral measure has dimension $-2$, so $d_W=r_W=1$, which yields the standard $R$ charge assignment $r_\q=1/2$.

The other G-analytic object we are
discussing here is $K^\a_- = D^\a_- W$ (\ref{16}). Unlike the superconformal primary $W$, $K^\a_-$ is a descendant and as such cannot be annihilated by all the generators $S, \bar S$. Indeed, the spinor derivative $D^\a_-$ is assimilated to the supersymmetry generator $Q^\a_-$, which does not anticommute with $S_\b^-$. As a consequence, we loose the consistency conditions following from \p{conantic'}. The remaining half \p{eqcha'} is in accord with the Lorentz spin $(1/2,0)$ of $K^\a_-$, and in addition fixes $d=r+i_0$. For $K^\a_- = D^\a_- W$ we have $d= 3/2$ and $r=1/2$, so we obtain $i_0=1$.

\subsubsection{Conformal supergravity}

The generalization to $\cN=2$ conformal supergravity is done by replacing the parameters
$\Lambda_{-}{}^{+}$ and $\Lambda$ in \p{46}, \p{47} by arbitrary G-analytic superfields. Poincar\'{e}
supergravity is obtained by coupling the Weyl multiplet to {\it two types of compensating multiplets}. The first is a vector multiplet
\begin{equation}\label{firstisav}
    W_0(x,\q) = \phi(x) + \mbox{$\q$ terms}\ .
\end{equation}
It transforms as a density\footnote{If rewritten in the G-analytic frame.},
\begin{equation}\label{ifrewr}
    \delta W_0 = \Lambda\, W_0\ .
\end{equation}
The second compensator is a hypermultiplet (cf. (\ref{01}))
\begin{equation}\label{48}
     y^+_{i}(x, \q^+, \bq_-, w) = \varphi_{i}{}^{k}(x) w_k{}{}^{+} + \mbox{$\q$ terms}\ .
\end{equation}
Here we see the $2\times 2$ matrix of compensating real scalars $\varphi_{i}{}^{k}$. Let us
consider the following projections of $y_{i}{}^+$ with the harmonics $w$:
\begin{align}
&y^{++} = \bar w_-{}^i\ y^+_{i}\,, && y_0 =  \bar w_+{}^i\ y^+_{i}\ .\label{49}
\end{align}
It is easy to check that they transform as follows:
\begin{align}
&\delta y^{++} = \Lambda_{-}{}^{+} \, y_0 + \Lambda\, y^+\,, && \delta y_0 = \Lambda\, y_0\ ,\label{49'}
\end{align}
so their ratio transforms as a {\it compensator} for the local superconformal
transformations:
\begin{equation}\label{50}
    \delta \left(\frac{y^{++}}{y_0} \right) =  \Lambda_{-}{}^{+}\ .
\end{equation}
Then, with the help of this compensator we can define new harmonics {\it inert under the
local superconformal transformations} (notice the similarity with (\ref{wharm}) and (\ref{trw})):
\begin{align}
& v_i{}^{+} = w_i{}^{+} - w_i{}^{-}\frac{y^{++}}{y_0}\,, && v_i{}^{-} = w_i{}^{-}\,, \nn\\
& \bar v_{+}{}^i = \bar w_{+}{}^i\,, &&  \bar v_{-}{}^i = \bar w_{-}{}^i + \frac{y^{++}}{y_0} \bar w_{+}{}^i\,, \label{vharm}\\
& \delta v = \delta \bar v = 0 \ . \nn
\end{align}

The role of the compensators is to completely absorb the local superconformal transformations. This allows us to use the parameter $\Lambda_{-}{}^{+}$ in (\ref{50}) to fix a gauge in which $y^{++}=0$, thus identifying the harmonics $v$ and $w$. This means, in particular, that the conformal $SU(2)$ (generators $I$ in (\ref{45})) is identified with $\widehat{SU(2)}$ (generators $T$ in (\ref{45})). By the same logic, we can use the parameter $\hat\lambda_-{}^+ $ of local $SU(2)$ transformations in (\ref{deltaz}) to gauge away the compensator $z_-{}^+$. This results in the identification of the harmonics $w$ with $u$. So, at the expense of manifest covariance, the different $SU(2)$ groups discussed above are reduced to a unique one, and the harmonics to the original ones (\ref{12'}). This gauge fixing procedure establishes a bridge between the S-analytic coset (\ref{30}) and the G-analytic coset (\ref{45}).

One further step in gauge fixing is to use the complex scalar parameter in $\Lambda$ (dilatations and R-symmetry $U(1)$) to gauge away two of the three real scalars in the compensators (the complex scalar $\phi$ in \p{firstisav} and the real $\det \varphi_{i}{}^{k}$ from \p{48}). The remaining real scalar matches the auxiliary field of the Weyl multiplet to form a Lagrange multiplier pair (see, e.g., \cite{Galperin:2001uw}).

Finally, we proceed to the conformal covariantization of the higher-derivative
term (\ref{19'}). It is achieved in three steps. Firstly, we replace the explicit
harmonics $u$ in $F(q,u)$ by the new inert ones $v$ (however, the superfields $q$ still
depend on the conformal harmonics $w$). Secondly, we introduce weightless G-analytic
superfields $q/y_0$. In this way the potential $F(q,v)$ becomes conformally invariant.
Thirdly, we use the compensating vector multiplet $W_0$ and  the G-analytic density $y_0$ to balance the R charge and the conformal weight  of the measure and of the
gaugino factor.
The result is
\begin{equation}\label{19bis}
S = \int \, d^4x \, du \, d^2\q^+ d^2\bq_{-} \, (K_{-}\cdot K_{-})^{g}  \, W_0^{-g}\, y_0^{-2(g-1)} \   F^{-2(g-2)}\left({q^{+}}/{y_0},v\right)\ .
\end{equation}
The factor $W_0^{-g}$ seems to break the G-analyticity of the integrand in \p{19bis}. Being chiral, $W_0$ is annihilated by $\bar D^+_{\da}$, but {\it a priori} not by $D^\a_-$. However, remember that $W_0$ is a compensating vector multiplet for $\cN=2$ Poincar\'e supergravity, whose `gaugino' component $D^\a_- W_0|_{\q=0}$ vanishes on shell (it forms a Lagrange pair with the auxiliary spinor from the $\cN=2$ Weyl multiplet). Thus, for our purposes we may consider that $D^\a_- W_0 = 0$, so the  integrand in \p{19bis} is G-analytic.     With the help of the table \p{tabu} it is easy to check the balance of the conformal weights $d$ and R charges $r$ in the action term \p{19bis}.
\begin{equation}\label{tabu}
\begin{tabular}{|c|c|c|c|c|}
  \hline
       & $d$ & $r$ & $i_0$ & $T_0$ \\
       \hline
  $y_0$  & 1 & 0 & 1 & 0  \\
  $W_0$  & 1 & 1 & 0 & 0  \\
  $K_{-}$& 3/2 & 1/2 & 1 & 1  \\
  $q^+$& 1 & 0 & 1  & 1 \\
  $d^4x d^2\q^+ d^2\bq_{-}$& -2 & 0 & -2 & -4  \\
  \hline
\end{tabular}
\end{equation}

At this point we recall the discussion from section \ref{Ganrep}  about the superconformal properties of the descendant $K_{-}$.  It is 1/2 BPS-like, i.e. it is annihilated by the supercharges $Q_-, \bar Q^+$ from the coset denominator in (\ref{45}). However, unlike the hypermultiplets $q^+$ and $y_0$, it is not a superconformal primary, being annihilate only by $\bar S^+$. This implies that only  $\bar S^+$ can be regarded as a symmetry of the term \p{19bis}. Further, the combined action of $Q_-, \bar Q^+$ and $\bar S^+$ in \p{19bis} fixes the  $I_0$ charge  $i_0 = d-r$ in terms of the conformal weight $d$ and $R$ charge $r$. In \p{tabu} we have listed the resulting charges $i_0$ of all objects appearing in \p{19bis}.

Now we can explain why in (\ref{modconstr}) we took the value  $\tz_0 F = -2(g-1) F$ of the charge $\tz_0$, different from that of the charge $T_0$. The local $SU(2)$ gauge-fixing procedure (elimination of the compensators) results in the identification of the $\Upsilon_0$ charge from (\ref{30}) with the $I_0$ charge from  (\ref{45}). We need to determine the value of the charge  $I_0$ for the  covariantized function $F$. We can consider the two densities $W_0$ and $y_0$ as parts of the {\it covariantized} function $F$. Thus, according to \p{tabu}, the $i_0$ charge of  $F$  in \p{19bis} should equal $-2(g-1)$.  The automorphism charge $T_0$ of $F$ remains independent and, indeed, takes a different value.
%%%%%%%%%%%%%%%%%%%%%%%%%%%%%%%%%%%%%%%%%%%%%%%%%%%%%%%%%%%%%%%%%%%%%%%
%%%%%%%%%%%%%%%%%%%%%%%%%%%%%%%%%%%%%%%%%%%%%%%%%%%%%%%%%%%%%%%%%%%%%%%
\section{Topological amplitudes in generic Calabi-Yau compactifications}\label{App:RelTopPhys}

In Sections \ref{Sect:TypeIamp} and \ref{Sect:hetF3} we considered a string amplitude stemming from the effective action (\ref{19'}) which involves two self dual field
strengths, $(2g-4)$ gauginos and two chiral vector multiplet scalars each
carrying one momentum. This amplitude computes the reduced function ${\mathcal F}_g$ of equation (\ref{24}). However, as pointed out in
these sections a direct computation of this term is complicated by the
presence of many terms in the  vertex operators. In this Section we therefore
compute another term that also comes from the action (\ref{19'})
and for which the string amplitude is easier to calculate. 
%%%%%%%%%%%%%%%%%%%%%%%%%%%%%%%%%%%%%%%%%%%%%%%%%%%%%%%%%%%%%%%%%%%
\subsection{Supersymmetrically related amplitude}
The effective action term in (\ref{19'}) which we are going to consider is
obtained by taking the lowest components from the superfield $K_-$ and
saturating the two $\theta$ and the two $\bar{\theta}$ from the
fermionic
components of the hypermultiplets. Thus we obtain
the term
\begin{equation}
(\lambda_-\cdot\lambda_-)^g (\Psi^{{  A}_1}\cdot
\Psi^{{  A}_2})(\bar{\Psi}^{{  A}_3}\cdot\bar{\Psi}^{{  A}_4}) (F^{-m})_{{  A}_1 {  A}_2
  {  A}_3 {  A}_4}\,,
\label{4der}
\end{equation}
where $\Psi^A \equiv \Psi^{\hat A a} = (\psi^{\hat A},\chi^{\hat A})$ (see \p{01})  and
$ (F^{-m})_{{  A}_1 {  A}_2 {  A}_3 {  A}_4}$ is four derivatives of $F^{-m}$ with respect to $q^{+{  A}_\kappa}$
with $\kappa=1,...,4$. Extracting the irreducible harmonic structure
$\bar{u}^{i_1}_+\cdots\bar{u}^{i_{2g}}_+$
needed to match the conjugate structure in
$(\lambda_-\cdot\lambda_-)^g$ we find the reduced function
\begin{equation}
({\mathcal F}_g)_{{  A}_1 {  A}_2 {  A}_3 {  A}_4}= \sum_{n\ge 4} \frac{n!}{(n-4)!} \xi_{(i_1  \cdots i_{2g+n-4})\ {{  A}_1 {  A}_2 {  A}_3 {  A}_4
  {  B}_1\cdots {  B}_{n-4} }}\  \bu_{+}^{i_1}
    \cdots \bu_{+}^{i_{2g}}  \ f^{i_{2g+1}\ {  B}_1} \cdots
    f^{i_{2g+n-4}\ {{  B}_{n-4}} } \ .
\label{redfn4der}
\end{equation}
It is easy to see that $({\mathcal F}_g)_{{  A}_1 {  A}_2 {  A}_3 {  A}_4}$ is just four
derivatives of ${\mathcal F}_g$ given in (\ref{24}) with respect to
$f^{+A_\kappa}$ for $\kappa=1,...,4$ where $\frac{\pa}{\pa f^{+  A}}=\bar{u}^i_+ \frac{\pa}{\pa
f^{iA}}$.  

It turns out that in this amplitude there is only one possible contraction
among the picture changing operators and we are therefore able to explicitly calculate $({\mathcal F}_g)_{{  A}_1 {  A}_2 {  A}_3 {  A}_4}$. Thereby we will show that the result is indeed four hyper derivatives of the topological expression (\ref{topamp2}).

Moreover, for the sake of putting the correspondence between string amplitudes and topological correlators on a broader basis we will generalize the compactification manifold from orbifolds (as considered in
Sections \ref{Sect:TypeIamp} and \ref{Sect:hetF3}) to the most general $(4,0)$ compactification.

Finally, (as has been done in this entire paper), in order not to complicate
the formulae, we will suppress all the vector indices: all the gauginos
come with vector indices and therefore the corresponding amplitudes will
carry these indices. However, we will keep track of the hypermultiplet
indices consistently.
%%%%%%%%%%%%%%%%%%%%%%%%%%%%%%%%%%%%%%%%%%%%%%%%%%%%%%%%%%%%%%%%%%%%%%%
\subsection{Generalities about Calabi-Yau compactifications}
We start by describing the essential points of the general $(4,0)$
compactification. The $SU(2)$ current algebra inside the $\cN=4$
superconformal algebra can be bosonized in terms of a free boson $H$ so
that:
\begin{align}
&J_{K3}=i\sqrt{2}\partial H\,,&& J_{K3}^{\pm \pm} = e^{i\pm \sqrt{2}
H}\,,&&G^{\pm}_{K3,i}= e^{\pm \frac{i}{\sqrt{2}} H} \hat{G}_{K3,i}\,,
\end{align}
where $\hat{G}_{K3,i}$ have dimension $5/4$, have non-singular OPE with
$H$ and have no spin structure dependence. The spin structure dependence
enters through the projections and shifts in the $U(1)$ charge lattice of
$J_{K3}$ which in turn is given by the momentum lattice of $H$.  Therefore
in the $(4,0)$ internal theory, only  correlation functions and the
partition function of $H$ depend on the spin-structure. The term in the
picture changing operator containing $(4,0)$ superconformal generators is
\begin{equation}
P = e^{\phi} ( G^+_{K3,+} + G^-_{K3,-} )+...
\end{equation}
where dots indicate the remaining terms and $\phi$ bosonizes the superghost.

The vertex operators contain a part that involves the space-time and torus
conformal theories which remain the same as in the discussion in the text.
We will here only point out their dependence on the $(4,0)$ conformal
field theory.
The vertex operators have the following dependence on $H$. The chiral
vector multiplet scalar in the $(-1)$-ghost picture is simply $\psi_3$ ie.
it does not depend on the $K3$ fields. The gaugino vertex, however,
involves,
besides the space-time and torus spin-field, also $e^{\pm
\frac{i}{\sqrt{2}} H}$. Thus the vertex operator for the gauginos
$\lambda_{\mp}$ and  $\bar{\lambda}^{\pm}$ in the $(-1/2)$-picture carry
$e^{\pm \frac{i}{\sqrt{2}} H}$.

Finally, the vertex operators (at zero momentum) for the hyperscalar
$f_A^{\pm}$ in $(-1)$ and $(0)$ ghost pictures are
\begin{align}
&V^{(-1)}_{f_A^{\pm}}= e^{-\phi}e^{\pm \frac{i}{\sqrt{2}} H}
\hat{V}_{A}\,,&&\text{and} &&V^{(0)}_{f_A^{\pm}}=P
V^{(-1)}_{f_A^{\pm}}=\lim_{z\to0} \sqrt{z}\hat{G}_{K3,\pm}(z) \hat{V}_{A}(0)\,,
\label{hypervertex}
\end{align}
where $\hat{V}_{A}$ have dimension $1/4$ and have non-singular OPE with
$H$.

The vertex operators in $(-1/2)$ ghost picture for the hyperfermions
$\chi_A$ and $\bar{\psi}_A$ are
\begin{eqnarray}
V^{(-1/2)}_{\chi^{\alpha}_A} &= & e^{-\frac{\phi}{2}} S^{\alpha}
e^{-i\frac{\phi_3}{2}} \hat{V}_{A},\nonumber\\
V^{(-1/2)}_{\bar{\psi}^{\dot{\alpha}}_A} &=& e^{-\frac{\phi}{2}}
S^{\dot{\alpha}} e^{i\frac{\phi_3}{2}} \hat{V}_{A}\,,
\end{eqnarray}
where $S^{\alpha}$ are the space-time spin fields.

As mentioned above the spin structure dependence enters only through the
super-ghost, spacetime and
torus fermions and the charge lattice of $H$. It does not depend on the
rest of the details of the $(4,0)$ superconformal theory. On the other
hand the topological theory (besides shifting the dimensions of  torus
fermion) involves precisely twisting  by adding an appropriate background
charge for the field $H$ and the rest of the internal $(4,0)$ theory is
insensitive to this twisting. This fact will enable us to show the
equivalence between the physical string amplitude and the topological
amplitude for an arbitrary $(4,0)$ internal theory.

Let $\Gamma$ be the $U(1)$ lattice of $H$ charges. The space-time and
torus fermions define an $SO(2)\times SO(2) \times SO(2)$ lattice. If one
takes an $SO(2)\times SO(2)$ sublattice of this and combines with
$\Gamma$, then, as has been shown in
\cite{Lerche:1988np,Lechtenfeld:1989be,Lust:1988yf}, the resulting
3-dimensional lattice is given by the coset $E_7/SO(8)$. The characters
are given by the branching functions $F_{\Lambda,s}(\tau)$ satisfying:
\begin{equation}
\chi_{\Lambda}(\tau)= \sum_{s} F_{\Lambda,s}(\tau) \chi_s(\tau)\,,
\end{equation}
where $\chi_{\Lambda}$ and $\chi_s$ are the $E_7$ and $SO(8)$ level one
characters respectively,
$\Lambda$ denotes the two conjugacy classes of $E_7$ and $s$ represent the
four conjugacy classes of $SO(8)$ in the spin structure basis. The
characters of the internal $(4,0)$ superconformal field theory times two
free complex fermions can therefore be expressed as $\sum_{\Lambda}
F_{\Lambda,s}(\tau) Ch_{\Lambda}(\tau)$ where $Ch_{\Lambda}(\tau)$ is the
contribution of the rest of the internal theory
and most importantly does not depend on the spin-structure. The
generalization to higher genus is obtained by assigning an $E_7$ conjugacy
class $\Lambda$ for each loop and we will denote this collection
by $\{\Lambda\}$. We can define a more general character
$F_{\{\Lambda\},s}(u_1,u_2,v)$ by introducing chemical potentials for the
three charges; $u_1$ and $u_2$ coupling to the two $SO(2)$ charges and $v$
to $H$-charge. For a genus $g$ surface, the couplings $u_1$, $u_2$ and $v$
each are $g$-dimensional vectors and represent the coupling to charges
going through each loop. In the calculation of the amplitudes
$(u_1,u_2,v)$ are related to the positions of various vertex operators
weighted by the corresponding charges via Abel map. The spin structure sum
is given by the formula:
\begin{equation}
\sum_{s} F_{\{\Lambda\},s}(u_1,u_2,v)=
F_{\{\Lambda\}}(\frac{1}{2}(u_1+u_2+\sqrt{2}v),\frac{1}{2}(u_1+u_2-\sqrt{2}v),
\frac{1}{\sqrt{2}}(u_1-u_2))\,,
\label{sssum}
\end{equation}
where
\begin{eqnarray}
F_{\{\Lambda\}}(u_1,u_2,v)&=&
\vartheta(\tau,u_1)\vartheta(\tau,u_2)\Theta(\tau,v)\,,\nonumber\\
\Theta(\tau,v)&=&\sum_{n_i \in Z} e^{2\pi i
(n_i+\frac{\lambda_i}{2})\tau_{ij}(n_j+\frac{\lambda_j}{2})+2\pi i
\sqrt{2}(n_i+\frac{\lambda_i}{2})v_i}\,,
\end{eqnarray}
where $\lambda_i$ (with $i=1,...,g$) are $0$ and $1$ for the $E_7$
conjugacy classes $(\underline{1})$ and $(\underline{56})$ respectively.
In fact, apart from the non-zero mode determinant of a scalar,  $\Theta$
is just the character valued genus $g$ partition function
of level one $SU(2)$, with the two classes above corresponding to the two
representations of level one $SU(2)$ Kac-Moody algebra based on $SU(2)$
representations $(\underline{1})$ and $(\underline{2})$.

%%%%%%%%%%%%%%%%%%%%%%%%%%%%%%%%%%%%%%%%%%%%%%%%%%%%%%%%%%%%%%%%%%%
\subsection{The amplitude $\mathcal{F}_g$}\label{App:CYtopGeng}
Now we are in a position to compute the amplitude involving $2 g$
gauginos, 2 chiral hyperinos $\chi$ and 2 anti-chiral
hyperinos $\bar{\psi}$ on a genus $g$ Riemann surface. First we consider
the $g>1$ case.  All these $2g+4$ fermions will be in the
$(-1/2)$-picture, therefore the total number of picture changing operators
is $(2g-2)+(g+2)=3 g$. For convenience we give the following table
containing the vertex operators and the picture-changing operators, their
fermion charges with respect to space-time and torus fermions (bosonized
in terms of scalars $\phi_1$, $\phi_2$ and $\phi_3$ respectively) and the
$H$-charge.

\begin{center}
\begin{tabular}{|c|c|c||c|c||c||c||c|}\hline
\textbf{field} & \textbf{pos.} & \textbf{number} &
\parbox{0.5cm}{\vspace{0.2cm}$\phi_1$\vspace{0.2cm}}&
\parbox{0.5cm}{\vspace{0.2cm}$\phi_2$\vspace{0.2cm}} &
\parbox{0.5cm}{\vspace{0.2cm}$\phi_3$\vspace{0.2cm}} &
\parbox{0.5cm}{\vspace{0.2cm}$H$\vspace{0.2cm}} &
\parbox{0.5cm}{\vspace{0.2cm}~\vspace{0.2cm}} \\\hline
gaugino & \parbox{0.35cm}{\vspace{0.2cm}$x_i$\vspace{0.2cm}} & $g$ &
\parbox{0.7cm}
{\vspace{0.2cm}$+\frac{1}{2}$\vspace{0.2cm}} & \parbox{0.7cm}
{\vspace{0.2cm}$+\frac{1}{2}$\vspace{0.2cm}} & \parbox{0.7cm}
{\vspace{0.2cm}$+\frac{1}{2}$\vspace{0.2cm}} & \parbox{0.7cm}
{\vspace{0.2cm}$+\frac{1}{\sqrt{2}}$\vspace{0.2cm}} & \parbox{0.4cm}
{\vspace{0.2cm}$\bar{J}$\vspace{0.2cm}} \\\hline
 & \parbox{0.35cm}{\vspace{0.2cm}$y_i$\vspace{0.2cm}} & $g$ & \parbox{0.7cm}
{\vspace{0.2cm}$-\frac{1}{2}$\vspace{0.2cm}} & \parbox{0.7cm}
{\vspace{0.2cm}$-\frac{1}{2}$\vspace{0.2cm}} & \parbox{0.7cm}
{\vspace{0.2cm}$+\frac{1}{2}$\vspace{0.2cm}} & \parbox{0.7cm}
{\vspace{0.2cm}$+\frac{1}{\sqrt{2}}$\vspace{0.2cm}} & \parbox{0.4cm}
{\vspace{0.2cm}$\bar{J}$\vspace{0.2cm}} \\\hline\hline
Hyperino $\chi_{A_1}$ & \parbox{0.35cm}{\vspace{0.2cm}$z_1$\vspace{0.2cm}}
& 1 & $+\frac{1}{2}$ & $+\frac{1}{2}$ & $-\frac{1}{2}$ & $0$ &
\parbox{0.7cm}
{\vspace{0.2cm}$\hat{V}_{A_1}$\vspace{0.2cm}} \\\hline
$\chi_{A_2}$& \parbox{0.35cm}{\vspace{0.2cm}$z_2$\vspace{0.2cm}} & 1 &
$-\frac{1}{2}$ & $-\frac{1}{2}$ & $-\frac{1}{2}$ & $0$  & \parbox{0.7cm}
{\vspace{0.2cm}$\hat{V}_{A_2}$\vspace{0.2cm}}\\\hline
$\bar{\psi}_{A_3}$ & \parbox{0.35cm}{\vspace{0.2cm}$z_3$\vspace{0.2cm}} &
1 & $-\frac{1}{2}$ & $+\frac{1}{2}$ & $+\frac{1}{2}$ & $0$  &
\parbox{0.7cm}
{\vspace{0.2cm}$\hat{V}_{A_3}$\vspace{0.2cm}}\\\hline
$\bar{\psi}_{A_4}$& \parbox{0.35cm}{\vspace{0.2cm}$z_4$\vspace{0.2cm}} & 1
& $+\frac{1}{2}$ & $-\frac{1}{2}$ & $+\frac{1}{2}$ & $0$  & \parbox{0.7cm}
{\vspace{0.2cm}$\hat{V}_{A_4}$\vspace{0.2cm}}\\\hline\hline

PCO & \parbox{0.4cm}{\vspace{0.2cm}$r_a$\vspace{0.2cm}} & $g$ & $0$ & $0$
& $-1$ & $0$ & \parbox{0.7cm}
{\vspace{0.2cm}$\partial X_3$\vspace{0.2cm}} \\\hline
& \parbox{0.4cm}{\vspace{0.2cm}$s_b$\vspace{0.2cm}} & $2 g$ & $0$ & $0$ &
$0$ & $-\frac{1}{\sqrt{2}}$  & \parbox{0.7cm}
{\vspace{0.2cm}$\hat{G}^-_{K3}$\vspace{0.2cm}}\\\hline

\end{tabular}
\end{center}
${}$\\[10pt]
In the last column we have indicated the  part of the operators that are
insensitive to the spin structures. The superghost part is not shown in
the table but it is understood that all the vertex operators are in
$(-1/2)$ ghost picture and hence come with $e^{-\phi/2}$ and the PCO come
with $e^{\phi}$.
Note that $g$ of the picture changing operators at $r_a$ contribute the
torus part and $2 g$ of them at $s_b$ the $K3$ part. Of course one needs
to take into account all possible distributions of the total number $3 g$
of the picture changing operators into these two classes which will be
important in cancelling various singularities. Note also that by charge
conservation this is the only possible contribution  coming from the
picture changing operators in contrast with the amplitude considered in
Sections \ref{Sect:TypeIamp} and \ref{Sect:hetF3}. The correlation
function in a given spin-structure $s$ can be  easily computed with the
result:
\begin{eqnarray}
(\mathcal{F}_{g,s})_{A_1A_2A_3A_4}&=& F_{\{\Lambda\},s}(u_1,u_2,v)
G_{\{\Lambda\}}(\{x_i,y_i,z_k,r_a,s_b\})\nonumber \\
&~&\cdot \frac{\vartheta_s(\frac{1}{2}\sum_i
(x_i-y_i)+\frac{1}{2}(z_1-z_2-z_3+z_4))}{\vartheta_s(\frac{1}{2}\sum_i(x_i+y_i)+\frac{1}{2}(z_1+z_2+z_3+z_4)
- \sum_a r_a-\sum_b s_b- 2 \Delta)}
\nonumber\\&~&
\cdot \frac{\prod_{i<j} E(x_i,x_j)E(y_i,y_j)}{\prod_i
E(x_i,z_2)E(y_i,z_1)}\frac{\prod_a
E(r_a,z_1)E(r_a,z_2)}{\prod_{a,b}E(r_a,s_b)}\nonumber\\&~&\cdot
\left(\frac{\prod_{b,k}E(s_b,z_k)}{\prod_{k<l}E(z_k,z_l)
\prod_{b<c}E(s_b,s_c)}\right)^{1/2}
\label{gamp}
\end{eqnarray}
where
\begin{eqnarray}
u_1 &=& \frac{1}{2}\sum_i
(x_i-y_i)+\frac{1}{2}(z_1-z_2+z_3-z_4))\,,\nonumber\\
u_2 &=&\frac{1}{2}\sum_i (x_i+y_i)+\frac{1}{2}(-z_1-z_2+z_3+z_4))-\sum_a
r_a\,, \nonumber\\
v &=& \frac{1}{\sqrt{2}}(\sum_i(x_i+y_i)-\sum_b s_b)\,.
\end{eqnarray}
In (\ref{gamp}), $G_{\{\Lambda\}}$ includes the correlation function of
the hatted fields, zero modes of $\partial X_3$, non zero mode
determinants and some factors involving $g/2$-differentials $\sigma$ that
have no zeroes and poles and which essentially make the above expression
transform correctly under conformal transformations and the
monodromies around various cycles. An important point here is that
$G_{\{\Lambda\}}$ is independent of
spin structures. We have only shown above the spin structure dependent
parts as well as the prime forms that come from the correlation functions
of the bosonized fermions and superghosts and the $H$ fields.

We can now choose the following gauge so that the theta functions in the
numerator and denominator of the second  line in (\ref{gamp}) cancel each
other:
\begin{equation}
\sum_{l=1}^{3g} p_l \equiv \sum_{a=1}^g r_a + \sum_{b=1}^{2g} s_b = \sum_i
y_i +z_2 + z_3 + 2 \Delta
\label{gaugechoice}
\end{equation}
Here $p_l$ for $l=1,...,3g$ are the positions of the picture changing
operators, of which $g$ contribute to $G^-_{T^2}$ (whose positions are
indicated by $r_a$) and $2g$ contribute to $G^-_{K3}$ at $s_b$. In
other words $r_a$ and $s_b$ are just partitioning of $p_l$.

We can perform the spin structure sum using  (\ref{sssum}) with the result
that $F_{\{\Lambda\},s}(u_1,u_2,v)$ is replaced by
$F_{\{\Lambda\}}(u'_1,u'_2,v')$ where, after using the gauge condition
(\ref{gaugechoice}),
\begin{equation}
u'_1= \sum_i x_i-z_2-\Delta, ~~~ u'_2= \sum_a r_a-z_3-\Delta, ~~~~v'=
\frac{1}{\sqrt{2}}(\sum_b s_b-z_1-z_2-z_3+z_4-2\Delta)\label{vprimerel}
\end{equation}
Note that the argument of $\Theta_{\{\Lambda\}}$ is now $v'$ which is
exactly the $U(1)$ charge lattice part of the correlation function
\begin{equation}
\langle \prod_b e^{-i\frac{1}{\sqrt{2}} H}(s_b) e^{-i\frac{1}{\sqrt{2}}
H}(z_4) e^{+i\frac{1}{\sqrt{2}} H}(z_1)e^{+i\frac{1}{\sqrt{2}}
H}(z_2)e^{+i\frac{1}{\sqrt{2}} H}(z_3)\rangle\,,
\end{equation}
of the $H$ fields. The presence of $-\sqrt{2} \Delta$ in $v'$ in (\ref{vprimerel}) indicates that this
correlation function includes a background charge for the $H$ field, with
the stress tensor of the $H$ field $T_H$ shifted to
$T_H \rightarrow -\frac{1}{2} (\partial H)^2 + i\frac{1}{2}\partial^2 H$.
With the modified stress tensor, the operator $e^{i\frac{p}{\sqrt{2}} H}$
has  conformal dimension $p(p-2)/4$. Taking into account that the operator
$\hat{G}_{K3}(s_b)$ has dimension $5/4$ and the operators $\hat{V}_A$ have
dimension $1/4$ we see that the total dimensions appearing from the
$(4,0)$ superconformal theory are: dimension $+2$ at $s_b$, dimensions $0$
at $z_1,z_2,z_3$ and dimension $1$ at $z_4$ (of course taking into account
the contribution coming from the rest of the conformal theory these
dimensions will change as we will see below).

Multiplying the resulting expression by identity (due to the gauge condition)
\begin{equation}
1= \frac{\vartheta(\sum_i y_i-z_1-\Delta)}{\vartheta(\sum_a r_a + \sum_b
s_b -z_1-z_2-z_3-3\Delta)}
\end{equation}
and using the bosonization formulae for spin $(1,0)$ and $(2,-1)$ systems,
we can identify this amplitude with the following one in the twisted
theory
\begin{eqnarray}
(\mathcal{F}_{g})_{A_1A_2A_3A_4} &=& \int_{{\cal{M}}_g} (\mu\cdot
b)^{3g-3}\,\frac{\text{det}\omega_i(x_j)
\text{det}\omega_i(y_j)}{\text{det}(\text{Im}\tau)^2}\,
\langle\prod_{a=1}^g G^-_{T^2}(r_a)  \psi_3(z_3)\rangle_{T^2}\cdot
\nonumber\\&~&
\cdot \frac{\langle \prod_{b=1}^{2g} G^-_{K3}(s_b)
e^{-\frac{i}{\sqrt{2}}H}\hat{V}_{A_4}(z_4)
\prod_{k=1}^{3}
e^{+\frac{i}{\sqrt{2}}H}\hat{V}_{A_k}(z_k)\rangle_{K3}}{\langle\prod_{l=1}^{3g}
b(p_l) \prod_{k=1}^3 c(z_k) \rangle_{b,c}}
\label{FgA4}
\end{eqnarray}
Note that the denominator above involves
correlation function of the spin $(2,-1)$ $(b,c)$ system. One can check
that the dimensions at all the points are the correct ones; at $x_i$ and
$y_i$ we have obviously one differential, at both $r_a$ and $s_b$ the
$G^-$ have dimension $2$ in the twisted theory but in the denominator $b$
are also dimension 2 at these points so that the net dimension is zero as
it should be. At $z_1$, $z_2$ and $z_3$ the operators in the twisted
$(4,0)$ superconformal theory, as shown above, carry dimension zero but
the denominator contain operators $c$ (of dimension $(-1)$ at these points
resulting in the net dimension 1 as required by the fact that these points
have to be integrated. Finally, at $z_4$ we have only the numerator part
which as argued above carries dimension $1$ in the twisted theory.

It is convenient to take three of the positions of the picture changing
operators (say $p_{3g-2}$, $p_{3g-1}$ and $p_{3g}$) and put them at the
positions $z_1$, $z_2$ and $z_3$ respectively. Note that there is no
obstruction in doing this and simultaneously having the gauge condition
(\ref{gaugechoice}), as there are still $3g-3$ positions $p_l$ available
while the gauge condition involves $g$ complex equations (recall that we
are here discussing the case $g>1$).
The denominator has poles (with residue $1$) when one of the $p$'s
approaches $z_1$, $z_2$ or $z_3$. This means that the result will vanish
unless there is a pole also in the numerator. By inspection it is clear
that this will happen only if $G^-_{K3}$ contributes at these three
points, i.e. these three points must be in the class $\{s_b\}$. The result
is just the residue of the poles in the numerator. As follows from
(\ref{hypervertex}) this residue is just the $0$-picture vertex operator
for the hypermultiplet scalar
$\prod_{k=1}^3 V^{(0)}_{f^+_{A_k}}(z_k)$. Now let us consider the dependence
of the amplitude on the remaining $(3g-3)$ positions $p_l$. Take $p_1$ for
instance. Both the numerator and denominator correlation functions have no
pole as $p_1$ approaches any other point. However the denominator has a
zero as $p_1$ approaches any one of the remaining $(3g-4)$ $p_l$ for
$l=2,...,3g-3$. As a function of $p_1$ the denominator is a holomorphic
quadratic differential and therefore must have a total of $4g-4$ points.
Let $q_m$ for $m=1,...,g$ be the positions of the remaining $g$ zeroes.
Then $\sum_{m=1}^g q_m = 4 \Delta-\sum_{l=2}^{3g-3}p_l$. These are $g$
complex equations for $g$ complex points $q_m$. Generically in the
world-sheet moduli space, there will be a unique solution for the $q_m$.
Now consider the numerator. By taking into account all the partitioning of
$p_l$ between $r_a$ and $s_b$ and the consequent anti-symmetrization, we
see that again as a function of $p_1$ it is a quadratic differential with
zeroes at $p_l$ for $l=2,...,3g-3$. By the uniqueness of the remaining $g$
zeroes the numerator must also vanish exactly at $q_m$. This proves that
the ratio has no zeroes or poles as a function of $p_1$ and similarly for
all $p_l$. Since the ratio is a zero differential as a function of $p_l$
and has no zero or pole it must be independent of $p_l$. This is to be
expected since the result must not depend on the positions of the picture
changing operators. This means that we can move the $3g-3$ $p_l$'s to
Beltrami differentials resulting in the cancellation of the $b$
correlators and finally we are left with Beltrami differentials folded
with $G^-= G^-_{T^2}+G^-_{K3}$.

We have so far not included the right-moving contribution which involves
the Kac-moody currents appearing in the gaugino vertices at $x_i$ and
$y_i$. As usual by taking suitable differences between different gauge
group factors we
can restrict the Kac-moody currents to contribute only the zero modes
which give abelian differentials $\omega_i$. One can now perform the
integrals over $x_i$ and $y_i$ which yields $(\text{det}(\text{Im}\tau))^2$
cancelling the term in the denominator that comes from zero mode integrals
of the non-compact space-time directions. The amplitude thus is
expressible entirely as a topological amplitude in the internal $T^2
\times (4,0)$ theory. The result is
\begin{equation}
(\mathcal{F}_{g})_{A_1 A_2 A_3 A_4} = D_{f^{+A_1}} D_{f^{+A_2}}
D_{f^{+A_3}} (\mathcal{F}_{g})_{ A_4}\,,
\end{equation}
where 
\begin{equation}
(\mathcal{F}_{g})_{ A_4}= \int_{{\cal{M}}_g} \int_{z_4}\langle (\mu\cdot 
G^-)^{3g-3} e^{-i\frac{1}{\sqrt{2}} H}(z_4)\hat{V}_{A_4}(z_4)
\psi_3(p)(\rm{det}Q)^2\rangle\,.
\end{equation}
The covariant derivatives with respect to the hypermultiplet moduli appearoperators of these moduli.
Integrating these positions therefore gives derivatives with respect to
these moduli. $(\rm{det}Q)^2$
appears from the zero modes of the right moving Kac-Moody currents from
the gaugino vertices. In
(\ref{FgA4}) $\psi_3$ is at $z_3$, however it provides a zero mode and
$\psi_3$ being a section of a trivial line bundle, the zero mode is
constant. This allowed us to move $\psi_3$ from $z_3$ to an arbitrary
point, say $p$.

We can further simplify this expression by noting that $g$ of the $G^-$
must contribute the torus part
and the remaining $2g-3$ the $K3$ part. Thus $(\mu G^-)^{3g-3}= (\mu
G^-_{T^2})^{g}(\mu G^-_{K3})^{2g-3}$.
We can now express one of the $G^-_{K3}$ as:
\begin{equation}
G^-_{K3}=\oint \tilde{G}^+_{K3}  J^{--}_{K3}\,.
\end{equation}
Deforming the contour and noting the fact that $\oint \tilde{G}^+_{K3}
~G^-_{K3}  =0$ we see that the only contribution comes from the contour
integral around the vertex operator at $z_4$ with the result:
\begin{eqnarray}
\oint \tilde{G}^+_{K3} ~~e^{-i\frac{1}{\sqrt{2}}
H}(z_4)\hat{V}_{A_4}(z_4)&= &\oint G^-_{K3}~( \oint J^{++}_{K3}
~e^{-i\frac{1}{\sqrt{2}} H}(z_4)\hat{V}_{A_4}(z_4) ~) \nonumber\\
&=&\oint G^-_{K3}~e^{+i\frac{1}{\sqrt{2}} H}(z_4)\hat{V}_{A_4}(z_4) =
V^{(0)}_{f^{+A_4}}(z_4)
\end{eqnarray}
Integrating $z_4$ now produces another derivative with respect to the
hypermultiplet moduli $f^{+A_4}$
so that
\begin{equation}
(\mathcal{F}_g)_{ A_4}=D_{f^{+A_4} } \mathcal{F}_g
\end{equation}
where $\mathcal{F}_g$ is :
\begin{equation}
\mathcal{F}_g=\int_{{\cal{M}}_g} \langle (\mu\cdot G^-_{T^2})^{g}\cdot (\mu\cdot
G^-_{K3})^{2g-4} \cdot(\mu J^{--}_{K3})
(\rm{det}Q)^2 \rangle\,.
\end{equation}
This is exactly the topological amplitude (\ref{topamp2}).

%%%%%%%%%%%%%%%%%%%%%%%%%%%%%%%%%%%%%%%%%%%%%%%%%%%%%5
\subsection{The $g=1$ case}\label{App:CYtopGen1}
Finally, we would like to comment on the $g=1$ case. In this case
$m=2(g-2)=-2$
and we do not have an analog of the term (\ref{21}) since extracting two $F_+$  from the two
$K_-$ will saturate only the two $\theta^+$. We
can
however saturate the two $\bar{\theta}_-$ by extracting two fermion
components $\bar{\Psi}$ from the hypermultiplets contained in
$F^{+2}$. Thus we have the term
\begin{equation}
F_+^2 (\bar{\Psi}^{{  A}}\cdot\bar{\Psi}^{  B})({F}^{+2})_{{  A}{  B}}
\label{g1term}\,,
\end{equation}
where the corresponding reduced function reads
\begin{equation}
({{\mathcal F}_1})_{{  A}{  B}}= \sum_{n}\frac{n!}{(n-2)!} \xi_{i_1
  \cdots i_{n-2}\ {{  A} {  B} {  A}_1 \cdots {  A}_{n-2}}} f^{i_1\ {  A}_1}\cdots f^{i_{n-2}\ {  A}_{n-2}}\,.
\label{g1calF}
\end{equation}
On the other hand for $g=1$ the analog of the term (\ref{4der}) exists
and the corresponding reduced function is again given by
(\ref{redfn4der}) with $g=1$. It is easy to see that
\begin{equation}
({{\mathcal F}_1})_{{  A}_1 {  A}_2 {  A}_3 {  A}_4}= \frac{\pa}{\pa f^{+{  A}_3}}\frac{\pa}{\pa f^{+{  A}_4}}
({{\mathcal F}_1})_{{  A}_1 {  A}_2}\,.
\end{equation}
One can also easily verify that $({{\mathcal F}_1})_{{  A}_1 {  A}_2}$ is
symmetric in ${  A}_1$ and ${  A}_2$ and moreover satisfies the condition
\begin{equation}
\frac{\pa}{\pa f^{i  C}} ({{\mathcal F}_1})_{{  A}{  B}}=\frac{\pa}{\pa f^{i  A}} ({{\mathcal F}_1})_{{  C}{  B}}\,.
\end{equation}
In fact, we can also verify these relations by a direct calculation of $ ({{\mathcal F}_1})_{{  A}{  B}}$ in string theory.

For $g=1$ we can repeat the arguments of Section \ref{App:CYtopGeng} all the way to
eq.(\ref{FgA4}) with two differences: 1) the number of Beltrami
differentials is changed from $3g-3$ to $1$
and as a result
$(\mu\cdot b)^{3g-3}$ is replaced by $(\mu\cdot b)$ and 2) since the total $(b,c)$
ghost charge must be zero on genus
$1$, there must be also a $c$ ghost field attached to one of the vertices
say at $z_3$. As a result the
dimension of the operator at $z_3$ becomes zero and $z_3$ is unintegrated.
This can also be seen from the
fact that there is a translational zero mode on
genus $1$ surface which is
taken care of by fixing a puncture.
However the subsequent discussion changes significantly for $g=1$ case. We
cannot put three picture changing
operators at points $z_1,z_2,z_3$ as that will contradict the gauge
condition (\ref{gaugechoice}). We can at
best put two of the picture changing operators at, say $z_1$
and $z_2$ respectively, which will again convert these operators into
$(0)$-picture ones, giving rise to
two derivatives.  This still leaves one position say $p_1$ of the PCO. In
order to soak the fermion zero mode for $\bar{\psi}_3$ (which is
constant on the genus $1$ world sheet) only the torus
part of the supercurrent $G^-_{T^2}$ must appear at $p_1$ in the numerator of (\ref{FgA4}).
In the denominator we have the correlation function of the $(b,c)$ system
$\langle b(p_1) c(z_3)\rangle$
which again gives only zero modes and is hence independent of both $p_1$
and $z_3$. Since the result is
independent of $p_1$ we can move it to the Beltrami differential. Thus the
final result for $g=1$ is:
\begin{equation}
(\mathcal{F}_1)_{A_1 A_2 A_3 A_4} = D_{f^{+A_1}} D_{f^{+A_2}}
(\mathcal{F}_{1})_{ A_3 A_4}\,,
\end{equation}
where
\begin{equation}
(\mathcal{F}_{1})_{ A_3 A_4}= \int_{{\cal{M}}_g} \int_{z_4} \langle(\mu\cdot
G^-_{T^2})  (c \psi_3 e^{+i\frac{1}
{\sqrt{2}} H}\hat{V}_{A_3})(z_3) (e^{-i\frac{1}{\sqrt{2}}
H}\hat{V}_{A_4})(z_4) Q^2\rangle\,.
\label{F1A2}
\end{equation}
Note that there are still two hyper insertions that are not $(0)$-picture
operators: one of them carries
$J_{K3}$ charge $+1$ and the other $-1$. As a result
this last expression cannot be simplified further.\footnote{We can however
put $\partial X_3$ from
$G^-_{T^2}$ (which only gives the zero mode $P_L$) together with one of
the $Q$ to convert it into a
derivative with respect to
a holomorphic vector modulus. This will leave just one $Q$ insertion. At
first sight the expression
(\ref{F1A2}) seems to have a singularity when $z_4$ approaches $z_3$,
however the residue is just
the identity operator. Together with the insertion of a single $Q$, the
singularity would be proportional
to $U(1)$ anomaly $tr Q$ which is in fact zero.} This can also be
understood from the effective action.
For $g=1$ there are only two superfields $K_-$. These can provide two self
dual field strengths that
saturate two $\theta^+$ from the superspace integral. To saturate the remaining two
$\bar{\theta}_-$ we need to
extract two hyperfermions $\bar{\psi}$
from $\mathcal{F}_1$. Thus the on-shell quantities that string theory
would probe will involve at least
two  hyper indices $(\mathcal{F}_{1})_{ A_3 A_4}$. In section (4.1.3) we
showed that this quantity is symmetric in $A_3$ and $A_4$ and
satisfies $D_{f^{i A}}(\mathcal{F}_{1})_{ B C}=D_{ f^{ i B}}
(\mathcal{F}_{1})_{A C}$. Symmetry can be seen from eq.(\ref{F1A2}) by
expressing $e^{-i \frac{1}{\sqrt{2}}H} =\oint  J^{--}_{K3} e^{-i
\frac{1}{\sqrt{2}}H}$
and deforming the contour. The differential condition can also be
shown by inserting a zero picture operator $V^{(0)}_{f^{\pm}_A}=\oint
G^{\mp}_{K3} e^{\pm \frac{i}{\sqrt{2}}H} \hat{V}_A$ and deforming the contour.
%%%%%%%%%%%%%%%%%%%%%%%%%%%%%%%%%%%%%%%%%%%%%%%%%%%%%%%%%%%%%%%%
%%%%%%%%%%%%%%%%%%%%%%%%%%%%%%%%%%%%%%%%%%%%%%%%%%%%%%%%%%%%%%%%
\section{Harmonicity relation from string theory}\label{Sect:StringHarmonicity}
In this section we will discuss special properties of the heterotic couplings $\FIz{g}$, which we have computed in section \ref{Sect:hetF3}, from a string theoretic point of view. On the one hand we will check the harmonicity relation (\ref{25}) by explicitly working out all derivatives. On the other hand, as we can see from the coupling (\ref{19'}), field theoretic considerations predict a dependence of $\FIz{g}$ only on the holomorphic vector multiplets $W$. The half-BPS nature of these couplings, however, suggests that there might be an anomalous dependence also on the anti-holomorphic vector multiplet moduli and that $\FIz{g}$ should satisfy a holomorphic anomaly equation along the lines of \cite{Bershadsky:1993cx}. We will explicitly verify that this is indeed the case.

In order to carry out this program, we first have to covariantize the amplitudes (\ref{topamp2}) with respect to the $SU(2)$ R-symmetry group by introducing harmonic coordinates. To this end it turns out to be useful to group the supercurrents in the following manner
\begin{align}
&G^+_{K3,i}\equiv\left(\begin{array}{c}\tilde{G}^+_{K3} \\ G^+_{K3}\end{array}\right),&&\text{and} &&G^-_{K3,i}\equiv \left(\begin{array}{c}G^-_{K3} \\ -\tilde{G}^-_{K3}\end{array}\right)\,.\label{CovarSupercurrents}
\end{align}
For later convenience we have written the relevant part of the superconformal algebra (see Appendix \ref{Sect:SCFT}) in this new notation in Appendix \ref{App:SCFTN4cov}. Following the procedure of section~\ref{Sect:HarmonicDescription} we can introduce the harmonic variables by projecting the $SU(2)\sim Sp(1)$ index $i$ with harmonic coordinates. In particular, we define
\begin{align}
&G^+_{K3}(\bpu_{\pm})=G^+_{K3,i}\bpu^{i}_{\pm},&&\text{and} &&G^-_{K3}(\bpu_{\pm})=G^-_{K3,i}\bpu^{i}_{\pm}.
\end{align}
This of course has consequences for the topological amplitude (\ref{topamp2}), which we can now generalize to
\begin{align}
\FIz{g}(u)=\int_{\mathcal{M}_{g}}(\mu\cdot G_{T^2}^-)^g (\mu\cdot G^-_{K3}(\bpu_{+}))^{2g-4} (\mu\cdot J^{--}_{K3} )\psi_3\rangle\cdot(\text{det}Q_I)(\text{det}Q_J).
\label{F3gnewnotation}
\end{align}
In what follows we will refrain from writing the factor $(\text{det}Q_I)(\text{det}Q_J)$ explicitly, since it will not be essential for our computations.
%%%%%%%%%%%%%%%%%%%%%%%%%%%%%%%%%%%%%%%%%%%%%%%%%%%%%%%%%%%%%%%%%%
%%%%%%%%%%%%%%%%%%%%%%%%%%%%%%%%%%%%%%%%%%%%%%%%%%%%%%%%%%%%%%%%%%
\subsection{Harmonicity relation}
We will now start discussing the harmonicity relation (\ref{25}). To this end, we first need to understand how the vertex operators of the internal $K3\times T^2$ moduli can be written in a language which is appropriate for our topological correlators. As explained in section \ref{Sect:HarmonicDescription}, in general the hypermultiplet moduli space will be a $4n$ dimensional quaternionic space with the tangent space at a given point being described by $e_{A i}^{\mu k} \frac{\partial}{\partial x^{\mu k}}$ where $i=\pm$ on which the R symmetry group $Sp(1)$ acts and $A=1,...,2n$ on which $Sp(n)$ acts and $x^{\mu k}$ are some local coordinates. $e_{A i}$ are therefore in the bi-fundamental representation of $Sp(n)\times Sp(1)$. In string theory, a given point in the moduli space corresponds to an $\N=4$ SCFT and the tangent space (i.e. small deformation) is given by the
vertex operators denoted by $\Phi_{A i}$ which have world-sheet conformal dimension $(1,1)$. In the $(-1)$ ghost picture they are related to doublets of the $SU(2)$ current algebra $(\Xi_{A},\bar{\Xi}_A)$ which are $\N=4$ primary operators in the sense that they are annihilated by half of the supercharges respectively
\begin{align}
&\oint G^+_{K3,i}\Xi_A=0\,,&&\text{and} &&\oint G^-_{K3,i}\bar{\Xi}_A=0\,.
\end{align}
These fields have left and right conformal dimension $(1/2,1)$ and are $SU(2)$ highest weight states
\begin{align}
&\oint J^{++}_{K3}\Xi_A=\oint J^{--}_{K3}\bar{\Xi}_A=0\,,&&\text{and}&&\oint J^{++}_{K3}\bar{\Xi}_A=\Xi_A\,,\label{HighestWeightRel}
\end{align}
with $U(1)$ charges $\pm 1$ respectively\footnote{Notice that here we are using a slightly non-standard normalization of $J_{K3}$ in the sense that $J^{\pm\pm}_{K3}$ have charges $\pm2$ respectively.}
\begin{align}
&\oint J_{K3}\Xi_A=1\,,&&\text{and} &&\oint J_{K3}\bar{\Xi}_A=-1\,.
\end{align}
The vertex operators $\Phi_{A i}$ are related to $\Xi_A$ as follows
\begin{equation}
\Phi_{A i} = \oint G^+_{K3,i} \bar{\Xi}_A = \oint G^-_{K3,i} \Xi_A\,,
\label{completecovariantization}
\end{equation}
where the second equality follows by using the above algebra and (\ref{HighestWeightRel}). A marginal perturbation of the action may then be written as
\begin{equation}
S \rightarrow S+ \int f^{Ai} \oint G^+_{K3,i} \bar{\Xi}_A\,,
\end{equation}
where $f^{Ai}$ are local coordinates near the given point in the hypermultiplet moduli space.
\subsubsection{Bulk equation}
\begin{figure}
\begin{center}
\epsfig{file=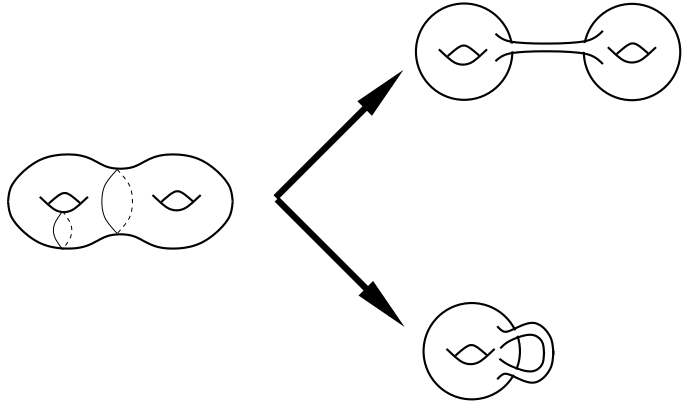, width=9cm}
\caption{Boundary contributions for a genus $g=2$ surface: The degeneration corresponds to pinching the surface along a non-contractible cycle. In this process, the surface develops a long and thin tube which eventually is replaced by just two punctures. For a compact Riemann surface there are generically two different possibilities, depending on which cycle is shrunk to zero: pinching of a dividing geodesic (top) or a handle (bottom).}
\label{degenerations}
\end{center}
\end{figure}

We are now ready to check the harmonicity relation (\ref{25}) on the string theory side. In order to do so, we consider
\begin{align}
&E_{A}=\epsilon^{ij}\frac{\partial^2\mathcal{F}_{g}}{\partial\bpusc{i}{+}\partial f^{Aj}}=\epsilon^{ij}\frac{\partial}{\partial\bpusc{i}{+}}\int_{\mathcal{M}_{g}}\langle(\mu\cdot G_{T^2}^-)^{g}(\mu\cdot G^-_{K3}(\bpu_+))^{2g-4}(\mu\cdot J^{--}_{K3})\psi_3\int \oint G^+_{K3,j} \bar{\Xi}_A\rangle\,.\nonumber
\end{align}
The first step is to contour-deform the $\oint G^+_{K3,j}$ integral, which yields two contributions
\begin{align}
E_{A}=\epsilon^{ij}\frac{\partial}{\partial\bpusc{i}{+}}&\bigg[(2g-4)\bpusc{k}{+}\epsilon_{kj}\int_{\mathcal{M}_{g}}\langle(\mu\cdot G_{T^2}^-)^{g}(\mu\cdot G^-_{K3}(\bpu_{+}))^{2g-5}(\mu\cdot T_{K3})(\mu\cdot J^{--}_{K3})\psi_3\int \bar{\Xi}_{A}\rangle\nonumber\\
&-\int_{\mathcal{M}_{g}}\langle(\mu\cdot G_{T^2}^-)^{g}(\mu\cdot G^-_{K3}(\bpu_+))^{2g-4}(\mu\cdot G^-_{K3,j})\psi_3\int \bar{\Xi}_{A}\rangle\bigg]\,.
\end{align}
Performing the differential with respect to $\bpusc{i}{+}$ we find
\begin{align}
E_{A}=&(2g-3)(2g-4)\int_{\mathcal{M}_{g}}\langle(\mu\cdot G_{T^2}^-)^{g}(\mu\cdot G^-_{K3}(\bpu_+))^{2g-5}(\mu\cdot T_{K3})(\mu\cdot J^{--}_{K3})\psi_3\int \bar{\Xi}_{A}\rangle-\nonumber\\
&-(2g-4)\epsilon^{ij}\int_{\mathcal{M}_{g}}\langle(\mu\cdot G_{T^2}^-)^{g}(\mu\cdot G^-_{K3}(\bpu_+))^{2g-4}(\mu\cdot G^-_{K3,i})(\mu\cdot G^-_{K3,j})\psi_3\int \bar{\Xi}_{A}\rangle.\label{harmonicityPRE}
\end{align}
The last term in this expression is zero due to the anti-symmetrization of the Beltrami differentials. We therefore only need to deal with the first term, which we can write as a boundary contribution in the moduli space $\mathcal{M}_{g}$. To see this, we complete $T_{K3}$ to a full energy momentum tensor
\begin{align}
\int_{\mathcal{M}_{g}}\langle(\mu\cdot &G_{T^2}^-)^{g}(\mu\cdot G^-_{K3}(\bpu_+))^{2g-5}(\mu\cdot T_{K3})(\mu\cdot J^{--}_{K3})\psi_3\int \bar{\Xi}_{A}\rangle=\nonumber\\
&=\int_{\mathcal{M}_{g}}\langle(\mu\cdot G_{T^2}^-)^{g}(\mu\cdot G^-_{K3}(\bpu_+))^{2g-5}(\mu\cdot T)(\mu\cdot J^{--}_{K3})\psi_3\int \bar{\Xi}_{A}\rangle-\nonumber\\
&-\int_{\mathcal{M}_{g}}\langle(\mu\cdot G_{T^2}^-)^{g}(\mu\cdot G^-_{K3}(\bpu_+))^{2g-5}(\mu\cdot T_{T^2})(\mu\cdot J^{--}_{K3})\psi_3\int \bar{\Xi}_{A}\rangle\,,\nonumber
\end{align}
and show that the last term is in fact zero. To this end, we use the OPE relation
\begin{align}
T_{T^2}=\frac{1}{2}\oint G^+_{T^2}G^-_{T^2}\,,
\end{align}
and pull off the $\oint G^+_{T^2}$ contour integral to find
\begin{align}
C_{A}&\equiv\int_{\mathcal{M}_{g}}\langle(\mu\cdot G_{T^2}^-)^{g}(\mu\cdot T_{T^2})(\mu\cdot G^-_{K3}(\bpu_+))^{2g-5}(\mu\cdot J^{--}_{K3} )\psi_3\int \bar{\Xi}_{A}\rangle=\nonumber\\
&=\frac{1}{2}\int_{\mathcal{M}_{g}}\langle(\mu\cdot G_{T^2}^-)^{g}\left[\oint G^+_{T^2}(\mu\cdot G^-_{T^2})\right](\mu\cdot G^-_{K3}(\bpu_+)^{2g-5}(\mu\cdot J^{--}_{K3} )\psi_3\int \bar{\Xi}_{A}\rangle=\nonumber\\
&=g\int_{\mathcal{M}_{g}}\langle(\mu\cdot G_{T^2}^-)^{g}(\mu\cdot T_{T^2})(\mu\cdot G^-_{K3}(\bpu_+))^{2g-5}(\mu\cdot J^{--}_{K3} )\psi_3\int \bar{\Xi}_{A}\rangle\,.\nonumber
\end{align}
This result can equivalently be expressed as $C_{A}=gC_{A}$, which for
$g>1$ implies
\begin{equation}
\int_{\mathcal{M}_{g}}\langle(\mu\cdot G_{T^2}^-)^{g}(\mu\cdot
T_{T^2})(\mu\cdot G^-_{K3}(\bpu_+))^{2g-5}(\mu\cdot J^{--}_{K3}
)\psi_3\int \bar{\Xi}_{A}\rangle=0.
\label{Ttorus}
\end{equation}
Therefore, we can rewrite (\ref{harmonicityPRE}) as
\begin{align}
E_A=(2g-3)(2g-4)\int_{\mathcal{M}_{g}}\langle&(\mu\cdot G_{T^2}^-)^{g}(\mu\cdot G^-_{K3}(\bpu_+))^{2g-5}(\mu\cdot T)(\mu\cdot J^{--}_{K3} )\psi_3\cdot\nonumber\\
&\cdot \int \bar{\Xi}_{A}(\text{det}Q_I)(\text{det}Q_J)\rangle.
\label{harmonicityPRE1}
\end{align}
where we have reinstated $(\text{det}Q_I)(\text{det}Q_J)$ factors that carry over from (\ref{F3gnewnotation}). This term is in fact a boundary contribution since it was shown in \cite{Bershadsky:1993cx,Antoniadis:1996qg} that the insertion of the (full) energy momentum tensor is equivalent to a total derivative in the moduli space of the Riemann surface. The right hand side of (\ref{harmonicityPRE1}) therefore gets contributions from degeneration limits of the genus $g$ Riemann surface, of which there are essentially two
\begin{align}
E_{A}=E^{\text{handle}}_{A}+E^{\text{geo}}_{A}\,.
\end{align}
These correspond to the pinching of either a handle or a dividing geodesic (for an example for genus $g=2$ see Figure \ref{degenerations}).

%%%%%%%%%%%%%%%%%%%%%%%%%%%%%%%%%%%%%%%%%%%%%%%%%%%5
\subsubsection{Boundary contributions}\label{Sect:deghand}
We will now consider the boundary contributions of (\ref{harmonicityPRE1}) by starting with the handle degeneration. Due to the presence of the $(\text{det}Q_I)(\text{det}Q_J)$ factors, it is clear  that only charged states can propagate through the node. However, as mentioned at the beginning of this section, we are restricting ourselves to a generic point in the moduli space. At a generic point in the vector moduli space (Coulomb branch) there are no charged massless vector or hypermultiplet states. Therefore the contribution due to the handle degeneration $E^{\text{handle}}_{A}$ vanishes.

\begin{figure}[htbp]
\begin{center}
\epsfig{file=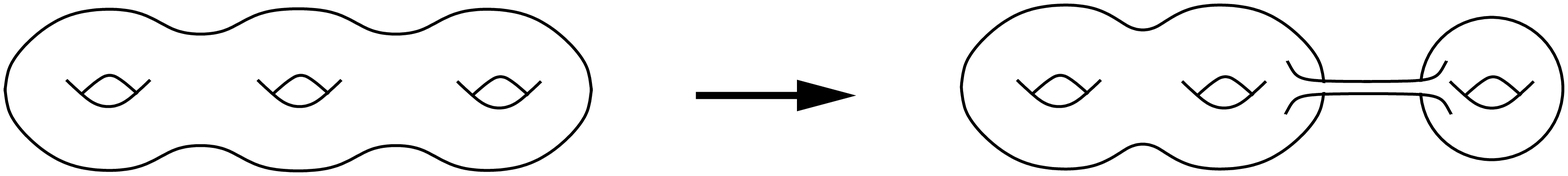, width=13cm}
\caption{Pinching of a dividing geodesic on a genus $3$ Riemann surface.}
\label{fig:pinchgeo}
\end{center}
\end{figure}

We are therefore left to consider the pinching of dividing geodesics. This essentially means that the original Riemann surface is split into two separated ones, each one of lower genus. An example for genus 3 is shown in figure \ref{fig:pinchgeo}.
\noindent
We will denote the genera of the new surfaces by $g_1$ and $g_2$ where $g=g_1+g_2$ and we have on the first surface $3g_1-3+1$ and on the second surface $3g_2-3+1$ Beltrami differentials where the additional one on each of the two surfaces corresponds to the integration of the position of the two punctures to which the thin tube degenerates.\footnote{This is true when both $g_1$ and $g_2$ are greater than one. When $g_1$ or $g_2$ is one then the number of Beltrami differentials on the genus 1 surface is just one and the corresponding puncture is not integrated, it merely fixes the translational invariance of the torus. In the following when not mentioned explicitly $g_1$ and $g_2$ will be greater than one and the case when either of them is one will be explicitly stated.} As discussed in detail in the context of $\N=1$ topological amplitudes in \cite{Antoniadis:1996qg}, the operators appearing at the two punctures  must carry twisted dimensions $(0,1)$ (recall that the twisted left moving dimension is the untwisted dimension minus half the $U(1)$ charge) in order for them to contribute to the degeneration limit as in this case their propagator behaves like $1/\bar{t}$ where $t$ is the plumbing-fixture coordinate; $\partial_t$ appearing from the total derivative then gives a delta function. Moreover the total $U(1)$ charge carried by these two operators must be $+3$. This is due to the fact that in the twisted theory, in order to balance the background charge, the total charge of the operators on the sphere must be $+3$. There are different ways of distributing this charge among the two operators:

\begin{enumerate}
\item The distribution with one operator carrying $0$ charge and the other $+3$ charge vanishes when $g_1$ and $g_2$ are greater than one due to the fact that upon the localization of the Beltrami differential (this corresponds to the moduli associated with the integration of the puncture) around charge $0$ operator the OPE of this operator with the  supercurrent has no singularity. The only exception is when one of the surfaces (say $\Sigma_2$) has genus 1. In this case the operator carrying zero charge can sit at the puncture on $\Sigma_2$. Taking into account the anomalies in various charges one obtains the following distribution of operators on the two surfaces:
\begin{center}
\begin{tabular}{c|c}
 \textbf{surface} $\Sigma_{1}$ & \textbf{surface} $\Sigma_{2}$\\\hline
&\\
$(\mu\cdot G^-_{T^2})^{g-1}$ & $(\mu\cdot G^-_{T^2})$ \\
&\\
$(\mu\cdot G^-_{K3}(\bpu_+))^{2g-5}$, $(\mu\cdot J^{--}_{K3} )$ & \\
&\\
$(\psi_3 J^{++}_{K3} \bar{J}_{\bar{K}})(p_1)$, $\int \bar{\Xi}_{A}$ & $\bar{J}_L(p_2)$, $\psi_3(x)$\\
&\\
$\text{det} Q_{I}$$\text{det} Q_{J}$ & $Q_{I_g}$$Q_{J_g}$\\
\end{tabular}
\end{center}
where $\text{det} Q_I$ on $\Sigma_1$ which is a genus $g-1$ surface refers to the determinant of the $(g-1)\times (g-1)$ matrix $Q_{I_i}^{a_j}$ where $a_j$ refers to the $j$-th $a$-cycle. It is understood that there is a total antisymmetrization between $I_i$ and $I_g$ indices (and similarly the corresponding $J$ indices). On $\Sigma_2$
this topological amplitude is just the new supersymmetry index \cite{Cecotti:1992qh} and physically corresponds to the one loop threshold correction to the gauge coupling $h^{(1)}_{I_g J_g}$
\begin{equation}
\langle F_{I_g} F_{J_g} \Phi_L\rangle_{\text{genus}=1} = D_L h^{(1)}_{I_g J_g}\,.
\end{equation}
The quantity on the genus $g-1$ surface is a new topological term and we shall denote it by
\begin{equation}
\mathcal{F}^{g-1}_{A,\bar{K}}=\int_{\mathcal{M}_{g-1}}\langle (\mu\cdot G^-_{T^2})^{g-1}(\mu\cdot G^-_{K3}(\bpu_+))^{2 g-5}(\psi_3
J^{++}_{K3} \bar{J}_{\bar{K}})(p_1)\int \bar{\Xi}_{A} \text{det} Q_{I}\text{det}
Q_{J}\rangle_{g-1}\,.
\label{FKA}
\end{equation}
In Appendix \ref{App:RelTopPhysD4} we will show that this quantity is just the first one in a series of new topological terms and we identify them with heterotic string amplitudes on a generic Calabi-Yau compactification. The final contribution to the anomaly in the harmonicity relation due to charge $0$ and $+3$ at the two punctures
is:
\begin{equation}
E_A^{\text{geo},(3,0)}=\mathcal{F}^{g-1}_{A,\bar{K}}G^{\bar{K} L } D_L h^{(1)}\,,
\label{vectorcont}
\end{equation}
where we have suppressed the chiral indices $I$ and $J$. The tree level vector multiplet propagator $G^{\bar{K}L}$ appears from the contribution of the thin tube.

\item Now we consider the case when the two operators at the puncture carry  charge $+1$ and $+2$ respectively. There are two different ways this can happen:
\begin{enumerate}
\item the operators are  $J^{++}_{K3} \bar{J}_{\bar{K}}$ and $\psi_3 \bar{J}_L$ on surfaces $\Sigma_1$ and $\Sigma_2$ respectively. The $\psi_3$ appearing in the original surface must now be on $\Sigma_1$ to balance the zero modes. The localization of the Beltrami differential at the puncture on $\Sigma_2$ again vanishes since the supercurrent has no singularity with $\psi_3$ which provides the zero mode. As in case 1.) above the only exception is when $\Sigma_2$ has genus 1. The result is the same as in case 1.) namely (\ref{vectorcont}) since $\psi_3$ merely provides the constant zero mode.

\item the operators are $\psi_3 \Xi_B$ and $\Xi_C$ where recall that $\Xi$ ($\bar{\Xi}$) carry $J_{K3}$ charge $+1$ ($-1$). The corresponding intermediate state propagating in the tube is a hypermultiplet. Moreover, the distribution of the operators on the new surfaces consistent with the requirement of the $\psi_3$ and $\bar{\psi}_3$ zero mode counting as well as the anomaly in the $J_{K3}$ charge on the two surfaces is in the following way
\begin{center}
\begin{tabular}{c|c}
 \textbf{surface} $\Sigma_1$ & \textbf{surface} $\Sigma_2$\\\hline
&\\
$(\mu\cdot G^-_{T^2})^{g_1}$ & $(\mu\cdot G^-_{T^2})^{g_2}$ \\
&\\
$(\mu\cdot G^-_{K3}(\bpu_+))^{2g_1-2}$ & $(\mu\cdot G^-_{K3}(\bpu_+))^{2g_2-3}$, $(\mu\cdot J^{--}_{K3}) $ \\
&\\
$\psi_3 \Xi_B(p_1)$, $\int \bar{\Xi}^{A}$ & $\Xi_C(p_2)$, $\psi_3(x)$\\
\end{tabular}
\end{center}
or equivalently the exchange of the operators on the two surfaces in the last row. Here $p_1$ and $p_2$ are the punctures on the two surfaces. There are also $\text{det} Q$ factors on the two surfaces which we have not shown
explicitly above. Note that in either case
$\psi_3$ at the puncture and at $x$ (which are on the two different surfaces) are replaced by zero-modes and
therefore we can split off the $\psi_3$ at the puncture to an arbitrary point in the same surface (say $y$). Let us first consider the case when $g_1 >1$ (Note that $g_2$ is greater than one since the number of supercurrents $G^-_{K3}(\bpu_+)$ on $\Sigma_2$ is $2 g_2-3$ which must be non-negative). Then on $\Sigma_1$ the localization of one of the $G^-_{K3}(\bpu_+)$ around $p_1$ converts $\Xi_B$ into $\bpu_{i+} \Phi_B^{i}$.
Writing one of the remaining $(2g_1-3)$ $G^-_{K3}(\bpu_+)$ on $\Sigma_1$ as
\begin{align}
G^{-}_{K3}(\bpu_+)=-\oint G^+_{K3}(\bpu_+)J^{--}_{K3} ,
\label{G-G+relation}
\end{align}
and deforming the contour the only contribution comes when the contour encircles $\bar{\Xi}_A$ converting the latter
into $\bpu_{j+} \Phi_A^{j}$. There are now $(2g_1-4)$ $G^-_{K3}(\bpu_+)$ and one $J^{--}_{K3} $ as well as $g_1$ $G^-_{T^2}$ folded with the $3g_1-3$ Beltrami differentials. This is exactly of the form of the original partition function. The difference now is that $\bpu_{i+} \Phi_A^{i}$ and $\bpu_{i+} \Phi_B^{i}$ are also inserted on $\Sigma_1$. Therefore the correlation function on $\Sigma_1$ is
\begin{align}
&D_+^A D_+^B \mathcal{F}_{g_1}, &&\text{with}  &&D_{A+} \equiv \bpu_{i+} D_A^{i}\,.
\label{sigma1}
\end{align}
When $g_1=1$ the correlation function on $\Sigma_1$ is simply
\begin{equation}
\langle(\mu\cdot G^-_{T^2})\psi_3 \Xi_B(p_1) \int \bar{\Xi}_{A}  Q_I Q_J \rangle_{g_1=1}\,.
\end{equation}
As discussed in Section \ref{App:CYtopGen1}, this is just the string amplitude $\langle F_I
F_J \bar{\psi}_{A}
\bar{\psi}_B\rangle_{g=1}$
where $\bar{\psi}$ are the anti-chiral hypermultiplet fermions. This
physical amplitude  comes from the
effective action term (\ref{g1term}) by extracting out two hypermultiplet
fermions. We denote this amplitude by
$\mathcal{F}_{1,AB}$.

On $\Sigma_2$, one of the Beltrami differentials must localize at the puncture $p_2$. If one of $(2g_2-3)$ $G^-(\bpu_+)$ localizes at $p_2$ then the result is simply
\begin{equation}
D_{C+} \mathcal{F}_{g_2}\,.\label{sigma2}
\end{equation}
If, on the other hand, $J^{--}_{K3} $ localizes at $p_2$ then it converts $\Xi_C$ to $\bar{\Xi}_C$. We can now write one of the $G^-(\bpu_+)$ as in (\ref{G-G+relation}) and follow the steps as above and the result is again (\ref{sigma2}).

We now need to include the propagator on the tube between the operators $\Xi_B$ and $\psi_3 \Xi_C$. Note that the total charge carried by these two operators is $J_{T^2}=+1$ and $J_{K3}=+2$ which is exactly what is
required on the sphere in the twisted theory. This propagator is simply equal to the $Sp(n)$ symplectic form
$\Omega_{BC}$. Combining with (\ref{sigma1}) and (\ref{sigma2}) we find that
\begin{align}
E^{\text{geo},(2,1)}_{A}=\sum_{g_1=2}^{g-2}D_{A+}D_{B+}\FIz{g_1}\Omega^{BC}D_{C+}\FIz{g-g_1} +\mathcal{F}_{1,AB}\Omega^{BC}D_{C+}\FIz{g-1} \,.
\end{align}
\end{enumerate}
\end{enumerate}
The complete contribution and therefore the full harmonicity relation is henceforth given by
\begin{align}
\epsilon^{ij}\frac{\partial^2\mathcal{F}_{g}}{\partial\bpusc{i}{+}\partial
f^{Aj}}&=E_A=E^{\text{geo},(3,0)}_{A}+
E^{\text{geo},(2,1)}_{A}=\nonumber\\
&=\mathcal{F}^{g-1}_{A,\bar{K}}G^{\bar{K} L } D_L h^{(1)}+
\sum_{g_1=2}^{g-2}D_{A+}D_{B+}\FIz{g_1}\Omega^{BC}D_{C+}\FIz{g-g_1}+\mathcal{F}_{1,AB}\Omega^{BC}D_{C+}\FIz{g-1}
\,.\label{StringHarmEqu}
\end{align}
Notice particularly that the right hand side of this equation is not zero in contrast to the prediction of the field theoretic considerations (see section \ref{Sect:HarmonicDescription}).
%%%%%%%%%%%%%%%%%%%%%%%%%%%%%%%%%%%%%%%%%%%%%%%%%%%%%%%%%%%%%
%%%%%%%%%%%%%%%%%%%%%%%%%%%%%%%%%%%%%%%%%%%%%%%%
%%%%%%%%%%%%%%%%%%%%%%%%%%%%%%%%%%%%%%%%%%%%%%%%
\subsection{Vector multiplet equation}\label{Sect:vectormultipletEQ}
We will now proceed and check the independence of the $\FIz{g}$ of the anti-holomorphic vector multiplet scalars. Similar to the procedure leading to the proof of the harmonicity relation, we first have to formulate the vertex operators in the topological theory which correspond to a perturbation of the action by a vector multiplet scalar. This is simply given by
\begin{align}
S\to S+\int \left[\bar{\varphi}^I\oint G^+_{T^2}\bar{\psi}_3 \bar{J}_I+\varphi^I  \bar{J}_I\oint G^-_{T^2}\psi_3\right]\,,
\end{align}
where $\bar{J}$ are the right-moving Kac-Moody currents and in the following we will not show them explicitly unless needed. The relation which we then have to check is
\begin{align}
E_3=\partial_{\bar{\varphi}^I}\mathcal{F}_{g}=\int_{\mathcal{M}_{g}}\langle(\mu\cdot G^-_{T^2})^{g}(\mu\cdot G^-_{K3}(\bpu_+))^{2g-2}(\mu\cdot J^{--}_{K3} )\psi_3(x)\int\oint G^+_{T^2}\bar{\psi}_3\bar{J}_I\rangle\,.
\end{align}
Deforming the contour integral $\oint G^+_{T^2}$, this expression is equal to
\begin{align}
E_3=g\int_{\mathcal{M}_{g}}\langle(\mu\cdot G^-_{T^2})^{g-1}(\mu\cdot T_{T^2})(\mu\cdot G^-_{K3}(\bpu_+))^{2g-4}(\mu\cdot J^{--}_{K3} )\psi_3(x)\int\bar{\psi}_3\bar{J}_I\rangle\,,
\end{align}
which can also be written as
\begin{align}
E_3=&g\int_{\mathcal{M}_{g}}\langle(\mu\cdot G^-_{T^2})^{g-1}(\mu\cdot T)(\mu\cdot G^-_{K3}(\bpu_+))^{2g-4}(\mu\cdot J^{--}_{K3} )\psi_3(x)\int\bar{\psi}_3\rangle-\nonumber\\
&-g\int_{\mathcal{M}_{g}}\langle(\mu\cdot G^-_{T^2})^{g-1}(\mu\cdot T_{K3})(\mu\cdot G^-_{K3}(\bpu_+))^{2g-4}(\mu\cdot J^{--}_{K3} )\psi_3(x)\int\bar{\psi}_3\bar{J}_I\rangle\,.
\end{align}
The first term is a boundary contribution and we will deal with it later. The second term is a genuine contribution, but we will now prove that it is actually zero. To this end we write
\begin{align}
T_{K3}=-\oint G^+_{K3}(\bpu_-)G^-_{K3}(\bpu_+)
\end{align}
and pull off the contour integral. We then find the following relation
\begin{align}
C\equiv&\int_{\mathcal{M}_{g}}\langle(\mu\cdot G^-_{T^2})^{g-1}(\mu\cdot T_{K3})(\mu\cdot G^-_{K3}(\bpu_+))^{2g-4}(\mu\cdot J^{--}_{K3} )\psi_3(x)\int\bar{\psi}_3\bar{J}_I\rangle=\nonumber\\
=&-(2g-4)\int_{\mathcal{M}_{g}}\langle(\mu\cdot G^-_{T^2})^{g-1}(\mu\cdot T_{K3})(\mu\cdot G^-_{K3}(\bpu_+))^{2g-4}(\mu\cdot J^{--}_{K3} )\psi_3(x)\int\bar{\psi}_3\bar{J}_I\rangle+\nonumber\\
&-\int_{\mathcal{M}_{g}}\langle(\mu\cdot G^-_{T^2})^{g-1}(\mu\cdot G^-_{K3}(\bpu_+))^{2g-3}(\mu\cdot G^-_{K3}(\bpu_-))\psi_3(x)\int\bar{\psi}_3\bar{J}_I\rangle=\nonumber\\
=&-(2g-4)C-\int_{\mathcal{M}_{g}}\langle(\mu\cdot G^-_{T^2})^{g-1}(\mu\cdot G^-_{K3}(\bpu_+))^{2g-3}(\mu\cdot G^-_{K3}(\bpu_-))\psi_3(x)\int\bar{\psi}_3\bar{J}_I\rangle.\nonumber
\end{align}
By writing one of the $G^-_{K3}(\bpu_+)$ in the last term above as
\begin{align}
G^-_{K3}(\bpu_+)=\oint G^+_{K3}(\bpu_+)J^{--}_{K3},
\end{align}
and pulling off the contour integral and using the fact that $\oint G^+_{K3}(\bpu_+) G^-_{K3}(\bpu_-) = T_{K3}$ we find that this term is also equal to $C$. This implies that $(2g-4) C=0$ which in turn means that for $g>2$, $C=0$ (for $g=2$ by explicit calculation we can also show that $C=0$). As a result we find
\begin{align}
E_3=&g\int_{\mathcal{M}_{g}}\langle(\mu\cdot G^-_{T^2})^{g}(\mu\cdot T)(\mu\cdot G^-_{K3}(\bpu_+))^{2g-2}(\mu\cdot J^{--}_{K3} )\psi_3(x)\int\bar{\psi}_3\bar{J}_I\rangle.
\end{align}
We now have to deal with the boundary term. As for the harmonicity relation, at a generic point in the Coulomb branch only degeneration along a dividing geodesic can contribute. Thus, we split the genus $g$ surface into two surfaces with genus $g_1$ and $g_2$ respectively. First let us consider the case when the states that propagate on the tube are the hypermultiplet states. The exact distribution of the operators on the two surfaces is given by
\begin{center}
\begin{tabular}{c|c}
 \textbf{surface} $\Sigma_1$ & \textbf{surface} $\Sigma_2$\\\hline
&\\
$(\mu\cdot G^-_{T^2})^{g_1}$ & $(\mu\cdot G^-_{T^2})^{g_2-1}$, $\int \bar{\psi}_3\bar{J}_I$ \\
&\\
$(\mu\cdot G^-_{K3}(\bpu_+))^{2g_1-4}$, $(\mu\cdot J^{--}_{K3}) $ & $(\mu\cdot G^-_{K3}(\bpu_+))^{2g_2-2}$ \\
&\\
$\psi_3 \Xi_B (p_1)$ & $\Xi_C(p_2)$, $\psi_3(x)$\\
\end{tabular}
\end{center}
This entails the following contribution
\begin{align}
E_3=&\int_{\mathcal{M}_{g_1}}\langle(\mu\cdot G^-_{T^2})^{g_1}(\mu\cdot G^-_{K3}(\bpu_+))^{2g_1-4}(\mu\cdot J^{--}_{K3} )\int_{\Sigma_1}\oint G^-_{K3}(\bpu_+)\psi_3 \Xi_B\rangle\cdot \Omega^{BC}\cdot\nonumber\\
\cdot&\int_{\mathcal{M}_{g_2}}\langle(\mu\cdot G^-_{T^2})^{g_2-1}(\mu\cdot G^-_{K3}(\bpu_+))^{2g_2-2}\psi_3(x)\int_{\Sigma_2}\oint G^-_{K3}(\bpu_+)\Xi_C \int\bar{\psi}_3\bar{J}_I\rangle.
\end{align}
In the first correlator, we can pull the $\psi_3$ out of the contour integral and put it at an arbitrary position $y$, since it may only contribute zero modes and cannot contract with any other operator. Moreover, we can rewrite
\begin{align}
E_3=&\int_{\mathcal{M}_{g_1}}\langle(\mu\cdot G^-_{T^2})^{g_1}(\mu\cdot G^-_{K3}(\bpu_+))^{2g_1-4}(\mu\cdot J^{--}_{K3} )\psi_3(y)\oint G^-_{K3}(\bpu_+) \Xi_B\rangle\cdot\Omega^{BC}\cdot\nonumber\\
\cdot&\int_{\mathcal{M}_{g_2}}\langle(\mu\cdot G^-_{T^2})^{g_2-1)}\mu\cdot (G^-_{K3}(\bpu_+)^{2g_2-3}\oint\left[G^+(\bpu_+), J^{--}_{K3} \right] \psi_3(x)\oint G^-_{K3}(\bpu_+)\Xi_C\int\bar{\psi}_3\bar{J}_I\rangle.
\end{align}
However, pulling off the contour integral in the last line, we find zero residue with any other operator inside the correlator. We therefore conclude that this boundary contribution vanishes identically.\\

Now let us consider the case when the intermediate state is a vector state. Due to the localization of one of the Beltrami's at the puncture this contribution will vanish when both $g_1$ and $g_2$ are greater than one. The only exception is when one of the two surfaces (say $\Sigma_1$) has genus one. The distribution of the operators is
\begin{center}
\begin{tabular}{c|c}
 \textbf{surface} $\Sigma_1$ & \textbf{surface} $\Sigma_2$\\\hline
&\\
$(G^-_{T^2})$ & $(\mu\cdot G^-_{T^2})^{g-1}$, $\int \bar{\psi}_3\bar{J}_I$ \\
&\\  & $(\mu\cdot G^-_{K3}(\bpu_+))^{2g-5}$ $(\mu\cdot J^{--}_{K3}) $ \\
&\\
$\psi_3 \bar{J}_L (p_1)$ & $J^{++}_{K3}  \bar{J}_K(p_2)$, $\psi_3(x)$\\
\end{tabular}
\end{center}
\noindent
Exactly as in the harmonicity equation for case 1.) and case 2.a) we find that
\begin{equation}\label{exactlyas}
\frac{\partial \mathcal{F}_g}{\partial \bar{\varphi}^I }= \mathcal{F}^{g-1,1}_{\bar{I},\bar{K}} G^{\bar{K} L}\partial_L h^{(1)}\,.
\end{equation}
Here again we have suppressed the chiral indices in $\mathcal{F}$ and the one-loop threshold correction to the gauge coupling $h^{(1)}$. The expression for $\mathcal{F}^{g-1,1}_{\bar{I},\bar{K}}$ is similar to the one for $ \mathcal{F}^{g-1,1}_{A ,\bar{K}}$ (\ref{FKA}) with $\bar{\Xi}^A$ replaced by $\bar{\psi}_3 \bar{J}^I$. In
Appendix \ref{App:RelTopPhysD4}, we show that the series $\mathcal{F}^{g,n}$ (where we
suppressed the vector indices) corresponds to
the effective action term $P\left({\hat F}_{g,n}{\hat
K}^{2(g-1)}{\hat{\bar{K}}}^{2(n-1)}\right)$. In Section 4
we argued that eliminating the auxiliary field in the latter gives rise to
a term of the form that appears in
the right hand side of (\ref{exactlyas}). \footnote{We should
therefore have used two different symbols for the
effective action term and the string amplitude however we chose to use the
same notation in order to indicate the
relation between the two quantities.} In other words the
$\mathcal{F}^g$ computed in string theory
is the sum of all connected graphs contributing to the amplitude in
question. These graphs include the effective
action term $\mathcal{F}^g$ but also $\mathcal{F}^{g-1,1}$ when auxiliary
fields are eliminated by the
equation of motion (which to the lowest order in string coupling is given
by the threshold correction term
$h^{(1)}$). While the effective action term $\mathcal{F}^g$ is
holomorphic, $\mathcal{F}^{g-1,1}$ is not.
However, in order to prove (\ref{exactlyas}) which is a relation
between two connected amplitudes
each of which receive contributions from several different terms in the
effective action we need to systematically
solve for the auxiliary fields.
The problem of exact elimination of the auxiliary fields is quite
complicated since in the presence of the
effective
action terms discussed in this paper auxiliary fields appear non-linearly.
However it should be possible to
solve the equation for auxiliary fields as a power series in string
coupling constant and check whether
the eq.(\ref{exactlyas}) is a consequence of supersymmetry.

%%%%%%%%%%%%%%%%%%%%%%%%%%%%%%%%%%%%%%%%%%%%%%%%%%%%%%%%%%%%%
%%%%%%%%%%%%%%%%%%%%%%%%%%%%%%%%%%%%%%%%%%%%%%%%%%%%%%%%%%%%%%%%%%%%%%%%%%%%%%%%%%%%%%%%%%%%%%%%%%%%%%%%%%%%%%%%%%%%
%%%%%%%%%%%%%%%%%%%%%%%%%%%%%%%%%%%%%%%%%%%%%%%%%%%%%%%%%%%%%%%%%%%%%%%%%%%%%%%%%%%%%%%%%%%%%%%%%%%%%%%%%%%%%%%%%%%%
%%%%%%%%%%%%%%%%%%%%%%%%%%%%%%%%%%%%%%%%%%%%%%%%%%%%%%%%%%%%%%%%%%%%%%%%%%%%%%%%%%%%%%%%%%%%%%%%%%%%%%%%%%%%%%%%%%%%
%%%%%%%%%%%%%%%%%%%%%%%%%%%%%%%

\subsection{Second order relation}\label{secondorderEQ}

In this subsection we will consider the second order relation (\ref{modconstr}). Although the analysis can be carried out for general string compactifications by inserting the vertex operators corresponding to the two moduli derivatives similar to the case of the harmonicity relation, there are several technically difficult issues which arise. Firstly one needs to subtract reducible diagrams corresponding to intermediate graviton-dilaton exchange and secondly one needs to deal with the cumbersome contact terms that appear. In order to avoid these problems we will focus on the special case when the hypermultiplet space is $SO(4,n)/SO(4) \times SO(n)$. A string realization of this is provided by an orbifold compactification on $T^2 \times (\frac{T^4}{\mathbb{Z}_2})$. More specifically, writing the $(6,22)$ lattice vectors in terms of $(2,22-n)$ sublattice vectors $(P_L,\bar{P}_L; \vec{P}_R)$ where $\mathbb{Z}_2$ acts only as a shift and $(4,n)$ sublattice vectors $(P_{ai}, P_{\hat{A}})$ for $i,a=+,-$ and ${\hat{A}}=1,...,n$. Here $P_{ia}$ is the left moving part where $SO(4) = SU(2)\times SU(2)$ acts on the index $i$ and $a$ respectively and $P_{\hat{A}}$ is the right moving part which transforms as a vector of $SO(n)$. The $\mathbb{Z}_2$ action on the $(4,n)$ lattice is $(P_{ai}, P_{\hat{A}})\rightarrow (-P_{ai}, -P_{\hat{A}})$ plus possibly a shift (the total shift in the $(2,22-n)$ and $(4,n)$ together must be a $Z_2$ shift in the $(6,22)$ lattice. We assume that $n$ and the shift vector is such that the level matching condition (and more generally modular invariance conditions for asymmetric orbifold) is satisfied. The resulting model has a gauge group of rank $22-n$ (besides the graviphotons). We further assume that we are sitting at a generic point in the vector moduli space (ie. we have turned on generic Wilson lines on $T^2$) such that the gauge group is abelian $U(1)^{22-n}$ and all the charged matter fields are massive. Hypermultiplets coming from the untwisted sector parametrize $SO(4,n)/SO(4) \times SO(n)$ space. The hypermultiplets coming from the twisted sectors will enlarge this space to a more general quaternionic space, however in the following we will restrict ourselves to taking derivatives only along the untwisted moduli and therefore only the quaternionic subspace $SO(4,n)/SO(4) \times SO(n)$ will be relevant.\footnote{Of course by choosing the orbifold $Z_2$ to include a suitable shift acting on $P_L$ of $(2,22-n)$ sublattice we can make all the twisted states massive, in which case the hypermultiplet moduli space will be precisely $SO(4,n)/SO(4) \times SO(n)$.} The vertex operators for these hypermultiplets in the (-1)-ghost picture are $\psi_{ia} \bar{\partial} X_{\hat{A}}$. In string amplitudes, the hypermultiplet moduli dependence enters through the $(4,n)$ sublattice vectors $(P_{ai}, P_{\hat{A}})$. Covariant derivatives w.r.t. moduli acting on the lattice vectors are given by:
\begin{align}
&D_{{\hat{A}},ai} P_{bj}= \epsilon_{ij} \epsilon_{ab} P_{\hat{A}}\,,&&\text{and} &&D_{{\hat{A}},ai} P_{\hat{B}} =  \delta_{{\hat{A}}{\hat{B}}} P_{ai}\,,\label{variation}
\end{align}
where we have chosen the $SO(4,n)$ invariant norm to be $\epsilon^{ij} \epsilon^{ab} P_{ai} P_{bj}
-\delta^{{\hat{A}}{\hat{B}}}P_{\hat{A}} P_{\hat{B}}$. It is easy to see that the above variation leaves the norm invariant as it should.
Furthermore, it is important to note that the normalizations in the
above variation is such that $D_{{\hat{A}},ai}$
satisfy the same algebra as $L_{{\hat{A}},ai}$ defined in eq.(\ref{algebra}) with the value of $Z_0$ for
$P_{a\pm}$ being $\pm 1/2$ as is appropriate for an $SU(2)$ doublet.

Let us denote the $N=4$ world-sheet supercurrent as
\begin{align}
&G^i_{K3,j} =\psi^{ai} \partial X_{aj}\,,
\end{align}
which we can project in the usual way with harmonic coordinates
\begin{align}
G^{\pm}_{K3}(\bar{u}_+)= G^{\pm}_j \bar{u}^j\,.
\end{align}
The OPE  for $\psi$ is
\begin{equation}
\psi^{ai}(z) \psi^{bj}(w) = \frac{\epsilon^{ab} \epsilon^{ij}}{z-w}\,,
\end{equation}
where we have used the same convention for the $\epsilon$-tensor as in (\ref{12}), i.e. $\epsilon^{+-}=-\epsilon_{+-}=1$. Moreover, $\epsilon$ will be used to raise or lower the indices, e.g. $\psi_a^i
= \epsilon_{ab}\psi^{bi}$.
The topological amplitude $F_g$ on a genus $g$ surface is given by
\begin{equation}
\FIz{g}=\int_{{\cal M}_g}\langle(\mu \cdot G^-_{T^2})^{g} (\mu \cdot
G^-_{K3}(\bar{u}_+))^{2g-4} (\mu \cdot J^{--}_{K3}) ~( {\rm Rightmovers})~ Z \rangle\,.
\end{equation}
The lattice part of the amplitude $Z$ is a product of the $T^2$ part and the $K3$ part. The $T^2$ part is the usual $(2,22-n)$ lattice sum
\begin{equation}
Z_{T^2}= \sum_{(P_L,\bar{P}_L;\vec{P}_R)\in \Gamma_{(2,22-n)}}e^{i\pi
  \bar{P}_L \tau P_L -
i\pi \vec{P}_R \bar{\tau} \vec{P}_R}\,,
\end{equation}
where $\tau$ is the usual  $g \times g$ period matrix and there are
$g$ $(2,22-n)$ vectors
$(P_L,\bar{P}_L;\vec{P}_R)$. The
$(4,n)$ lattice contribution is given by
\begin{equation}
Z_{K3}=\sum_{(P_{ai};P_{\hat{A}})\in\Gamma_{(4,n)}}e^{i\pi \epsilon^{ij}
  \epsilon^{ab}P_{ai}^{\alpha} t_{\alpha \beta}
P_{bj}^{\beta} - i\pi P_{\hat{A}}^{\alpha} \bar{t}_{\alpha \beta}
P_{\hat{A}}^{\beta}} \equiv \sum_{(P_{ai};P_{\hat{A}})\in\Gamma_{(4,n)}}
Z_{K3}(P_{ai};P_{\hat{A}})\,,
\label{lattice}
\end{equation}
where $\alpha, \beta= 1,...,g-1$ and $t_{\alpha \beta}$ are $(g-1)\times (g-1)$ symmetric period matrix associated with the $(g-1)$ twisted differentials $\omega_{\alpha}$ and  $\bar{\omega}_{\alpha}$. These twisted differentials are normalized with respect to $g-1$ cycles $A_{\alpha}$ and the periods $t_{\alpha \beta}= \int_{B_{\alpha}} \omega_{\beta}$ with $B$ being dual cycles.\footnote{$A$ and $B$ cycles are not the usual homology cycles, instead they are $2g-2$ independent cycles around which the twisted differentials are single valued. Details of this and a particular choice of these cycles can be found in \cite{Narain:1990mw}. The lattices $\Gamma_{2,22-n}$ and $\Gamma_{4,n}$ are not exactly sublattices of $\Gamma_{6,22}$ but instead they are related to them as explained in \cite{Narain:1990mw}. However the precise form of these lattices will not be important in the following. The only important property of the lattices will be their invariant quadratic forms which is the usual one i.e. the square of the left moving part minus the square of the right moving part.}

In the following we will need the formulae for the $(\mu\cdot Q)$ when $Q=\omega_{\alpha}\omega_{\beta}$
and $Q= \omega_I \omega_J$ respectively where $\omega_{I}$ and $\omega_J$ are untwisted abelian differentials
with $I,J=1,...,g$:
\begin{align}
&(\mu\cdot \omega_{\alpha} \omega_{\beta})= dm_N \frac{\partial t_{\alpha \beta}}{\partial m_N}\,,&&\text{and}&&(\mu\cdot \omega_I \omega_J)= dm_N \frac{\partial \tau_{IJ}}{\partial m_N}\,.
\label{muomega}
\end{align}
In the topological amplitude all the $\partial X$ in the supercurrents are replaced by the zero modes. Therefore throughout the remaining part of this section we will use the following shorthand notation
\begin{align}
&G^i_{K3,j}= \psi^{ai} P_{aj}\,, &&G^-_{K3}(\bar{u}_+)=\psi^{a-} P_{aj}\bar{u}^j_+\,,&&P_{bj}=P_{bj}^{\alpha} \omega_{\alpha}\,.
\end{align}
Note that the symbol $P_{bj}$ is a twisted holomorphic 1-form on the
Riemann  surface $\Sigma$.
$\psi^{ai}$ in the amplitude are also replaced by their zero modes $\zeta^{a-}_{\alpha} \omega_{\alpha}$
with $\zeta^{a-}_{\alpha}$ being Grassman numbers. However in the following it is convenient to keep
$\psi^{a-}$ as a conformal field since in the various manipulations carried out below it simplifies
keeping track of various combinatorics and (anti-) symmetrization properties.

Variation with respect to hypermultiplet moduli can act on either $Z_{K3}$ or the $G^-$. Using (\ref{variation})
we find
\begin{align}
D_{{\hat{A}},ai} Z_{K3}(P_{ai};P_{\hat{A}})=2 i \pi (t-\bar{t})_{\alpha \beta} P_{ai}^{\alpha} P_{\hat{A}}^{\beta}
=-2i\pi \int_{\Sigma_g} P_{ai} P_{\hat{A}}^{\beta} \bar{\omega}_{\beta}= - \int_{\Sigma_g}\oint G^+_{K3,i}  (\psi_{a}^{-} \bar{P}_{\hat{A}})\,,
\end{align}
where we have used the fact that $2i\pi P_{ai}= \oint
G^+_{K3,i} \psi_{a}^-$. Here and in the
following we use the shorthand notation
\begin{align}
&P_{\hat{A}} = P_{\hat{A}}^{\beta} \omega_{\beta}\,,&&\text{and} &&\bar{P}_{\hat{A}}= P_{\hat{A}}^{\beta} \bar{\omega}_{\beta}\,.
\end{align}
Similarly we will use the notation
\begin{equation}
D_{{\hat{A}},ai} G^-_{K3}(\bar{u}_+)= \epsilon_{ik} \bar{u}^k_+ \psi_{a}^- P_{\hat{A}}^{\alpha} \omega_{\alpha}\,.
\end{equation}
Thus we find
\begin{align}
D_{{\hat{A}},ai} \FIz{g} = \int_{{\cal M}_g} \langle &(\mu \cdot G^-_{T^2})^g \bigl{[}
(2g-4) (\mu \cdot G^-_{K3}(\bar{u}_+))^{2g-5} (\mu \cdot \epsilon_{ik}\bar{u}^k_+
\psi_{a}^- P_{\hat{A}})(\mu \cdot J^{--}_{K3})\nonumber\\
&-(\mu \cdot G^-_{K3}(\bar{u}_+))^{2g-4} (\mu \cdot J^{--}_{K3})\int_{\Sigma_g}
\oint G^{+}_{K3,i}  (\psi_{a}^-\bar{P}_{\hat{A}})\bigr{]} Z\rangle\,.
\end{align}
By deforming the contour integral and using the world-sheet $N=4$
superconformal algebra, the second
term on the right hand side can be written as
\begin{align}
\int_{{\cal M}_g} \langle(\mu \cdot G^-(T^2))^g &\bigl{[}\frac{2g-4}{2}\epsilon_{ik}\bar{u}^k_+ (\mu \cdot G^-_{K3}(\bar{u}_+))^{2g-5} (\mu \cdot (\epsilon^{cd}\epsilon^{mn} P_{cm} P_{dn}+P_L
\bar{P}_L))(\mu \cdot J^{--}_{K3})\nonumber\\
&+(\mu \cdot G^-_{K3}(\bar{u}_+))^{2g-4}
(\mu \cdot G^{-}_{K3,i})\bigr{]} \int_{\Sigma_g} (\psi_{a}^-\bar{P}_{\hat{A}})Z\rangle\,,
\label{dlattice}
\end{align}
where in the first term we have added $P_L \bar{P}_L \equiv P_L^I \bar{P}_L^J \omega_I \omega_J$ with $\omega_I$ and $\omega_J$ being the untwisted holomorphic 1-differentials, $I,J=1,...,g$. This additional term can be shown to vanish by repeating the argument leading to (\ref{Ttorus}). Now using (\ref{lattice}) one obtains
\begin{align}
i\pi(\mu (\epsilon^{cd} \epsilon^{mn} P_{cm} P_{dn}+ P_L \bar{P}_L)) Z
&= dm_N [\frac{\partial
t_{\alpha \beta}}{\partial m_N} \frac{Z_{K3}}{\partial t_{\alpha
  \beta}} Z_{T^2}+ \frac{\partial \tau_{IJ}}
{\partial m_N}
\frac{\partial Z_{T^2}}{\partial \tau_{IJ}} Z_{K3}] \nonumber\\
& = d m_N \frac{\partial Z}{\partial m_N} = \partial Z\,,
 \end{align}
where in the last equality $\partial$ is the holomorphic exterior derivative in ${\cal M}_g$. We also have the relation:
\begin{equation}
(\mu \cdot \psi_{a}^- P_{\hat{A}}) = dm_N \frac{\partial t_{\alpha \beta}}{\partial m_N} {\psi_{a}^-}^{\alpha} P_{\hat{A}}^{\beta}
=-dm_N \frac{\partial}{\partial m_N} \int_{\Sigma_g} \psi_{a}^i
\bar{P}_{\hat{A}}= -\partial \int_{\Sigma_g} \psi_{a}^i\bar{P}_{\hat{A}}\,.
\label{relation}
\end{equation}
Combining the above four equations we find
\begin{align}
D_{{\hat{A}},ai} \FIz{g}&=(2g-4) \epsilon_{ik}u^k \int_{{\cal M}_g} \partial \langle
(\mu \cdot G^-_{T^2})^g (\mu \cdot G^-_{K3}(\bar{u}_+))^{2g-5}(\mu \cdot J^{--}_{K3})
\int_{\Sigma_g} \psi_{a-} \bar{P}_{\hat{A}} Z \rangle
\nonumber\\
&-2i\pi \int_{{\cal M}_g} \langle(\mu \cdot G^-_{T^2})^g (\mu \cdot G^-_{K3}(\bar{u}_+))^{2g-4}
(\mu \cdot G^-_{K3,i}) \int_{\Sigma_g} \psi_{a-} \bar{P}_{\hat{A}} Z\rangle\,.
\end{align}
In obtaining the first term we have used the fact that the holomorphic
exterior  derivative of $(\mu \cdot G^-_{T^2})$,
$(\mu \cdot G^-_{K3}(\bar{u}_+))$ and $(\mu \cdot J^{--}_{K3})$ vanish due to (\ref{muomega}).
At this stage, we can again verify the harmonicity equation (\ref{StringHarmEqu}) by operating with
$\epsilon^{ji}\frac{\partial}{\partial \bar{u}^j_+}$. As we can see, in this case the second term above drops out
and we are left with only the first term which is a total derivative in ${\cal M}_g$ and hence gets contribution only from the degeneration limits.\\

\noindent
Now let us take a second derivative with respect to hypermultiplet moduli. In order not to clutter the equations we will ignore the total derivative terms, so the equations below are true only up to total derivatives in
${\cal M}_g$
\begin{align}
D_{{\hat{B}},bj}&D_{{\hat{A}},ai} \FIz{g}=\nonumber\\
& -2i\pi \int_{{\cal M}_g} \langle(\mu \cdot G^-_{T^2})^g
\bigg[ \big[(2g-4) (\mu \cdot G^-_{K3}(\bar{u}_+))^{2g-5}
(\mu\cdot\epsilon_{jk}\bar{u}^k_+ \psi_{b}^- P_{\hat{B}})(\mu \cdot G^{-}_{K3,i})\nonumber\\
&+ (\mu \cdot G^-_{K3}(\bar{u}_+))^{2g-4}
(\mu \cdot \epsilon_{ji}\psi_{b}^-
P_{\hat{B}})\nonumber\\
&-(\mu \cdot G^-_{K3}(\bar{u}_+))^{2g-4} (\mu \cdot G^{-}_{K3,i})\int_{\Sigma_g} \oint G^{+}_{K3,j}
(\psi_{b}^- \bar{P}_{\hat{B}})\big]\int_{\Sigma_g} \psi_{a}^-
\bar{P}_{\hat{A}} \nonumber\\
&+ (\mu \cdot G^-_{K3}(\bar{u}_+))^{2g-4} (\mu \cdot G^{-}_{K3,i})\int_{\Sigma_g} \delta_{{\hat{A}}{\hat{B}}}\psi_{a}^-\bar{P}_{bj}\bigg]~ Z\rangle\,.
\end{align}
Contracting by $\epsilon^{ji}$ we obtain
\begin{align}
\epsilon^{ji}D_{{\hat{B}},bj}D_{{\hat{A}},ai} \FIz{g}=& -2i\pi \int_{{\cal M}_g}\langle(\mu
\cdot G^-_{T^2})^g \bigg[\big[(2g-2)
(\mu \cdot G^-_{K3}(\bar{u}_+))^{2g-4} (\mu \cdot \psi_{b}^- P_{\hat{B}})\nonumber\\
&-(\mu \cdot G^-_{K3}(\bar{u}_+))^{2g-4} \epsilon^{ji}(\mu \cdot G^{-}_{K3,i})\int_{\Sigma_g} \oint
G^{+}_{K3,j} (\psi_{b}^- \bar{P}_{\hat{B}})\big]
\int_{\Sigma_g} \psi_{a}^-\bar{P}_{\hat{A}}\nonumber\\
&+ (\mu \cdot G^-_{K3}(\bar{u}_+))^{2g-4} \epsilon^{ji}(\mu \cdot G^{-}_{K3,i})\int_{\Sigma_g} \delta_{{\hat{A}}{\hat{B}}}\psi_{a}^-\bar{P}_{bj}\bigg]~ Z\rangle\,.
\label{20form}
\end{align}
We are interested in antisymmetrizing the pairs $({\hat{A}},a)$ and $({\hat{B}},b)$, since symmetrization gives a commutator
and is trivially governed by the algebra (\ref{algebra}). In the following
therefore we restrict ourselves to
the antisymmetric part. By deforming the contour in the second term  and carrying out the steps as above we
find for the antisymmetric part:
\begin{align}
&-\int_{{\cal M}_g} \langle(\mu \cdot G^-_{T^2})^g (\mu \cdot G^-_{K3}(\bar{u}_+))^{2g-4}
\epsilon^{ji}(\mu \cdot G^{-}_{K3,i})\int_{\Sigma_g} \oint G^{+}_j
(\psi_{b}^- \bar{P}_{\hat{B}})]\int_{\Sigma_g} \psi_{a}^-\bar{P}_{\hat{A}} ~Z\rangle=\nonumber\\
&=-\frac{\pi i}{2} (2g-2)\int_{{\cal M}_g}\langle (\mu \cdot G^-_{T^2})^g(\mu
\cdot G^-_{K3}(\bar{u}_+))^{2g-4}(\mu \cdot \epsilon^{cd}
\epsilon^{mn}P_{cm}P_{dn})\int_{\Sigma_g}
\psi_{b}^-\bar{P}_{\hat{B}}  \int_{\Sigma_g} \psi_{a}^-\bar{P}_{\hat{A}}
~Z\rangle\nonumber\\
&=-\frac{1}{2} (2g-2)\int_{{\cal M}_g} \langle(\mu \cdot G^-_{T^2})^g(\mu
\cdot G^-_{K3}(\bar{u}_+))^{2g-4} \int_{\Sigma_g} \psi_{b}^-\bar{P}_{\hat{B}}
\int_{\Sigma_g} \psi_{a}^-\bar{P}_{\hat{A}}  ~\partial Z\rangle\nonumber\\
&=-\frac{1}{2} (2g-2)\int_{{\cal M}_g} \partial\langle(\mu \cdot
G^-_{T^2})^g(\mu \cdot G^-_{K3}(\bar{u}_+))^{2g-4} \int_{\Sigma_g} \psi_{b}^-
\bar{P}_{\hat{B}}  \int_{\Sigma_g} \psi_{a}^-\bar{P}_{\hat{A}}
~Z\rangle\nonumber\\&\hspace{0.5cm}-(2g-2)\int_{{\cal M}_g}
\langle(\mu \cdot G^-_{T^2})^g(\mu \cdot G^-_{K3}(\bar{u}_+))^{2g-4} (\mu \cdot \psi_{b}^- P_{\hat{B}})\int_{\Sigma_g} \psi_{a}^-\bar{P}_{\hat{A}}  ~Z\rangle\,,
\end{align}
where in order to obtain the third equality we have again added the
$T^2$ part of the left moving momentum
square for the same reason as in eq.(\ref{dlattice}). In the last equality we
have used eq.(\ref{relation}). Note that the
last term in the last equation exactly cancels the first term in
eq.(\ref{20form}) giving, up to total derivatives
in ${\cal M}_g$:
\begin{align}
&\epsilon^{ji}D_{{\hat{B}},bj}D_{{\hat{A}},ai} \FIz{g}
=-2i\pi \delta_{{\hat{A}}{\hat{B}}}\int_{{\cal M}_g}\langle (\mu \cdot G^-_{T^2})^g (\mu
\cdot G^-_{K3}(\bar{u}_+))^{2g-4} \epsilon^{ji}(\mu \cdot G^{-}_{K3,i})
\int_{\Sigma_g} \psi_{a}^-\bar{P}_{bj} ~Z\rangle\nonumber\\
&= \frac{1}{2}\delta_{{\hat{A}}{\hat{B}}}\int_{{\cal M}_g} \langle(\mu \cdot G^-_{T^2})^g
(\mu \cdot G^-_{K3}(\bar{u}_+))^{2g-4} \epsilon^{ji}
(\mu \cdot G^{-}_{K3,i})\int_{\Sigma_g} \oint G^{+}_{K3,j}(\psi_{a}^-\Pi \psi_{b}^-) ~Z\rangle\,,
\end{align}
where $(\Pi \psi_{b}^-)(\bar{z})= \bar{\omega}_{\alpha}(\bar{z})(t-\bar{t})^{-1}_{\alpha \beta}\int_{\Sigma_g}
\bar{\omega}_{\beta}(\bar{w}) \psi_{b}^-(w)$. Basically $\Pi$ takes a
holomorphic 1-form to an
anti-holomorphic 1-form. By deforming the contour of $G^+$ and going through the steps as before results in:
\begin{align}
\epsilon^{ji}D_{{\hat{B}},bj}D_{{\hat{A}},ai} \FIz{g}&=
\delta_{{\hat{A}}{\hat{B}}}(g-1)\int_{{\cal M}_g}\langle(\mu \cdot G^-_{T^2})^g
(\mu \cdot G^-_{K3}(\bar{u}_+))^{2g-4} \int_{\Sigma_g} (\psi_{a}^-\Pi \psi_{b}^-)~\partial Z\rangle\nonumber\\
&= \delta_{{\hat{A}}{\hat{B}}}(g-1)\int_{{\cal M}_g}\partial\langle(\mu \cdot
G^-_{K3})^g (\mu \cdot G^-_{K3}(\bar{u}_+))^{2g-4} \int_{\Sigma_g}
(\psi_{a}^-
(\Pi \psi_{b}^-)~Z\rangle\nonumber\\
&\hspace{0.5cm}-(g-1)\delta_{{\hat{A}}{\hat{B}}}\int_{{\cal
    M}_g}\langle(\mu \cdot G^-_{T^2})^g (\mu \cdot G^-_{K3}(\bar{u}_+))^{2g-4}
(\mu \cdot J^{--}_{K3}) ~Z\rangle\,.
\label{finaleq}
\end{align}
In the last line we have used the fact that
\begin{equation}
\partial \left(\int_{\Sigma_g} (\psi_{a}^-
\Pi \psi_{b}^-)\right)=\partial ({\psi_{a}^-}^{\alpha}{\psi_{b}^-}^{\beta}(t-\bar{t})_{\alpha \beta})
=(\mu \cdot J^{--}_{K3})\,.
\end{equation}
The first term in the last equality of (\ref{finaleq}) is a total derivative in
${\cal M}_g$, while the second term is proportional to $F_g$. Thus
up to total derivative terms (which give rise to anomaly) the second order
relation is
\begin{equation}
\epsilon^{ji}D_{{\hat{B}},bj}D_{{\hat{A}},ai} \FIz{g}=\delta_{{\hat{A}}{\hat{B}}} \epsilon_{ab}(g-1) \FIz{g}\,.
\end{equation}
This is indeed the eq.(\ref{modconstr}) obtained in Section \ref{Sect:HarmonicDescription}.

\section{Concluding remarks}\label{Sect:Conclusions}

In this work, we have analyzed a new series of ${\cal N}=2$ topological amplitudes associated to a particular class of higher dimensional 1/2-BPS terms in the low energy effective superstring action, involving both vector multiplets and neutral hypermultiplets. We computed these amplitudes on both type I and heterotic string side and studied the moduli dependence of the corresponding couplings in string theory, as well as in the effective supergravity using harmonic superspace. Their BPS character implies holomorphicity with respect to vector moduli and harmonicity with respect to the hypermultiplet ones, together with a second order differential equation, similarly to the couplings of an ${\cal N}=4$ series we had found in the past~\cite{Antoniadis:2007cw}. All conditions are violated by anomalies due to world-sheet boundary contributions, that in the effective field theory seem to be associated to connected graphs upon elimination of the auxiliary fields. This is an interesting open question that needs further work in the future. A disturbing property is that the anomalies bring new `semi-topological' objects, similarly to the ${\cal N}=1$ case studied in the past~\cite{Antoniadis:1996qg, Antoniadis:2005sd}. It would be also interesting to clarify the relation between our analysis and the conditions imposed on the open topological amplitudes of ref.~\cite{Walcher:2007tp} that avoid the appearance of extra quantities in the anomaly equation. Another open question is to study the partial/soft breaking of ${\cal N}=2\to{\cal N}=1$ supersymmetry, as well as the possible relevance of such couplings in the entropy correction of black holes.

%%%%%%%%%%%%%%%%%%%%%%%%%%%%%%%%%%%%%%%%%%%%%%%%%%%%%%%%%%%%%%%%%%%%%
\section*{Acknowledgements}
We would like to thank Atish Dabholkar, Sergio Ferrara, Boris Pioline and Johannes Walcher for
helpful discussions. This work was supported in part by the European Commission under the ERC Advanced Grant 226371 and in part by the French Agence Nationale de la Recherche, contract ANR-06-BLAN-0142. The research of S.H. has been supported by the Swiss National Science Foundation.
%%%%%%%%%%%%%%%%%%%%%%%%%%%%%%%%%%%%%%%%%%%%%%%%%%%%%%%%%%%%%%%%%%%%%

\appendix

\section{Superconformal algebras}\label{append:SCA}
The world-sheet theory for string theories on $K3\times T^2$ is a product of an $\N=2$ and an $\N=4$ superconformal field theory representing $T^2$ and $K3$ respectively. In this appendix we will describe both of these theories and also discuss their topologically twisted versions.
%%%%%%%%%%%%%%%%%%%%%%%%%%%%%%%%%%%%%%%%%%%%%%%%%%%%%%%%%%%%%%%%%
\subsection{The $\N=2$ superconformal algebra}\label{reviewopentopI}
The (untwisted) $\N=2$ SCFT of the $T^2$ contains besides the energy momentum tensor $T_{T^2}$ two  supercurrents $G^\pm_{T^2}$ which are positively and negatively charged under a $U(1)$ Kac-Moody current $J_{T^2}$. The conformal weights of these operators are given by
\begin{align}
&h_{T_{T^2}}=2\,,&&h_{G^\pm_{T^2}}=\frac{3}{2}\,, &&h_{J_{T^2}}=1\,.
\end{align}
The non-trivial operator product expansions (OPE) of these objects are given by
{\allowdisplaybreaks
\begin{align}
&T_{T^2}(z)T_{T^2}(w)=\frac{2T_{T^2}(w)}{(z-w)^2}+\frac{\partial_wT_{T^2}(w)}{z-w}\,,\label{N2SCAa}\\
&T_{T^2}(z)G_{T^2}^{\pm}(w)=\frac{3G_{T^2}^\pm(w)}{2(z-w)^2}+\frac{\partial_wG_{T^2}^\pm(w)}{z-w}\,,\\
&T_{T^2}(z)J_{T^2}(w)=\frac{J_{T^2}(w)}{(z-w)^2}+\frac{\partial_wJ_{T^2}(w)}{z-w}\,,\\
&G_{T^2}^+(z)G_{T^2}^-(w)=\frac{6}{(z-w)^3}+\frac{2J_{T^2}(w)}{(z-w)^2}+\frac{2T_{T^2}(w)+\partial_wJ_{T^2}(w)}{z-w}\,,\\
&J_{T^2}(z)G_{T^2}^\pm(w)=\pm\frac{G_{T^2}^\pm(w)}{z-w}\,,\\
&J_{T^2}(z)J_{T^2}(w)=\frac{2}{(z-w)^2}\,.\label{N2SCAf}
\end{align}}
An explicit representation of this algebra in terms of the free boson $X_3$ and fermion $\psi_3$ living on $T^2$ is given by
\begin{align}
&T_{T^2}=\frac{1}{2}\psi_3{\leftrightarrow\atop\displaystyle{\partial\atop~}}\bar{\psi}_3+\partial X_3\partial \bar{X}_3,&&G^-_{T^2}=\bar{\psi}_3\partial X_3, &&G^{+}_{T^2}=\psi_3\partial\bar{X}_3, &&J_{T^2}=\psi_3\bar{\psi}_3\ .
\end{align}
A twisted version of the $\N=2$ super-conformal algebra is given by redefining the energy momentum tensor in the following manner
\begin{align}
T_{T^2}\to T_{T^2}-\frac{1}{2}\partial J_{T^2}\,.
\end{align}
This in particular has the effect of shifting the conformal weight of all operators by half their $U(1)$ charge. In this way, we find
\begin{align}
&h^{\text{twist}}_{T_{T^2}}=2\,,&&h^{\text{twist}}_{G^-_{T^2}}=2\,, &&h^{\text{twist}}_{G^+_{T^2}}=1\,, &&h^{\text{twist}}_{J_{T^2}}=1\,.
\end{align}
The new conformal weights of the supercurrents allow us to identify $G^+_{T^2}$ with the BRST operator of the twisted theory, while $G^-_{T^2}$ becomes the reparametrization anti-ghost thereby defining the measure of the topological string.
%%%%%%%%%%%%%%%%%%%%%%%%%%%%%%%%%%%%%%%%%%%%%%%%%%%%%%%%%%%%%%%%%
\subsection{The $\N=4$ superconformal algebra}\label{Sect:SCFT}
The $\N=4$ SCFT representing $K3$ contains an energy momentum tensor $T_{K3}$ which is accompanied by two doublets of supercurrents $(G^+_{K3},\tilde{G}^-_{K3})$ and $(\tilde{G}^+_{K3},G^-_{K3})$ which transform under an $SU(2)$ Kac-Moody current algebra formed by $(J^{\pm\pm}_{K3},J_{K3})$. The conformal weights of these operators are given to be
\begin{align}
&h_{T_{K3}}=2\,,&&h_{G^\pm_{K3}}=h_{\tilde{G}^\pm_{K3}}=\frac{3}{2}\,,&&h_{J^{\pm\pm}_{K3}}=h_{J_{K3}}=1\,.
\end{align}
The non-trivial OPEs of these objects are given by
{\allowdisplaybreaks
\begin{align}
&J^{--}_{K3}(z)G^+_{K3}(0)\sim\frac{\tilde{G}^-_{K3}(0)}{z}, &&J_{K3}^{++}(z)\tilde{G}_{K3}^-(0)
\sim-\frac{G^+_{K3}(0)}{z}\,,\nonumber\\
&J_{K3}^{++}(z)G^-_{K3}(0)\sim \frac{\tilde{G}^+_{K3}(0)}{z}, &&J_{K3}^{--}(z)
\tilde{G}^+_{K3}(0)\sim-\frac{G^-_{K3}(0)}{z}\,,\nonumber
\end{align}
\begin{align}
&G^+_{K3}(z)G^-_{K3}(0)\sim\frac{J_{K3}(0)}{z^2}-\frac{T^B_{K3}(0)-\frac{1}{2}\partial J_{K3}(0)}{z}\,,
\nonumber\\
&\tilde{G}^+_{K3}(z)\tilde{G}^-_{K3}(0)\sim\frac{J_{K3}(0)}{z^2}-\frac{T^B_{K3}(0)-\frac{1}{2}
\partial J_{K3}(0)}{z}\,,\nonumber\\
&\tilde{G}^+_{K3}(z)G^+_{K3}(0)\sim\frac{2J^{++}_{K3}(0)}{z^2}+\frac{\partial J^{++}_{K3}(0)}{z}\,,
\nonumber\\
&\tilde{G}^-_{K3}(z)G^-_{K3}(0)\sim\frac{2J^{--}_{K3}(0)}{z^2}+\frac{\partial J^{--}_{K3}(0)}{z}\,,
\nonumber
\end{align}
}
while for any operator $O^q_{K3}$ of $U(1)$ charge $q$, one has:
\begin{align}
J_{K3}(z)O^q_{K3}(0)\sim q\frac{O^q_{K3}(0)}{z}\,.
\nonumber
\end{align}
A representation in terms of free bosons $X_{4,5}$ and fermions $\psi_{4,5}$ living on a torus-orbifold realization of $K3$ is given by
\begin{align}
T_{K3}=\partial X_4\partial{\bar X}_4+\partial X_5\partial{\bar X}_5+
{1\over 2}(\psi_4{\leftrightarrow\atop\displaystyle{\partial\atop~}}{\bar\psi}_4+
\psi_5{\leftrightarrow\atop\displaystyle{\partial\atop~}}{\bar\psi}_5)\ ,
\end{align}
${}$\vspace{-0.8cm}
\begin{align}
&J_{K3}=\psi_4\bar{\psi}_4+\psi_5\bar{\psi}_5, && J^{++}_{K3}=\psi_4\psi_5, &&J^{--}_{K3}
=\bar{\psi}_4\bar{\psi}_5\,,
\label{N4curs}
\end{align}
\begin{align}
&G^+_{K3}=\psi_4\partial \bar{X}_4+\psi_5\partial\bar{X}_5, &&\tilde{G}^+_{K3}=-\psi_5\partial X_4+\psi_4\partial X_5\,,\label{K3scurrent1}\\
&G^{-}_{K3}=\bar{\psi}_4\partial X_4+\bar{\psi}_5\partial X_5, &&\tilde{G}^-_{K3}=-\bar{\psi}_5\partial \bar{X}_4+\bar{\psi}_4\partial \bar{X}_5\,.
\label{K3scurrent2}
\end{align}
The topologically twisted theory can be defined after specifying an $\N=2$ subalgebra inside the $\N=4$. We can then similarly redefine the energy momentum tensor
\begin{align}
T_{K3}\to T_{K3}-\frac{1}{2}\partial J_{K3}\ .
\end{align}
In this way, again the conformal dimensions of all operators are shifted by half their charge with respect to $J_{K3}$
\begin{align}
&h^{\text{twist}}_{T_{K3}}=2\,,&&h^{\text{twist}}_{G^-_{K3}}=h^{\text{twist}}_{\tilde{G}^-_{K3}}=2\,, &&h^{\text{twist}}_{G^+_{K3}}=h^{\text{twist}}_{\tilde{G}^+_{K3}}=1\,,\nonumber\\
&h^{\text{twist}}_{J^{--}_{K3}}=2\,, &&h^{\text{twist}}_{J^{++}_{K3}}=0\,, &&h^{\text{twist}}_{J_{K3}}=1\,.
\end{align}
%%%%%%%%%%%%%%%%%%%%%%%%%%%%%%%%%%%%%%%%%%%%%%%%%%%%%%%%%%%%%%%%%
\subsection{The $\N=4$ superconformal algebra in covariant basis}\label{App:SCFTN4cov}
In this section we would like to rewrite the $\N=4$ superconformal algebra of Appendix \ref{Sect:SCFT} in an $SU(2)$ covariant manner. To this end, we simply have to specify the OPE relations of the doublet supercurrents of (\ref{CovarSupercurrents}). They are given by
\begin{align}
&G^+_{K3,i}(z)G^+_{K3,j}(0)\sim\epsilon_{ij}\left(\frac{2J^{++}_{K3} (0)}{z^2}+\frac{\partial J^{++}_{K3} (0)}{z}\right),\label{algebraCOV1}\\
&G^-_{K3,i}(z)G^-_{K3,j}(0)\sim \epsilon_{ij}\left(\frac{2J^{--}_{K3} (0)}{z^2}+\frac{\partial J^{--}_{K3} (0)}{z}\right),\\
&G^+_{K3,i}(z)G^-_{K3,j}(0)\sim-\epsilon_{ij}\left(\frac{J(0)}{z^2}-\frac{T_{K3}(0)-\frac{1}{2}\partial J(0)}{z}\right),
\end{align}
\begin{align}
&J^{++}_{K3} (z)G^+_{K3,i}(0)\sim0, &&J^{++}_{K3} (z)G^-_{K3,i}\sim\frac{G^+_{K3,i}}{z},\\
&J^{--}_{K3} (z)G^+_{K3,i}(0)\sim-\frac{G^-_{i}}{z}, &&J^{--}_{K3} (z)G^-_{K3,i}\sim0.\label{algebraCOV5}
\end{align}
The relation between $G^{\pm}_{K3,i}$ and ($G^{\pm}_{K3},\tilde{G}^{\pm}_{K3}$) is $G^{\pm}_{K3}=G^{\pm}_{K3,\pm}$ and $\tilde{G}^{\pm}_{K3}=\pm G^{\pm}_{K3,\mp}$. In particular
the picture changing operator contains the term $e^{\varphi} (G^{+}_{K3,+}+ G^-_{K3,-})$.
%%%%%%%%%%%%%%%%%%%%%%%%%%%%%%%%%%%%%%%%%%%%%%%%%%%%%%%%%%%%%%%%%
%%%%%%%%%%%%%%%%%%%%%%%%%%%%%%%%%%%%%%%%%%%%%%%%%%%%%%%%%%%%%%%%%
%%%%%%%%%%%%%%%%%%%%%%%%%%%%%%%%%%%%%%%%%%%%%%%%%%%%%%%%%%%%%%%%%

\section{Vertex operators}\label{App:VertexOperators}
In this appendix we list the relevant vertex operators for calculating the topological amplitudes in the heterotic theory. Our convention for the latter will be that the left-moving sector is supersymmetric, while the right moving one is made up from just a bosonic theory.\\

\noindent
With this convention the vertex operator for a gauge field in the 0-picture is given by
\begin{align}
V_{(F)}^A(p)=:\left(\partial X_\mu+ip\cdot\psi\psi_\mu\right)\bar{J}^Ae^{ip\cdot X}:\,.
\end{align}
Here $X_\mu$ are complex bosonic space-time coordinates with $\psi$ their fermionic partners. Moreover, $\bar{J}^A$ are the right moving Kac-Moody currents. In a similar way we can define the 0-picture vertex operator for (the derivative of) a scalar field
\begin{align}
V_{(S)}^A(p)=:\left(\partial X_I+ip\cdot\psi\psi_I\right)\bar{J}^Ae^{ip\cdot X}:\,,
\end{align}
where $X_I$ denote the complex coordinates of the internal theory with $\psi_I$ their fermionic partners.\\

\noindent
It remains to introduce the vertex operator for the gauginos. We will take the latter in the $-\frac{1}{2}$-picture, where the vertex reads
\begin{align}
V_{(\la)}^{A\alpha}(p)=:e^{-\frac{1}{2}\varphi}S^\alpha\Sigma\bar{J}^Ae^{ip\cdot X}:\,.
\end{align}
Here $\varphi$ is the scalar field bosonizing the superghost, $S^\alpha$ is a space-time spin field and $\Sigma$ a spin-field of the internal theory. Notice in particular the appearance of the spinor index $\alpha$.
%%%%%%%%%%%%%%%%%%%%%%%%%%%%%%%%%%%%%%%%%%%%%%%%%
\section{Theta functions and prime forms}\label{App:mathstruct}
In this appendix, we review some of the mathematical quantities which we need for the calculation of topological amplitudes. The material is fairly standard and we adopt the notation of \cite{Verlinde:1986kw}.\\

\noindent
Points (``coordinates'') on a Riemann surface $\Sigma_g$ of genus $g$ are defined using the Jacobi map. To this end, we define a base point $P_0$ on the surface and cut $\Sigma_g$ open along the homology cycles defined in section \ref{Sect:Involution}. The coordinates of a point $P$ different from $P_0$ are then defined as\footnote{In computations, for notational simplicity, we drop the distinction between the point on the Riemann surface and the corresponding Jacobi map.}
\begin{align}
\mathcal{I}:\ P\to z_i(P)=\int_{P_0}^P\omega_i\,,
\end{align}
where $\omega_i$ are the $g$ holomorphic 1-differentials (see (\ref{periodmatrix})) on $\Sigma_g$. $z$ can be thought of as an element of the complex torus
\begin{align}
J(\mathcal{M}_g)=\mathbb{C}^g/(\mathbb{Z}^g+\tau \mathbb{Z}^g)\,.
\end{align}
Furthermore, for $z\in J(\mathcal{M}_g)$ one can define the Riemann theta function \cite{Fay}
\begin{align}
\vartheta (z,\tau)=\sum_{n\in \mathbb{Z}^g}e^{i\pi n_i\tau_{ij}n_j+2\pi i n_i z_i}\,,\label{thetadefinition}
\end{align}
where $\tau_{ij}$ is the period matrix of the Riemann surface, which we have already defined in (\ref{periodmatrix}). $\vartheta (z,\tau)$ satisfies the following identity for shifts under the lattice $\mathbb{Z}_g+\tau \mathbb{Z}_g$
\begin{align}
\vartheta(z+\tau n+m,\tau)=e^{-i\pi n\tau n-2\pi imz}\vartheta(z,\tau)\,.
\end{align}
An important result, which we only state without derivation is the Riemann vanishing theorem: There exists a divisor class $\Delta$ of degree $g-1$ such that $\vartheta(z,\tau)=0$ if and only if there are $g-1$
points $p_1,\ldots, p_{g-1}$ on $\mathcal{M}_g$ such that
\begin{align}
z=\Delta-\sum_{i=1}^{g-1}p_i\,.
\end{align}
Moreover, based on the definition (\ref{thetadefinition}), we can define $\vartheta$-functions with non-trivial characteristics (``spin structures'') $(\alpha_1,\alpha_2)\in\left(\frac{1}{2}\mathbb{Z}^g/\mathbb{Z}^g\right)$
\begin{align}
\vartheta_\alpha(z,\tau)=e^{i\pi \alpha_1\tau\alpha_1+2\pi i\alpha_1(z+\alpha_2)}
\vartheta(z+\tau\alpha_1+\alpha_2,\tau)\,.\label{thetadef}
\end{align}
Besides $\vartheta$-functions we also need to discuss prime forms $E(x,y)$. The latter can be viewed as a generalization of the holomorphic function $x-y$  on the Riemann sphere. The precise definition is
\begin{align}
E(x,y)=\frac{f_\alpha(x,y)}{h_\alpha(x)h_\alpha(y)}\,,
\end{align}
where $h_\alpha$ is a holomorphic $\frac{1}{2}$-differential and $f_\alpha$ is defined to be
\begin{align}
f_\alpha(x,y)=\vartheta_\alpha(\int_x^y\omega)\,.
\end{align}
One can show, that $E$ is independent of $\alpha$ and as a true generalization of the function $x-y$, it is antisymmetric under the exchange $x\leftrightarrow y$. Besides that it has a simple root at $x=y$:
\begin{align}
&E(x,y)=-E(y,x),&&\text{and} &&\lim_{x\to y}E(x,y)=\mathcal{O}(x-y)\,.
\end{align}
These identities will turn out to be very important for the calculations in the main body of this work.
%%%%%%%%%%%%%%%%%%%%%%%%%%%%%%%%%%%%%%%%%%%%%%%%%%%%%%%%%%%%%%%%%%%%%%%%%%%
%%%%%%%%%%%%%%%%%%%%%%%%%%%%%%%%%%%%%%%%%%%%%%%%%%%%%%%%%%%%%%%%%%%%%%%%%%%

%%%%%%%%%%%%%%%%%%%%%%%%%%%%%%%%%%%%%%%%%%%%%%%%%%%%
\section{A generalization of $\mathcal{F}_g$ to $\mathcal{F}_{g,n}$}\label{App:RelTopPhysD4}
In this appendix we will compute a more general amplitude than those in Sections \ref{Sect:TypeIamp} and \ref{Sect:hetF3}, which involves $2g$ $K_-$ and $2 n$ $\bar{K}^+$ in the effective action:
\begin{equation}
S_{g,n} = \int \ d^4x \ du \ d^2\q^+ d^2\bq_{-} \ (D_-\cdot
D_-)(\bar{D}^+\cdot \bar{D}^+)(K_{-} \cdot K_{-})^{g-1}\,(\bar{K}^+ \cdot
\bar{K}^+)^{n-1} F_{g,n}(W,\bar{W},q^{+A},u)\ ,
\label{Fgn}
\end{equation}
Such a term will give couplings of the type $2g$
chiral gauginos $\lambda_-$ (where we allow the $(D_-\cdot D_-)$ to act on
two $W$'s inside $\mathcal{F}_{g,n}$) and $2n$ anti-chiral gauginos
$\bar{\lambda}^+$ (where again $(\bar{D}^+\cdot \bar{D}^+)$ on two
$\bar{W}$ inside $\mathcal{F}_{g,n}$), together with two chiral hyperino
$\chi$ and two anti-chiral hyperino $\bar{\psi}$ to soak the two
$\theta^+$ and two $\bar{\theta}_-$ in the superspace integral. It will
turn out that these generalized amplitudes mix with $\mathcal{F}_g$ via
the holomorphic anomaly (see Section \ref{Sect:vectormultipletEQ} where we
show that an anti-holomorphic derivative of $\mathcal{F}_g$ gives a
boundary term proportional to $\mathcal{F}_{g-1,1}$).

We now want to explicitly compute these amplitudes in string theory and therefore we add $2n$ anti-chiral gauginos $\bar{\lambda}^+$ in the $(-1/2)$-picture to the amplitude considered in Section \ref{App:CYtopGeng}. Here we take $g>n$ and the total number of PCO's is $3g+n$. The corresponding operators have the following content
\begin{center}
\begin{tabular}{|c|c|c||c|c||c||c||c|}\hline
\textbf{field} & \textbf{pos.} & \textbf{number} &
\parbox{0.5cm}{\vspace{0.2cm}$\phi_1$\vspace{0.2cm}}&
\parbox{0.5cm}{\vspace{0.2cm}$\phi_2$\vspace{0.2cm}} &
\parbox{0.5cm}{\vspace{0.2cm}$\phi_3$\vspace{0.2cm}} &
\parbox{0.5cm}{\vspace{0.2cm}$H$\vspace{0.2cm}} &
\parbox{0.5cm}{\vspace{0.2cm}~\vspace{0.2cm}} \\\hline
gaugino ${\lambda}_-$& \parbox{0.35cm}{\vspace{0.2cm}$x_i$\vspace{0.2cm}}
& $g$ & \parbox{0.7cm}
{\vspace{0.2cm}$+\frac{1}{2}$\vspace{0.2cm}} & \parbox{0.7cm}
{\vspace{0.2cm}$+\frac{1}{2}$\vspace{0.2cm}} & \parbox{0.7cm}
{\vspace{0.2cm}$+\frac{1}{2}$\vspace{0.2cm}} & \parbox{0.7cm}
{\vspace{0.2cm}$+\frac{1}{\sqrt{2}}$\vspace{0.2cm}} & \parbox{0.4cm}
{\vspace{0.2cm}$\bar{J}$\vspace{0.2cm}} \\\hline
 & \parbox{0.35cm}{\vspace{0.2cm}$y_i$\vspace{0.2cm}} & $g$ & \parbox{0.7cm}
{\vspace{0.2cm}$-\frac{1}{2}$\vspace{0.2cm}} & \parbox{0.7cm}
{\vspace{0.2cm}$-\frac{1}{2}$\vspace{0.2cm}} & \parbox{0.7cm}
{\vspace{0.2cm}$+\frac{1}{2}$\vspace{0.2cm}} & \parbox{0.7cm}
{\vspace{0.2cm}$+\frac{1}{\sqrt{2}}$\vspace{0.2cm}} & \parbox{0.4cm}
{\vspace{0.2cm}$\bar{J}$\vspace{0.2cm}} \\\hline\hline
gaugino $\bar{\lambda}^+$&
\parbox{0.35cm}{\vspace{0.2cm}$w_j$\vspace{0.2cm}} & $n$ & \parbox{0.7cm}
{\vspace{0.2cm}$+\frac{1}{2}$\vspace{0.2cm}} & \parbox{0.7cm}
{\vspace{0.2cm}$-\frac{1}{2}$\vspace{0.2cm}} & \parbox{0.7cm}
{\vspace{0.2cm}$-\frac{1}{2}$\vspace{0.2cm}} & \parbox{0.7cm}
{\vspace{0.2cm}$+\frac{1}{\sqrt{2}}$\vspace{0.2cm}} & \parbox{0.4cm}
{\vspace{0.2cm}$\bar{J}$\vspace{0.2cm}} \\\hline
 & \parbox{0.35cm}{\vspace{0.2cm}$v_j$\vspace{0.2cm}} & $n$ & \parbox{0.7cm}
{\vspace{0.2cm}$-\frac{1}{2}$\vspace{0.2cm}} & \parbox{0.7cm}
{\vspace{0.2cm}$+\frac{1}{2}$\vspace{0.2cm}} & \parbox{0.7cm}
{\vspace{0.2cm}$-\frac{1}{2}$\vspace{0.2cm}} & \parbox{0.7cm}
{\vspace{0.2cm}$+\frac{1}{\sqrt{2}}$\vspace{0.2cm}} & \parbox{0.5cm}
{\vspace{0.2cm}$c \bar{J}$\vspace{0.2cm}} \\\hline\hline
Hyperino $\chi_{A_1}$ & \parbox{0.35cm}{\vspace{0.2cm}$z_1$\vspace{0.2cm}}
& 1 & $+\frac{1}{2}$ & $+\frac{1}{2}$ & $-\frac{1}{2}$ & $0$ &
\parbox{0.7cm}
{\vspace{0.2cm}$\hat{V}_{A_1}$\vspace{0.2cm}} \\\hline
$\chi_{A_2}$& \parbox{0.35cm}{\vspace{0.2cm}$z_2$\vspace{0.2cm}} & 1 &
$-\frac{1}{2}$ & $-\frac{1}{2}$ & $-\frac{1}{2}$ & $0$  & \parbox{0.7cm}
{\vspace{0.2cm}$\hat{V}_{A_2}$\vspace{0.2cm}}\\\hline
$\bar{\psi}_{A_3}$ & \parbox{0.35cm}{\vspace{0.2cm}$z_3$\vspace{0.2cm}} &
1 & $-\frac{1}{2}$ & $+\frac{1}{2}$ & $+\frac{1}{2}$ & $0$  &
\parbox{0.7cm}
{\vspace{0.2cm}$\hat{V}_{A_3}$\vspace{0.2cm}}\\\hline
$\bar{\psi}_{A_4}$& \parbox{0.35cm}{\vspace{0.2cm}$z_4$\vspace{0.2cm}} & 1
& $+\frac{1}{2}$ & $-\frac{1}{2}$ & $+\frac{1}{2}$ & $0$  & \parbox{0.7cm}
{\vspace{0.2cm}$\hat{V}_{A_4}$\vspace{0.2cm}}\\\hline\hline

PCO & \parbox{0.4cm}{\vspace{0.2cm}$r_a$\vspace{0.2cm}} & $g-n$ & $0$ &
$0$ & $-1$ & $0$ & \parbox{0.75cm}
{\vspace{0.2cm}$\partial X_3$\vspace{0.2cm}} \\\hline
& \parbox{0.4cm}{\vspace{0.2cm}$s_b$\vspace{0.2cm}} & $2 g+2 n$ & $0$ &
$0$ & $0$ & $-\frac{1}{\sqrt{2}}$  & \parbox{0.75cm}
{\vspace{0.2cm}$\hat{G}^-_{K3}$\vspace{0.2cm}}\\\hline

\end{tabular}
\end{center}
${}$\\[10pt]
Note that, for later convenience, we have inserted $c$ ghosts at positions
$v_j$ so the operators at $v_j$ have dimension zero and are hence
unintegrated. This means there will be  $n$ additional Beltrami
differentials corresponding to
integration over these $n$ punctures. These additional Beltrami
differentials are folded with $b$ ghosts. Therefore the total number of
$b$'s is $(3g-3+n)$ which balances the $(b,c)$ ghost number.

The spin structure dependent part of the amplitude is
\begin{eqnarray}
(\mathcal{F}_{g,n,s})_{ A_1 A_2 A_3 A_4}&=& F_{\{\Lambda\},s}(u_1,u_2,v)
G_{\{\Lambda\}}(\{x_i,y_i,w_j,v_j,z_k,r_a,s_b\})\nonumber \\
&& \hskip -3cm\cdot \frac{\vartheta_s(\frac{1}{2}\sum_i
(x_i-y_i)+\frac{1}{2}\sum_i
(w_j-v_j)+\frac{1}{2}(z_1-z_2-z_3+z_4))}{\vartheta_s(\frac{1}{2}\sum_i(x_i+y_i)+\frac{1}{2}\sum_i
(w_j+v_j)+\frac{1}{2}(z_1+z_2+z_3+z_4) - \sum_a r_a-\sum_b s_b- 2 \Delta)}
\nonumber\\&& \hskip -3cm
\cdot \frac{\prod_{i<j} E(x_i,x_j)E(y_i,y_j)}{\prod_i
E(x_i,z_2)E(y_i,z_1)}\frac{\prod_a
E(r_a,z_1)E(r_a,z_2)}{\prod_{a,b}E(r_a,s_b)}\nonumber\\&& \hskip -3cm
\cdot \frac{(\prod_{i<j} E(w_i,w_j)E(v_i,v_j))(\prod_{j,a}E(w_j,r_a)
E(v_j,r_a))}{\prod_j E(w_j,z_3)E(v_j,z_4)}\frac{\prod_a
E(r_a,z_1)E(r_a,z_2)}{\prod_{a,b}E(r_a,s_b)}\nonumber\\&& \hskip -3cm
\cdot \left(\frac{\prod_{b,k}E(s_b,z_k)}{\prod_{k<l}E(z_k,z_l)
\prod_{b<c}E(s_b,s_c)}\right)^{1/2}\,,
\label{gamp1}
\end{eqnarray}
where now
\begin{eqnarray}
u_1 &=& \frac{1}{2}\sum_i
(x_i-y_i)-\frac{1}{2}\sum_j(w_j-v_j)+\frac{1}{2}(z_1-z_2+z_3-z_4))\,,\nonumber\\
u_2 &=&\frac{1}{2}\sum_i
(x_i+y_i)-\frac{1}{2}\sum_j(w_j+v_j)+\frac{1}{2}(-z_1-z_2+z_3+z_4))-\sum_a
r_a\,, \nonumber\\
v &=& \frac{1}{\sqrt{2}}(\sum_i(x_i+y_i)+\sum_j(w_j+v_j)-\sum_b s_b)\,.
\end{eqnarray}

To cancel the theta functions in the second line above we choose the gauge
condition:
\begin{equation}
\sum_l p_l \equiv \sum_a r_a +\sum_b s_b = \sum_i y_i +\sum_j v_j +z_2 +
z_3 + 2 \Delta\,.
\end{equation}
We can now perform the spin structure sum as before with the result that
$(u_1,u_2,v)$ are replaced by $(u'_1,u'_2,v')$ where
\begin{eqnarray}
u'_1&=&\sum_i x_i-z_2-\Delta, ~~~ u'_2= \sum_a r_a+\sum_j
w_j-z_3-\Delta,\nonumber\\v'&=& \frac{1}{\sqrt{2}}(\sum_b s_b-2\sum_j
v_j-z_1-z_2-z_3+z_4-2\Delta)\,.
\end{eqnarray}
Multiplying the resulting expression by identity (due to the gauge condition)
\begin{equation}
1= \frac{\vartheta(\sum_i y_i-z_1-\Delta)}{\vartheta(\sum_a r_a + \sum_b s_b
-\sum_j v_j-z_1-z_2-z_3-3\Delta)}\,,
\end{equation}
and proceeding as before we find
\begin{eqnarray}
&& \hskip -0.7cm (\mathcal{F}_{g,n})_{A_1 A_2 A_3 A_4} \!=\!
\int_{{\cal{M}}_{g,n}}\!\! (\mu\cdot b)^{3g-3+n}\frac{\det(\omega_i(x_j))
\det(\omega_i(y_j))}{\det(\rm{Im}\tau)^2}\langle\prod_{a=1}^{g-n}
G^-_{T^2}(r_a) \prod_{j=1}^n\bar{\psi}_3(w_j) \psi_3(z_3)\rangle_{T^2}
\nonumber\\&&
\frac{\langle\prod_{b=1}^{2g+n} G^-_{K3}(s_b)
e^{-\frac{i}{\sqrt{2}}H}\hat{V}_{A_4}(z_4)
\prod_{k=1}^{3}  e^{+\frac{i}{\sqrt{2}}H}\hat{V}_{A_k}(z_k)\prod_{j=1}^n
(c e^{i \sqrt{2} H})(v_j)\rangle_{K3}}{\langle\prod_{l=1}^{3g+n}
b(p_l)\prod_{j=1}^n c(v_j) \prod_{k=1}^3 c(z_k)\rangle_{b,c}}\,.
\label{FgA4n}
\end{eqnarray}
One can check that all the dimensions are correct: at $w_j$ we have
$\bar{\psi}_3$ that has dimension $1$ in twisted theory and at $v_j$ in
the numerator we have $c J^{++}_{K3}$ that has dimension $-1$ (since
$J^{++}_{K3}$ in twisted theory has dimension $0$) while in the
denominator also there is dimension $(-1)$ so the total dimension at $v_j$
is zero. We can again take three of the PCO positions to $(z_1,z_2,z_3)$
converting these three operators into $(0)$-picture vertex operators of
hyper moduli.
Finally, in the denominator, the remaining $p_l$ have zeroes when they
approach the $(3g-4+n)$ other $p$'s but have a simple pole when they
approach $n$ $v_j$. Given that it is a meromorphic quadratic differential
with divisor class $4g-4$ it must have $g$ other zeroes say at points
$q_m$ with $m=1,...,g$. The  points $q_m$ are uniquely determined as a
function of other $p$'s and $v_j$. Now the numerator also (after taking
all partitions of $p_l$ into $r_a$ and $s_b$)  as a function of $p_l$ has
zeroes at other $(3g-3+n)$ $p$'s and simple poles at $v_j$ as seen from
the OPE of $G^-_{K3}$ with $J^{++}_{K3}$. This means that zeroes and poles
exactly cancel and the result is independent of $p_l$. We can therefore
move the remaining $(3g-3+n)$ $p_l$ to the Beltrami differentials which
results in the cancellation of $(b,c)$ correlators between numerator and
denominator with the result:
\begin{equation}
(\mathcal{F}_{g,n})_{A_1 A_2 A_3 A_4} = D_{f^{+A_1}} D_{f^{+A_2}}
D_{f^{+A_3}} (\mathcal{F}_{g,n})_{ A_4}
\end{equation}
where
\begin{eqnarray}
(\mathcal{F}_{g,n})_{ A_4}&=& \int_{{\cal{M}}_{(g,n)}}
\int_{z_4}\langle(\mu\cdot G^-_{T^2})^{g-n} (\mu\cdot G^-_{K3})^{2g-3+2
n}\prod_{j=1}^n[\int_{w_j} (\bar{\psi}_3 \bar{J}(w_j)(J^{++}_{K3}
\bar{J})(v_j)]
\nonumber\\ &~& (e^{-i\frac{1}{\sqrt{2}} H}
\hat{V}_{A_4})(z_4) \psi_3(p)(\rm{det}Q)^2\rangle\,.
\end{eqnarray}
We can further  simplify this expression as before by writing one of the
$G^-_{K3}$ as contour integral
of $\tilde{G}^+_{K3}$ around $J^{--}_{K3}$ and deforming the contour. The
final result is:
\begin{equation}
(\mathcal{F}_{g,n})_{ A^+_4}=D_{f^{+A_4} } \mathcal{F}_{g,n}\,,
\end{equation}
where $\mathcal{F}_{g,n}$ is the topological amplitude :
\begin{equation}
\mathcal{F}_{g,n}=\int_{{\cal{M}}_{g,n}}\! \langle(\mu\cdot
G^-_{T^2})^{g-n}(\mu\cdot G^-_{K3})^{2g-4+2n}(\mu\cdot J^{--}_{K3})
\prod_{j=1}^n[\int_{w_j} (\bar{\psi}_3 \bar{J}(w_j)(J^{++}_{K3}
\bar{J})(v_j)]\psi_3(p)(\rm{det}Q)^2 \rangle
\end{equation}
Note that there are two sets of $n$ operator insertions here as compared
to $F_g$: at $w_j$ there
are dimension one operators $\bar{\psi}_3 $ which are integrated while at
$v_j$ we have dimension $0$ operators $J^{++}_{K3}$, integration over
$v_j$ is provided by the additional $n$ Beltrami differentials.
In the above we have not explicitly shown the right moving sector (apart
from the Kac-Moody currents)
to avoid complicating the equations. But the right moving sector is the
standard bosonic string correlator
including $(\bar{b},\bar{c})$ ghost system as well as space-time part.

In this Appendix we have mainly focussed on the heterotic string
calculation but it can be extended to the type II
case in a straightforward way by treating the right moving sector exactly
as the left moving sector discussed here.
This would then give a correct derivation of the relation between the
string amplitude and the topological
quantity in the type II side discussed in \cite{Antoniadis:2006mr}. An
involution of the latter gives a derivation
of the result in open (Type I) string quoted in Section 2.

%%%%%%%%%%%%%%%%%%%%%%%%%%%%%%%%%%%%%%%%%%%%%%%%

\end{document}